\documentclass[lettersize,journal]{IEEEtran}
\usepackage{amsmath,amsfonts}
\usepackage{algorithm}
\usepackage{array}
\usepackage{textcomp}
\usepackage{stfloats}
\usepackage{hyperref}
\usepackage{url}
\usepackage{verbatim}
\usepackage{graphicx}
\usepackage{svg}
\usepackage{cite}
\usepackage{amsmath,amssymb,amsfonts}
\usepackage{bm}
\usepackage{orcidlink}
\usepackage[table,xcdraw]{xcolor}
\usepackage{subcaption}
\usepackage{algorithm}
\usepackage{algpseudocode} 
\algrenewcommand\algorithmicrequire{\textbf{Input:}}
\algrenewcommand\algorithmicensure{\textbf{Output:}}
\algnewcommand\algorithmicparameters{\textbf{Parameters:}}
\algnewcommand\Parameters{\item[\algorithmicparameters]}
\usepackage{float} 
\usepackage{booktabs}

\usepackage{draftwatermark}
\SetWatermarkText{IEEE/ASME Transactions on Mechatronics}
\SetWatermarkScale{0.07}
\SetWatermarkLightness{0.4} 
\SetWatermarkAngle{0} 
\SetWatermarkHorCenter{300pt} 
\SetWatermarkVerCenter{10pt}

\makeatletter
\newenvironment{supplementary}[1][Supplementary Material]{%
  \clearpage
  \onecolumn
  \pagenumbering{arabic}\setcounter{page}{1}%
  \begingroup
  \section*{#1}%
  \setcounter{section}{0}\setcounter{subsection}{0}%
  \setcounter{figure}{0}\setcounter{table}{0}\setcounter{equation}{0}%
  \setcounter{algorithm}{0}
  \renewcommand{\thesection}{S\arabic{section}}%
  \renewcommand{\thesubsection}{S\arabic{section}.\Alph{subsection}}%
  \@addtoreset{figure}{section}%
  \@addtoreset{table}{section}%
  \@addtoreset{equation}{section}%
  \renewcommand{\thefigure}{S\arabic{section}.\arabic{figure}}%
  \renewcommand{\thetable}{S\arabic{section}.\arabic{table}}%
  \renewcommand{\theequation}{S\arabic{section}.\arabic{equation}}%

  \@addtoreset{algorithm}{section}

  \@ifundefined{theHsection}{}{%
    \renewcommand{\theHequation}{supp.S\arabic{section}.\Alph{subsection}.\arabic{equation}}%
  }%
  \@ifundefined{theHalgorithm}{}{%
}%
}{%
  \par\bigskip
  \endgroup
}
\makeatother

\begin{document}

\title{A Sliced Learning Framework for Online Disturbance Identification in Quadrotor SO(3) Attitude Control}

\author{Tianhua Gao$^{1}$\,\orcidlink{0009-0008-5929-3545},~\IEEEmembership{Graduate Student Member,~IEEE}, Masashi Izumita$^{2}$, Kohji Tomita$^{3}$\,\orcidlink{0000-0001-9796-0443},~\IEEEmembership{Member,~IEEE}, Akiya Kamimura$^{4}$\,\orcidlink{0000-0002-0046-5639},~\IEEEmembership{Member,~IEEE}
\thanks{
The authors are with the Intelligent Systems Research Institute, National
Institute of Advanced Industrial Science and Technology (AIST), Japan (\{$^{2}$m.izumita, $^{3}$k.tomita, $^{4}$kamimura.a\}@aist.go.jp). $^{1}$T. Gao is also with
the Graduate School of
Systems and Information Engineering, University of Tsukuba, Japan ($^{1}$gao.tianhua@ieee.org, gao.tianhua.tkb\_gb@u.tsukuba.ac.jp). Corresponding author: Akiya Kamimura.
Contact address:
Tsukuba Central2, 1-1-1 Umezono, Tsukuba, Ibaraki 305-8568, Japan. Phone:
+81 80-2309-1517.

This paper has been accepted for publication in IEEE/ASME Transactions on Mechatronics (TMECH) on April 23, 2026.
}}

\maketitle

\begin{abstract}
This paper introduces a dimension-decomposed geometric learning framework called \textit{Sliced Learning} for disturbance identification in quadrotor geometric attitude control. Instead of conventional \textit{learning-from-states}, this framework adopts a \textit{learning-from-error} strategy by using the Lie-algebraic error representation as the input feature, enabling axis-wise space-decomposition (\textit{``slicing"}) while preserving the $\mathbf{SO}(3)$ structure. This is highly consistent with the geometric mechanism of cognitive control observed in neuroscience, where neural systems organize adaptive representations within structured subspaces to enable cognitive flexibility and efficiency.
Based on this framework, we develop a lightweight and structurally interpretable \textit{Sliced Adaptive-Neuro Mapping} (SANM) module.
The high-dimensional mapping for online identification is axially ``sliced" into multiple low-dimensional submappings (\textit{``slices"}), implemented by shallow neural networks and adaptive laws. These neural networks and adaptive laws are updated online via Lyapunov-based adaptation within their respective shared subspaces. To enhance interpretability, we prove exponential convergence despite time-varying disturbances and inertia uncertainties.  To our knowledge, \textit{Sliced Learning} is among the first frameworks to demonstrate lightweight online neural adaptation at 400 Hz on resource-constrained microcontroller units (MCUs), such as \textit{STM32}, with real-world experimental validation.
\end{abstract}


\section*{Supplementary Videos}

The supplementary videos of this work are available online at:
\url{https://www.youtube.com/watch?v=wadR-C_ZXIU&list=PLH6K4qEzDA9ATy3C-jUomYAyk1xR_ZLjK}. For further information, please contact T. Gao by email.
Website:
\url{https://tianhuagao.github.io/}.

\section{Introduction}

\IEEEPARstart{Q}{uadrotor} attitude control has received considerable attention given its essential role in ensuring stable flight performance. Due to the underactuated nature \cite{2018 A review of quadrotor: An underactuated mechanical system} of quadrotors and the coupling between position and attitude dynamics \cite{2019 Position and attitude control of multi-rotor aerial vehicles: A survey}, attitude control with robustness to complex disturbances is crucial for accurate trajectory tracking and overall flight stability. However, the attitude dynamics of a quadrotor inherently evolves on a nonlinear differential manifold known as the special orthogonal group $\mathbf{SO}(3)$. This feature poses significant challenges in attitude control since the topology of the state space $\mathbf{SO}(3)$ precludes the existence of globally asymptotically stable equilibrium points under continuous feedback control \cite{1998 Application of a new Lyapunov function to global
adaptive attitude tracking, 2000 A topological obstruction to continuous global stabilization of rotational motion and the unwinding phenomenon}. 

In existing approaches, Euler angle methods \cite{2020 Learning-Based Robust Tracking Control of Quadrotor With Time-Varying and Coupling Uncertainties}, \cite{2021 Adaptive Integral Sliding Mode Control Using Fully Connected Recurrent Neural Network for Position and Attitude Control of Quadrotor}, \cite{2023 Discrete-Time Adaptive Neural Tracking Control and Its Experiments for Quadrotor Unmanned Aerial Vehicle Systems}, \cite{2023 Active Wind Rejection Control for a Quadrotor UAV Against Unknown Winds}, \cite{2023 Antisaturation fixed-time attitude tracking control based low-computation
learning for uncertain quadrotor UAVs with external disturbances}, \cite{2024 Predictor-Based Neural Attitude Control of A Quadrotor With Disturbances}, \cite{2025 Antiwindup Finite-Time Attitude Control for a Quadrotor System: Multistage Semiimplicit Euler Implementation} are widely used due to their intuitive physical interpretation and simplicity in implementation. However, Euler angles inherently suffer from singularities, commonly known as gimbal lock, where two of the three rotation axes align and one degree of rotational freedom is lost. To overcome this limitation, quaternions have been widely adopted as an alternative representation that avoids the singularities associated with Euler angles. However, quaternion methods (e.g., \cite{2015 Quaternion-Based Robust Attitude Control for Uncertain Robotic Quadrotors}, \cite{2022 Performance Precision and Payloads Adaptive Nonlinear MPC for Quadrotors}, \cite{2025 Fixed-Time Disturbance Observer-Based MPC Robust Trajectory Tracking Control of Quadrotor}) introduce their own limitation, known as double coverage, where two different quaternions represent the same orientation. This ambiguity can cause the so-called unwinding phenomenon \cite{2000 A topological obstruction to continuous global stabilization of rotational motion and the unwinding phenomenon}, in which the control system unnecessarily performs large-angle rotations instead of following the shortest path. Therefore, alternative coordinate-free approaches are required to effectively resolve these issues. 

In contrast, geometric methods \cite{2010 Geometric tracking control of a quadrotor UAV on SE(3)}, \cite{2011 Geometric tracking control of the attitude dynamics of a rigid body on SO(3)}, \cite{2013 Geometric nonlinear PID control of
a quadrotor UAV on SE(3)}, \cite{2013 Robust Adaptive Attitude Tracking on SO3 With an Application to a Quadrotor UAV}, \cite{2021 Geometric Adaptive Control With Neural Networks
for a Quadrotor in Wind Fields}, \cite{2023 Robust observer-based visual servo control for quadrotors tracking unknown moving targets}, \cite{2024 Neural Moving Horizon Estimation for Robust Flight Control}, \cite{2024 Meta-Learning Augmented MPC for Disturbance-Aware Motion Planning and Control of Quadrotors}, \cite{2025 High Maneuverability and Efficiency Control for Hybrid Quadrotor With All-Moving Wings in SE(3) Based on Deep Reinforcement Learning}, \cite{2025 L1Adaptive Augmentation of Geometric Control for Agile Quadrotors With Performance Guarantees} leverage Lie algebra-induced representations of rotational errors on the Lie group $\mathbf{SO}(3)$, thus avoiding the singularities and ambiguities in traditional attitude representations. As a result, geometric control has become a widely adopted baseline framework for advanced aerial robotic systems \cite{2026 Hand-like autonomous flying robot for airborne grasping and interaction}. Due to these advantages, the current state of the art has increasingly focused on enhancing the robustness and adaptivity of geometric control methods.

In \cite{2013 Robust Adaptive Attitude Tracking on SO3 With an Application to a Quadrotor UAV}, T. Lee proposed a robust adaptive attitude tracking control that achieves asymptotic attitude tracking without requiring prior knowledge of the inertia matrix. To compensate for wind-generated aerodynamic disturbances,  M. Bisheban \textit{et al.} \cite{2021 Geometric Adaptive Control With Neural Networks for a Quadrotor in Wind Fields} further developed a geometric adaptive control using multilayer neural networks. In \cite{2024 Neural Moving Horizon Estimation for Robust Flight Control}, B. Wang \textit{et al.} adopted a multilayer perceptron
(MLP) network for the existing geometric baseline controller, and the feasibility was verified by numerical and physical experiments. The MLP architecture is not guided by Lyapunov-based design principles and therefore does not offer theoretical guarantees or interpretability with respect to system stability. Furthermore, the representational capacity of feedforward neural networks depends on both their ``width" (i.e., the number of neurons in a hidden layer) and ``depth" (i.e., the number of layers), while depth has been shown to be exponentially more valuable \cite{2016 The Power of Depth for Feedforward Neural Networks}. Therefore, recent works such as \cite{2025 High Maneuverability and Efficiency Control for Hybrid Quadrotor With All-Moving Wings in SE(3) Based on Deep Reinforcement Learning} leveraged deep neural networks (DNN) for better representation ability. However, the data collection and inference processes involved in training DNN are inherently opaque and structurally complex, which makes these approaches essentially black-box in nature. This lack of transparency raises significant concerns \cite{2019 Stop explaining black box machine learning models for high stakes decisions and use interpretable models instead} about interpretability and trustworthiness, particularly in safety-critical applications. In addition, although these methods \cite{2021 Geometric Adaptive Control With Neural Networks for a Quadrotor in Wind Fields}, \cite{2024 Neural Moving Horizon Estimation for Robust Flight Control}, \cite{2025 High Maneuverability and Efficiency Control for Hybrid Quadrotor With All-Moving Wings in SE(3) Based on Deep Reinforcement Learning}  are built upon geometric control, the neural network input relies on quadrotor states such as Euler-angle coordinates, which break the $\mathbf{SO}(3)$ geometric structure and may reduce representational efficiency. Recent works \cite{2024 Learning to Fly in Seconds}, \cite{2025 Equivariant Reinforcement Learning Frameworks for Quadrotor Low-Level Control} also explore reinforcement learning for quadrotor control, but focus on policy optimization. Consequently, a structurally interpretable and geometry-consistent learning approach with effective representation is warranted in the state of the art.

In this paper, we propose a geometric learning framework, termed \textit{Sliced Learning}, designed to provide structural interpretability and sufficient representational capacity, while enabling lightweight online learning for unseen disturbances. 
This approach is inspired by the Lie-algebra–based error representation in geometric control \cite{2010 Geometric tracking control of a quadrotor UAV on SE(3)} and its consistency with the geometric mechanisms observed in biological neural systems \cite{2019 Cortical Areas Interact through a Communication Subspace}, \cite{2019 High-dimensional geometry of population responses in visual cortex}, \cite{2025 Multiplexed subspaces route neural activity across brain-wide networks}. A key property of this formulation is that
attitude and angular velocity errors are naturally mapped from the Lie algebra $\mathfrak{so}(3)$ to the Euclidean space $\mathbb{R}^3$. This mapping reveals a structural property that enables the representation of rotational error dynamics in a vector space, which motivates using these geometric errors directly as neural network inputs. This, in turn, enables an axis-wise decomposition into independent, parallel, and low-dimensional learning processes—a subspace neural adaptation strategy that aligns closely with the multiplexed subspace networks and geometry-dependent routing of neural activity recently reported in large-scale mouse cortex recordings \cite{2025 Multiplexed subspaces route neural activity across brain-wide networks}. To realize this idea, the high-dimensional mapping for disturbance identification is ``sliced" into multiple low-dimensional submappings (\textit{``slices"}, as shown in Fig.~\ref{SANM_Structure}), each solved by a shallow neural network (SNN). We then demonstrate the extensibility of the \textit{Sliced Learning} framework by constructing adaptive laws directly within the axis-wise shared subspaces, thereby establishing an innovative \textit{Sliced Adaptive-Neuro Mapping} (SANM) module.
Our contributions in this work are summarized as follows:

 \textbf{\textit{(1)}} Proposed a \textit{Sliced Learning} framework paradigm with the following novel properties:
 \begin{itemize}
 \item \textbf{\textit{$\mathbf{SO}(3)$-Preserving}}-In contrast to conventional \textit{learning-from-states} (e.g., Euler angle \cite{2021 Geometric Adaptive Control With Neural Networks
for a Quadrotor in Wind Fields}, \cite{2024 Neural Moving Horizon Estimation for Robust Flight Control}), this framework adopts a \textit{learning-from-error} strategy by using the Lie-algebraic error representation as the input feature to preserve the $\mathbf{SO}(3)$ structure.
\item \textbf{\textit{Geometry Consistency}}-The Lie-algebraic error ensures that \textit{``slices"} operate in a subspace 
aligned with the tangent space of $\mathbf{SO}(3)$. Since $\mathfrak{so}(3)\cong\mathbb{R}^3$, the three 
axis-aligned 1-dimensional subspaces of the tangent space provide natural 
locations for placing the \textit{``slices"}, enabling dimension-decomposed 
geometric learning on the manifold.
 \end{itemize}

 \textbf{\textit{(2)}} Developed a SANM-augmented geometric attitude control based on \textit{Sliced Learning} with the following benefits:

\begin{itemize}
\item \textbf{\textit{Universal Identification}}-Since neural network \textit{``slices"} target disturbance features at the acceleration level, the universal approximation theorem \cite{1989 Multilayer feedforward networks are universal approximators} ensures that they can theoretically capture any continuous nonlinear acceleration effect in a compact domain (e.g., multi-agent coupling effects \cite{2025 Robustness Enhancement for Multi-Quadrotor Centralized Transportation System via Online Tuning and Learning}, \cite{2025 Online Identification using Adaptive Laws and Neural Networks for Multi-Quadrotor Centralized Transportation System}). Combined with the Lyapunov-based online adaptation, this theoretical universality further guarantees 
bounded weight estimation in unseen environments without requiring the persistent excitation (PE) condition 
or any offline training. 
\item \textbf{\textit{Efficient Learning}}-The \textit{$\mathbf{SO}(3)$-Preserving} property provides each neural network \textit{``slice"} with geometry-consistent inputs that efficiently encode the underlying rotational geometry, thereby reducing the learning complexity.
\item \textbf{\textit{Flexibility}}-The \textit{Geometry Consistency} property provides independent and parallel online neural adaptation, thereby allowing each \textit{``slice"} to be enabled or disabled as needed, and each \textit{``slice"} can be individually customized. 
\item \textbf{\textit{Exponential Convergence}}-SANM ensures almost-global exponential attractiveness of the rotational error dynamics to a bounded residual set, and local exponential convergence to an arbitrarily small ball, even under bounded time-varying disturbances and inertia uncertainties.
\item \textbf{\textit{Axis-wise Tunable Convergence Rate}}-Since the exponential convergence rate is governed by parameters such as learning rates, SANM allows axis-wise adjustment of the rotational convergence rate. This tunability enables preferential moment allocation to a specific axis, making direction-dependent disturbance handling feasible.
\item \textbf{\textit{Structural Interpretability}}-A rigorous Lyapunov analysis that explicitly considers neural network approximation errors supports the structural interpretability.
\item \textbf{\textit{Lightweight}}-Without relying on ground computers or high-performance onboard processors 
(e.g., \textit{NVIDIA Jetson}), SANM can run in real time at 400 Hz on microcontroller units (MCUs), such as \textit{STM32} processors.
\item $\mathbf{SE}(3)$-\textbf{\textit{Compatibility}}-SANM (for $\mathbf{SO}(3)$) can be integrated into the existing geometric control on $\mathbf{SE}(3)$\cite{2010 Geometric tracking control of a quadrotor UAV on SE(3)}. 
\end{itemize}

A comprehensive comparison with representative geometric controllers \cite{2013 Robust Adaptive Attitude Tracking on SO3 With an Application to a Quadrotor UAV},\cite{2021 Geometric Adaptive Control With Neural Networks
for a Quadrotor in Wind Fields}, \cite{2025 L1Adaptive Augmentation of Geometric Control for Agile Quadrotors With Performance Guarantees} is provided in Table~\ref{tab:comparison_with_SOTA}.

\textbf{\textit{(3)}} Proved the almost-global exponential convergence and \emph{Input-to-State Practical Stability (ISpS)} under sampled-data (discrete-time) implementation.

\textbf{\textit{(4)}} Conducted comprehensive validation through numerical simulation 
(\textit{MATLAB Simulink}), high-fidelity physics simulation (\textit{Gazebo Harmonic}), 
testbed ablation and variant experiments (\textit{Real-world}), and flight experiments (\textit{Real-world}).

This paper is organized as follows. Section \uppercase\expandafter{\romannumeral 2} describes the problem formulation. Section \uppercase\expandafter{\romannumeral 3} introduces \textit{Sliced Learning} and the design of SANM-augmented attitude controller. Section \uppercase\expandafter{\romannumeral 4} presents the results of the simulation and real-world experiments. Finally, Section \uppercase\expandafter{\romannumeral 5} concludes the paper and discusses future work. The stability proof is provided in the supplementary material, along with additional figures, tables, implementation guidance, video resources, and other supplementary information.

\section{Problem Formulation}
\subsection{Attitude Kinematics and Dynamics with Disturbances}

\begin{figure}[!t]
      \centering
      \includegraphics[scale=0.19]{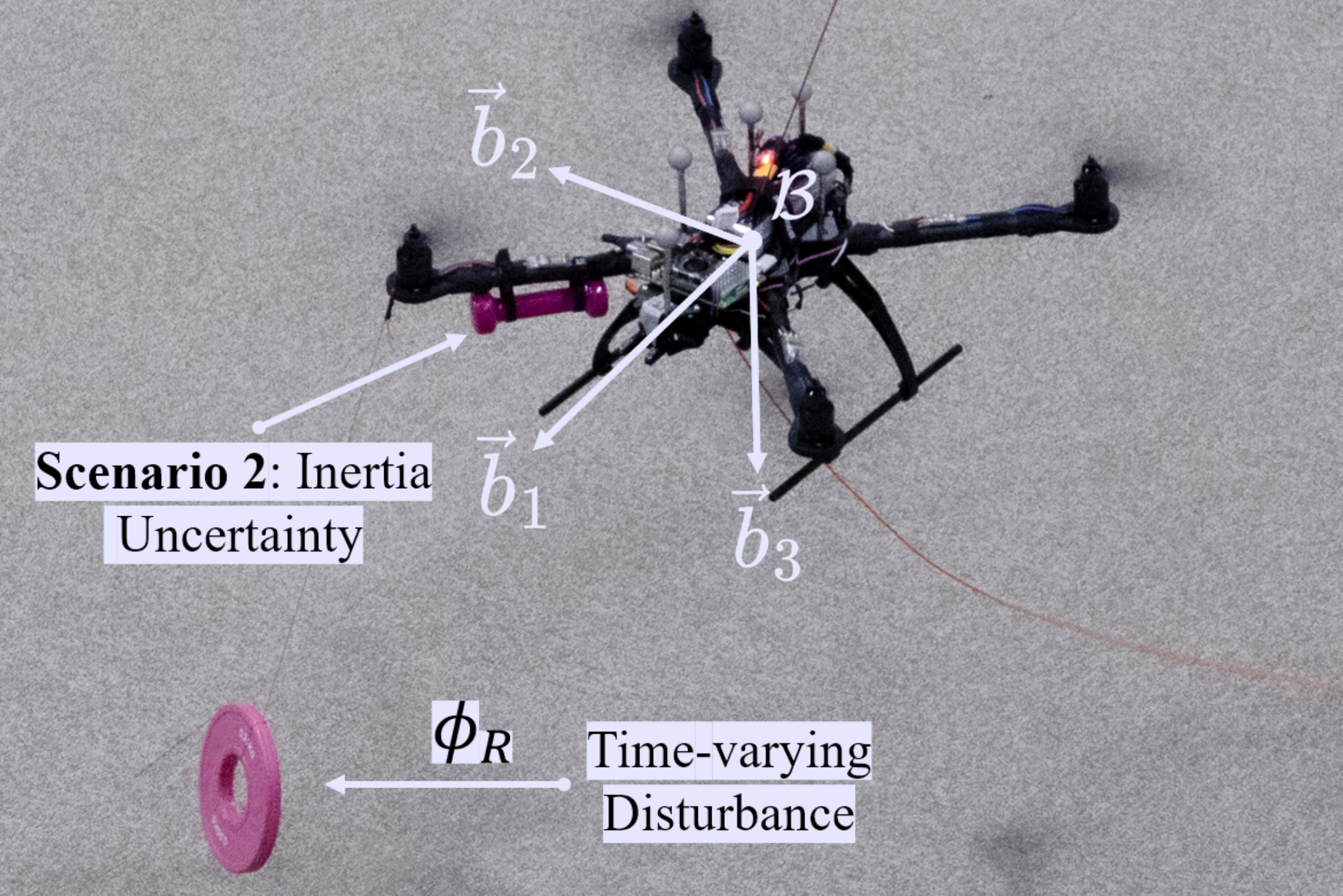}
      \caption{\footnotesize Quadrotor under time-varying disturbances and unknown inertia. }
      \label{Quadrotor}
\end{figure}

This subsection introduces the attitude kinematics and dynamics of the quadrotor augmented with disturbances. The orientation of the quadrotor is defined in a North-East-Down (NED) body-fixed frame $\mathcal{B}\triangleq\{\bm{\vec{b}}_{j}\}_{j=1,2,3}$, fixed at the center of mass of the rigid body structure, as illustrated in Fig.~\ref{Quadrotor}. The attitude is represented by a rotation matrix $\bm{R}\in \mathbf{SO}(3) = \{\bm{R}\in\mathbb{R}^{3\times3}\mid\bm{R}^{\top}\bm{R} = \mathbf{I}^{3\times3}, \mathrm{det}(\bm{R}) = 1\}$, which describes the rotation of $\mathcal{B}$ relative to an inertial reference. 
For disturbance modeling, we consider two scenarios.

\textbf{\textit{Scenario 1: ($\bm{J}$ is known)}} If the inertia tensor $\bm{J}\!\in\!\mathbb{R}^{3\times3}$ is known, we augment the standard attitude dynamics with unknown time-varying dynamics  term of rotational disturbance, $\bm{\phi}_{\bm{\textit{R}}}\in \mathbb{R}^3$ at the acceleration level:
\begin{equation}
    {
    \begin{aligned}
        \bm{\dot{R}}= \bm{R}[\bm{\Omega}]_{\times},
    \end{aligned}
    }
\end{equation}
\begin{equation}
    {
    \begin{aligned}
       \bm{\dot{\Omega}} = \bm{J}^{\scalebox{0.6}{$-1$}}\left(\bm{\mathrm{M}}-[\bm{\Omega}]_{\times}\bm{J}\bm{\Omega}\right)+\bm{\phi}_{\bm{\textit{R}}},
    \end{aligned}
    }
    \label{Dynamics with Augmented Disturbance}
\end{equation}
where $\bm{\Omega}\in\mathbb{R}^3$ is the angular velocity in the body-fixed frame and $\bm{\mathrm{M}}\in\mathbb{R}^3$ denotes the control moment. 

\textbf{\textit{Notation 1: (Skew-symmetric Map) }}
The symbol $[\,\bullet\,]_{\times}\!:\!\mathbb{R}^3\!\to\!\mathfrak{so}(3)$ represents the skew-symmetric map defined by the condition that $[\mathfrak{a}]_{\times}\mathfrak{b}=\mathfrak{a}\times\mathfrak{b}, \forall\mathfrak{a},\mathfrak{b}\in\mathbb{R}^3$.

\textbf{\textit{Scenario 2: ($\bm{J}$ is unknown)}} If the inertia tensor is unknown (see Fig.~\ref{Quadrotor}), the term $\bm{J}^{-1}[\bm{\Omega}]_{\times}\bm{J}\bm{\Omega}$ in Eq.~\eqref{Dynamics with Augmented Disturbance} cannot be compensated for in the attitude control introduced later in Eq.~\eqref{Md}. However, since this term also represents an unknown time-varying dynamic component, we can treat it as an internal disturbance and incorporate it into the  $\bm{\phi}_{\bm{\textit{R}}}$ term:
\begin{equation}
{
\begin{aligned}
     \bm{\dot{\Omega}} = \bm{J}^{\scalebox{0.6}{$-1$}}\bm{\mathrm{M}}+\bm{\phi}_{\bm{\textit{R}}}(\bm{J}, \bm{\Omega}),
\end{aligned}
}
    \label{Dynamics with Augmented Disturbance2}
\end{equation}
where $\bm{\phi}_{\bm{\textit{R}}}(\bm{J}, \bm{\Omega})$ represents a universal rotational disturbance term that absorbs the acceleration-level effects induced by both inertia uncertainties and external disturbances.

\textbf{\textit{Remark 1: (Universal Disturbance)}}
Since $\bm{\phi}_{\bm{\textit{R}}}(\cdot)$ acts at the acceleration level, it serves as a universal disturbance term in theory, capable of representing any continuous nonlinear unknown acceleration effect.
In such cases where dynamics are not explicitly known, recent studies (e.g., \cite{2025 Robustness Enhancement for Multi-Quadrotor Centralized Transportation System via Online Tuning and Learning}, \cite{2025 Online Identification using Adaptive Laws and Neural Networks for Multi-Quadrotor Centralized Transportation System}, \cite{2025 Kinematics-informed neural network control on SO(3)}) have employed neural networks to implicitly capture dynamic information. In this study, we also employ neural networks in a similar vein to approximate the universal disturbance term $\bm{\phi}_{\bm{\textit{R}}}(\bm{J}, \bm{\Omega})$. 

\subsection{Attitude Control Problem Formulation on $\mathbf{SO}(3)$}
 The complete quadrotor geometric control pipeline on $\mathbf{SE}(3)$ inherently couples position control in $\mathbb{R}^3$ and attitude control on $\mathbf{SO}(3)$.
 We first define the mapping deviation of the control moment $\bm{\Delta_{\mathrm{M}}}\in\mathbb{R}^{3}$ as follows:
\begin{equation}
  \bm{\Delta_{\mathrm{M}}}\triangleq \bm{\mathrm{M}}-\bm{\mathrm{M}_d},
   \label{mapping deviations}
\end{equation}
where $\|\bm{\Delta_{\mathrm{M}}}\|$ is bounded (see Supp. \ref{supp:SE(3)}). If the aerodynamic coefficients in the allocation mapping are chosen precisely through experiments, the upper bound $\varepsilon_{\mathbf{M}}\in\mathbb{R}$ of $\|\bm{\Delta_{\mathrm{M}}}\|$ can be regarded as converging to zero. 
 
 In this work, our objective is to design the desired resultant control moment $\bm{\mathrm{M}_d}$, given a desired attitude $\bm{R}_{{\bm{d}}}\in \mathbf{SO}(3)$. The aim is to achieve exponential convergence of the rotational state errors $\bm{e}_{\bm{R}}$, $\bm{e}_{\bm{\Omega}}\in\mathbb{R}^3$ even in the presence of unknown inertia tensor $\bm{J}$  and bounded time-varying rotational disturbance dynamics $\bm{\phi}_{\bm{\textit{R}}}$. 
The rotational state errors of the quadrotor, including the attitude and angular velocity errors $\bm{e}_{\bm{R}}$ and $\bm{e}_{\bm{\Omega}}$, are defined in Lie-algebra-induced representations as follows:
\begin{equation}
\begin{aligned}
     \bm{e}_{\bm{R}}:=&\frac{1}{2}(\bm{R}_{{\bm{d}}}^{\top}\bm{R}-\bm{R}^{\top}\bm{R}_{{\bm{d}}})^\vee , \bm{e}_{\bm{\Omega}}:=\bm{\Omega}-\bm{R}^{\top}\bm{R}_{{\bm{d}}}\bm{\Omega}_{{\bm{d}}},
    \label{errors}
\end{aligned}
\end{equation}
where the desired angular velocity $\bm{\Omega}_{{\bm{d}}}\in\mathbb{R}^3$ is obtained by:
\begin{equation}
    \bm{\Omega}_{{\bm{d}}}:=(\bm{R}^{\top}_{{\bm{d}}}\dot{\bm{R}_{{\bm{d}}}})^{\vee}.
    \label{Omegac}
\end{equation}

\textbf{\textit{Notation 2: (Vee Map)}} $\bullet^\vee: \mathfrak{so}(3)\to\mathbb{R}^3$ denotes the inverse of skew-symmetric map $[\,\bullet\,]_{\times}$. The Lie algebra $\mathfrak{so}(3)$ enables a Euclidean vector representation of rotational errors in $\mathbb{R}^3$.

\section{Sliced Adaptive-Neuro Mapping for Geometric Attitude Control on $\mathbf{SO}(3)$}

\begin{table}[!t]
  \centering
  \caption{List of notations: maps, subscripts and superscripts}
  \begin{tabular}{>{\columncolor{gray!20}}c|l}
    \hline
    \hline
    $[\,\bullet\,]_{\times}$ & Skew-symmetric map: $\mathbb{R}^3\to\mathfrak{so}(3)$\\
   $\bullet^{[\cdot]}$ &  Subspace map: $ (\mathbb{R}^3\cup\mathbb{R}^{3\times 3})\times\mathbb{N}\to\mathbb{R}$\\
   $\bullet^\vee$& Vee map: $\mathfrak{so}(3)\to\mathbb{R}^3$\\
   $\bullet_{\bm{d}}$& Desired value\\
   $\bullet^{\text{vec}}$ & Feature vector formed by diagonal elements of a matrix\\
   $\bar{\bullet}$ & Estimation value\\
   $\widetilde{\bullet}$ & Estimation error\\
   $\bullet^{\mathrm{nom}}$ & Nominal value  \\
   $\bullet^*$ & Optimal value\\
    $\lambda_{\min}(\bullet)$ &Minimum eigenvalue of a matrix \\
  $\lambda_{\max}(\bullet)$ &Maximum eigenvalue of a matrix\\
    \hline
    \hline
    \multicolumn{2}{{p{225pt}}}{
For a list of symbol references, see Table \ref{TABLE_Symbol_References}.

    }
    
  \end{tabular}
  \label{table1}
\end{table}

This section introduces the \textit{Sliced Learning} framework and
the design of the geometric attitude control augmented with the
\textit{Sliced Adaptive-Neuro Mapping} (SANM) module. 


\subsection{Sliced Learning Framework}

This subsection introduces the conceptual perspective that motivates the \textit{Sliced Learning} paradigm.
The existing neural adaptation methods exhibit high-dimensional coupling and often rely on computationally expensive structures such as multilayer perceptron (MLP) networks \cite{2021 Geometric Adaptive Control With Neural Networks
for a Quadrotor in Wind Fields}, \cite{2024 Neural Moving Horizon Estimation for Robust Flight Control}.
Inspired by geometric mechanisms observed in biological neural systems \cite{2019 Cortical Areas Interact through a Communication Subspace}, \cite{2019 High-dimensional geometry of population responses in visual cortex}, \cite{2025 Multiplexed subspaces route neural activity across brain-wide networks}, we are motivated to decompose the high-dimensional disturbance mapping into a set of low-dimensional submappings on axis-wise subspaces, enabling independent and parallel neural identification. 

To formalize this idea, we begin by considering a continuous nonlinear mapping from the desired resultant control moment $\bm{\mathrm{M}_d}$, inertia tensor $\bm{J}$ and unknown rotational disturbance dynamics $\bm{\phi}_{\bm{\textit{R}}}$ to the rotational state error vector $\mathbf{E}_{\bm{\textit{R}}}\triangleq\big{(} \bm{e}^{\top}_{\bm{R}},\bm{e}^{\top}_{\bm{\Omega}}\big{)}^{\top}\!\!\in\!\mathbb{R}^{6}$, denoted as $\mathbf{E}_{\bm{\textit{R}}}\triangleq\bm{\mathcal{S}}(\bm{\mathrm{M}_d},\bm{J},\bm{\phi}_{\bm{\textit{R}}})\!:\!\mathbb{R}^3\times\mathbb{R}^{3\times3}\times\mathbb{R}^3\!\to\!\mathbb{R}^{6}$.

\label{Design of the Adaptive Law Slice}
\begin{figure}[!t]
      \centering
      \includegraphics[scale=0.1]{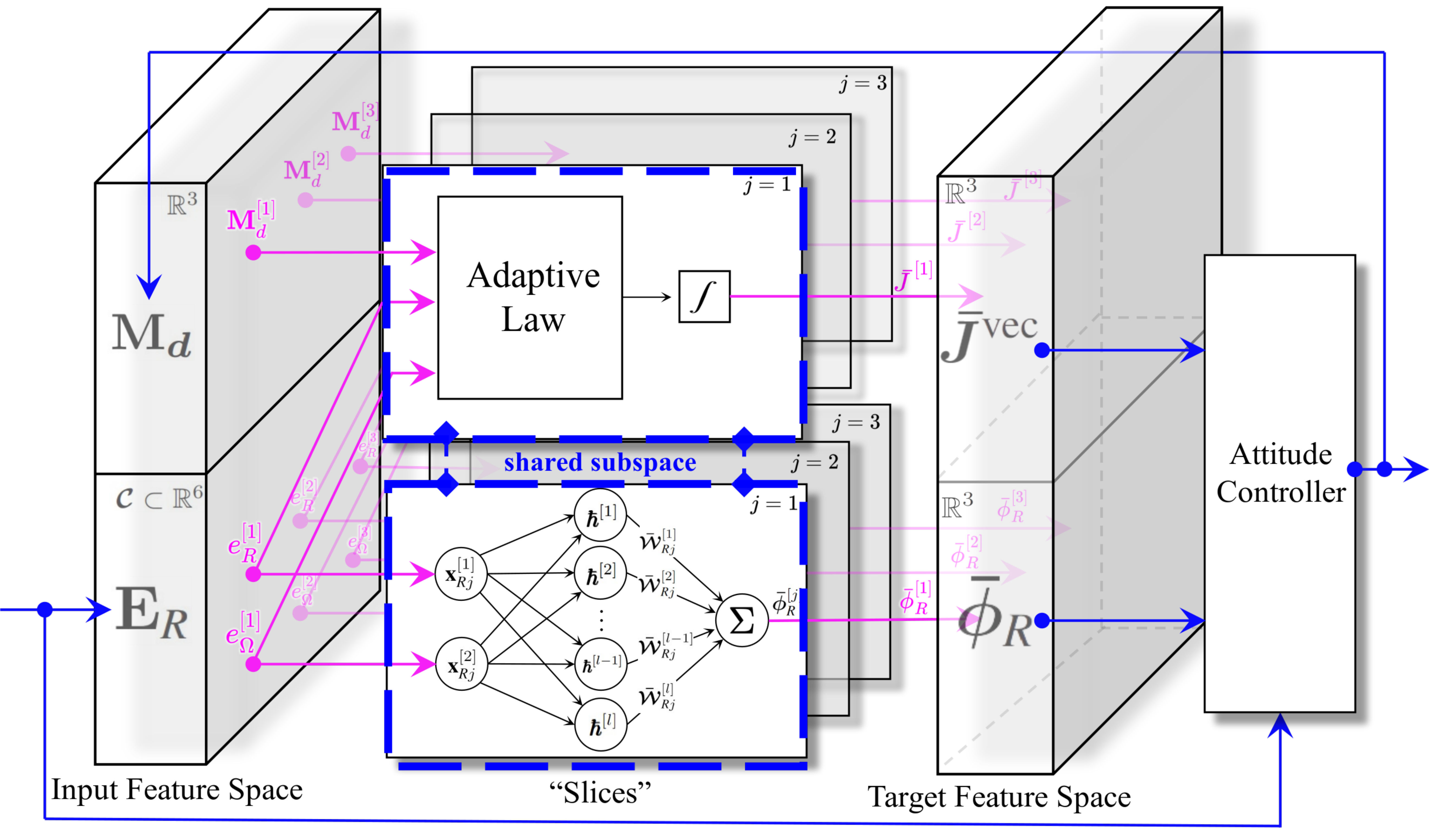}
      \caption{\footnotesize The  structure of \textit{Sliced Adaptive-Neuro Mapping} (SANM) module. The high-dimensional mapping for disturbance and uncertainty identification is axially ``sliced" into multiple low-dimensional submappings (\textit{``slices"}). This module is a clear embodiment of the proposed \textit{Sliced Learning} paradigm. Each \textit{``slice"} learns directly from the geometric error representation induced on $\mathfrak{so}(3)$, thereby preserving the $\mathbf{SO}(3)$ structure while decomposing high-dimensional features into low-dimensional, axis-wise \textit{``slices"}. This yields independent and parallel online neural adaptation, and the axis-aligned subspace structure enables direct \emph{subspace sharing} with other adaptive mechanisms, which in turn provides strong modularity and extensibility. }
      \label{SANM_Structure}
\end{figure}

\textbf{\textit{Assumption 1: (Local Existence of the Pseudo-Inverse Mapping)}}
Since $\bm{\mathcal{S}}$ is not bijective, its global inverse mapping does not exist. Nevertheless, we assume that its pseudo-inverse mapping, denoted as $(\bm{\mathrm{M}_d},\bm{J}^{\text{vec}},\bm{\phi}_{\bm{\textit{R}}})\triangleq\bm{\mathcal{S}}^\dagger(\mathbf{E}_{\bm{\textit{R}}}):\mathcal{C}\to\mathbb{R}^3\times\mathbb{R}^3\times\mathbb{R}^3$, exists locally with the input $\mathbf{E}_{\bm{\textit{R}}}$ bounded in a compact set $\mathcal{C}$ (defined in \textbf{\textit{Proposition 3}}). Here, $\bm{J}^{\text{vec}}\triangleq\left(\bm{J}^{[1]},\bm{J}^{[2]},\bm{J}^{[3]}\right)\in\mathbb{R}^3$ is a vector composed of the principal moments of inertia, obtained by diagonalizing the inertia tensor $\bm{J}$.  This assumption is an existence assumption and is validated through numerical simulation, physics simulation, and real-world experiments in Section \ref{Experiments Validation}.

\textbf{\textit{Notation 3: (Subspace Map)}} The superscript $\bullet^{[\cdot]}$ denotes a subspace map $\bullet^{[\cdot]}: (\mathbb{R}^3\cup\mathbb{R}^{3\times 3})\times\mathbb{N}\to\mathbb{R}$ which extracts the $\cdot^{th}$ axis-wise element from either a vector or the main diagonal of a matrix.

Building upon \textbf{\textit{Assumption~1}} and inspired by neuroscience evidence \cite{2019 Cortical Areas Interact through a Communication Subspace}, \cite{2019 High-dimensional geometry of population responses in visual cortex}, \cite{2025 Multiplexed subspaces route neural activity across brain-wide networks}, we introduce:

  \textbf{\textit{Hypothesis 1: (Pseudo-Inverse Mapping is ``Sliceable'')}} The pseudo-inverse mapping 
$\bm{\mathcal{S}}^{\dagger}(\mathbf{E}_{\bm{\textit R}}):\mathcal{C}\to\mathbb{R}^{3}\times\mathbb{R}^{3}\times\mathbb{R}^{3}$, is \textit{``sliceable"}, i.e., it can be dimensionally decomposed into a set of mutually independent submappings (\textit{``slices"}) along body-frame axes:  
\begin{equation}
\bm{\mathcal{S}}^{\dagger}(\mathbf{E}_{\bm{\textit R}})
= \bigoplus_{j=1}^{3}\bm{\mathcal{S}}_{j}^{\dagger}(\bm{e}_{\bm{R}}^{[j]}, \bm{e}_{\bm{\Omega}}^{[j]}),
\label{eq:sliceable_mapping}
\end{equation}
where each $\bm{\mathcal{S}}_{j}^{\dagger}(\cdot): \mathcal{C}_j \to (\mathbb{R})^{3} $ represents a low-dimensional 
submapping (\textit{``slice"}) corresponding to the $j^{th}$ body-frame axis $\bm{\vec{b}}_{j}$. The $\mathcal{C}_j\subset\mathbb{R}^2$ is a compact axis-wise subset of $\mathcal{C}$.  
This hypothesis follows from the fact that the geometric error representation is locally linearized in $\mathbb{R}^3$, making the axis-wise decomposition natural. Under this decomposition, each \textit{``slice"} is placed on the geometric subspace aligned with the 
$\bm{\vec{b}}_{j}$-axis, which leads to the following hypothesis.

\textbf{\textit{Hypothesis 2: (Subspace Sharing)}} The geometric subspaces aligned with the body-frame axes 
can be \textit{``shared"} to construct additional axis-aligned submappings (\textit{``slices"}).

At this point, the structural foundation of the proposed \textit{Sliced Learning} framework becomes evident, enabling the construction of the following 
\textit{Sliced Adaptive-Neuro Mapping} (SANM) module for axis-wise subspace index $j=1,2,3$:
\begin{equation}
\footnotesize{
\begin{aligned}
\bigoplus_{j=1}^{3}\Bigg{\{}\Big{(}\!\!\!\!\!\!\!\!\!\!\!\!\!\underbrace{\bm{\bar{J}}^{[j]},\bm{\bar{\phi}}_{\bm{\textit{R}}}^{[j]}}_{\textbf{Sliced Target Feature Space}}\!\!\!\!\!\!\!\!\!\!\!\!\!\Big{)}\triangleq \bm{\mathcal{S}}^{AN}_{j}\Big{(}\!\!\underbrace{\bm{\mathrm{M}_d}^{[j]},(\bm{e}_{\bm{R}}^{[j]}, \bm{e}_{\bm{\Omega}}^{[j]})}_{ \textbf{Sliced Input Feature Space}}\!\!\Big{)}\!:\!\mathbb{R}\!\times\!\mathcal{C}_j\!\!\to\!\mathbb{R}\!\times\!\mathbb{R}\Bigg{\}},
\end{aligned}
}\notag
\label{Sliced Adaptive-Neuro Mapping Equation}
\end{equation}
where $\{\bm{\bar{J}}^{[j]}\in\mathbb{R}\}_{j=1,2,3}$ are the estimated values to the moment of inertia along the $\bm{\vec{b}}_{j}$-axis. The  $\{\bm{\bar{\phi}}_{\bm{\textit{R}}}^{[j]}\in\mathbb{R}\}_{j=1,2,3}$ are the estimated dynamics of the rotational disturbance decomposed along  the $\bm{\vec{b}}_{j}$-axis. 
The  structure of this mapping is shown in Fig.~\ref{SANM_Structure}.  The 
$\bm{\bar{J}}^{\text{vec}}\!\!=\!\!\left(\bm{\bar{J}}^{[1]},\bm{\bar{J}}^{[2]},\bm{\bar{J}}^{[3]}\right)^{\top}\!\!\!\!\!\!\in\!\!\mathbb{R}^{3}$ denotes the estimated inertia  feature vector composed of the principal moments of inertia extracted from the estimated diagonalized  inertia tensor $\bm{\bar{J}}\in\mathbb{R}^{3\times3}$.  The  $\bm{\bar{\phi}}_{\bm{\textit{R}}} \!\!=\!\!(\bm{\bar{\phi}}_{\bm{\textit{R}}}^{[1]},\bm{\bar{\phi}}_{\bm{\textit{R}}}^{[2]},\bm{\bar{\phi}}_{\bm{\textit{R}}}^{[3]})^{\top}\!\!\in\mathbb{R}^3$ represents the estimated rotational disturbance dynamics feature vector. The submappings (\textit{``slices"}) in SANM are approximated by adaptive laws and neural networks, whose design is provided in the following subsection. 

\subsection{SANM-augmented Geometric Attitude Control}

\begin{figure}[!t] 
 \centering
 \includegraphics[scale=0.125]{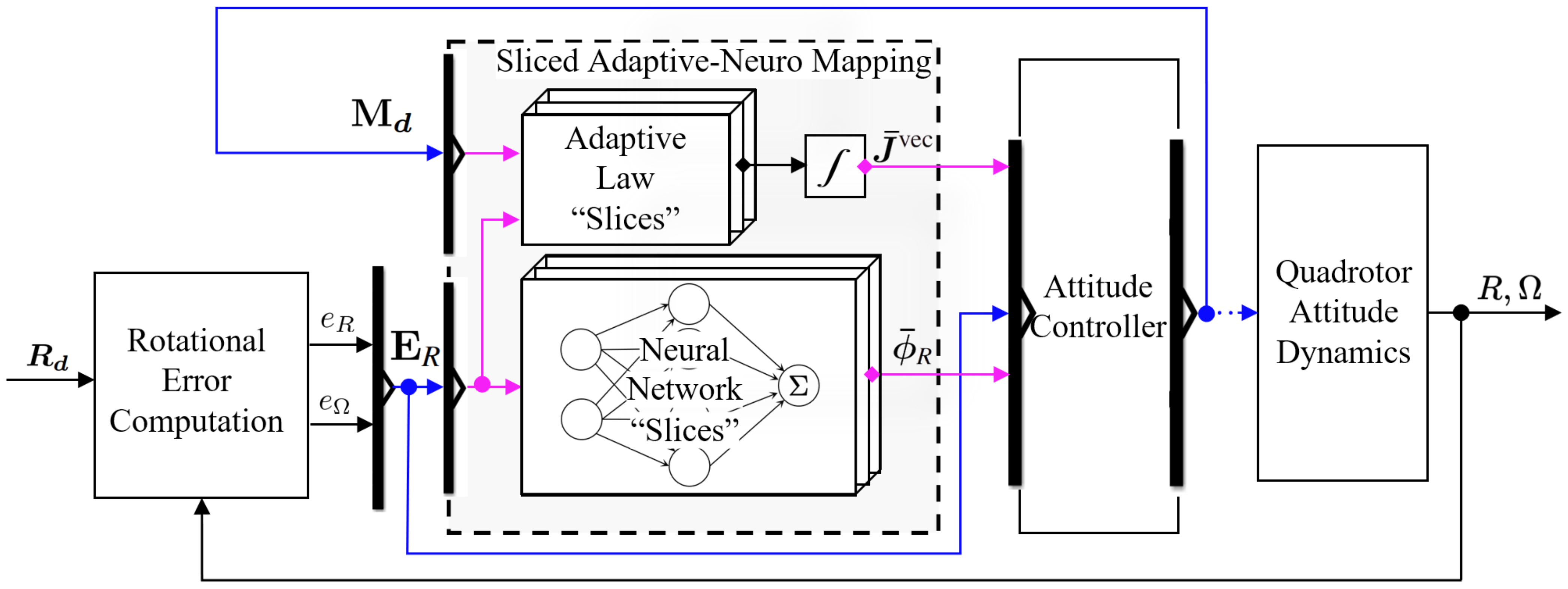}
 \caption{\footnotesize The architecture of the SANM-augmented geometric attitude control system based on \textit{Sliced Learning}. The SANM module  serves as a feedforward compensator for the geometric attitude controller. } 
      \label{Control_System_Structure}
\end{figure}

The architecture of the SANM-augmented closed-loop control system is illustrated in Fig.~\ref{Control_System_Structure}. The design of the attitude controller and \textit{``slices"} is detailed as follows.
\subsubsection{\textbf{Design of the Attitude Controller}}
\label{Design of the Attitude Controller}
First, we consider using the foregoing target features $\{\bm{\bar{J}}^{[j]},\bm{\bar{\phi}}_{\bm{\textit{R}}}^{[j]}\}_{j=1,2,3}$ from SANM to compensate for the disturbances and uncertainties.  To apply these values, 
the desired resultant control moment $\bm{\mathrm{M}_d}$ is axially decomposed into individual components $\{\bm{\mathrm{M}_d}^{[j]}\in\mathbb{R}\}_{j=1,2,3}$. These components are designed as follows:
\begin{equation}
{\small
\begin{aligned}
\bm{\mathrm{M}_d}^{[j]}:=&\bm{\bar{J}}^{[j]}
\Big{\{}-k_{R}\bm{e}^{[j]}_{\bm{R}}-k_{\Omega}\bm{e}^{[j]}_{\bm{\Omega}
}-\left([\bm{\Omega}]_{\times}\bm{R}^{\top}\bm{R}_{{\bm{d}}}\bm{\Omega}_{{\bm{d}}}\right)^{[j]}\\
&\!\!\!+\big{(}\bm{R}^{\top}\bm{R}_{{\bm{d}}}\bm{\dot{\Omega}}_{{\bm{d}}}\big{)}^{[j]}
-\bm{\bar{\phi}}_{\bm{\textit{R}}}^{[j]}+\underbrace{(\bm{J}^{-1}[\bm{\Omega}]_{\times}\bm{J}\bm{\Omega})^{[j]}}_{\text{if $\bm{J}$ is known}}\Big{\}},\\[-20pt]
\label{Md}
\end{aligned}
}
\end{equation}
where $k_{R}$ and $k_{\Omega}\in \mathbb{R}$ are positive gains for rotational Proportional-Derivative (PD) control. In \textbf{\textit{Scenario 1 ($\bm{J}$ is known)}}, the term $(\bm{J}^{-1}[\bm{\Omega}]_{\times}\bm{J}\bm{\Omega})^{[j]}$ can be augmented as a compensation term. In \textbf{\textit{Scenario 2 ($\bm{J}$ is unknown)}}, $(\bm{J}^{-1}[\bm{\Omega}]_{\times}\bm{J}\bm{\Omega})^{[j]}$ is omitted and the neural networks intervene to learn and compensate for the universal disturbance described in Eq.~\eqref{Dynamics with Augmented Disturbance2}, i.e., $\bm{\bar{\phi}}_{\bm{\textit{R}}}^{[j]}\to\bm{\phi}_{\bm{\textit{R}}}(\bm{J}, \bm{\Omega})^{[j]}$.

\subsubsection{\textbf{Design of the $j^{th}$ Adaptive Law ``Slice''}} 

Substituting the above designed $\{\bm{\mathrm{M}_d}^{[j]}\}_{j=1,2,3}$ into the rotational error dynamics in Supp.~\ref{Rotational Error Dynamics}, the adaptive laws for updating the estimated inertia feature vector $\bm{\bar{J}}^{\text{vec}}$ are derived based on Lyapunov analysis in Supp.~\ref{Proofs of Proposition 1 and 2} to ensure system stability.  Its components $\{\bm{\bar{J}}^{[j]}\}_{j=1,2,3}$ are updated online by their respective adaptive laws, following the inherently bounded update rule below:
\begin{equation}
{
\footnotesize
\bm{\dot{\bar{\mathit{J}}}}^{[j]}\!\!:=\!\!
\begin{cases}
   \frac{-\bm{\bar{J}}^{[j]^2}}{\eta_{j}}\!\left(\!\bm{e}^{[j]}_{\bm{\Omega}}\!\!+\!c_R\bm{e}^{[j]}_{\bm{R}}\!\right)\!\bm{\mathrm{M}_d}^{[j]}\!, \,\, \left(\!\bm{e}^{[j]}_{\bm{\Omega}}\!\!+\!c_R\bm{e}^{[j]}_{\bm{R}}\!\right)\bm{\mathrm{M}_d}^{[j]}\!>\!0,\\[2pt]
 \frac{-\bm{\bar{J}}^{[j]^2}}{\eta_{j}}\!\left(\!\bm{e}^{[j]}_{\bm{\Omega}}\!\!+\!c_R\bm{e}^{[j]}_{\bm{R}}\!\right)\bm{\mathrm{M}_d}^{[j]}\!, \,\,\left(\!\bm{e}^{[j]}_{\bm{\Omega}}\!\!+\!c_R\bm{e}^{[j]}_{\bm{R}}\!\right)\!\bm{\mathrm{M}_d}^{[j]}\!\!\leq\! 0, \,\bm{\bar{J}}^{[j]}\!\!<\!\!\overset{\tiny \text{max}}{J}_{j}\\[2pt]
 \,\mathfrak{s}_{j}\frac{-\bm{\bar{J}}^{[j]^2}}{\eta_{j}}, \,\,\,\,\,\,\,\,\,\,\,\,\,\,\,\,\,\,\,\,\,\,\,\,\,\,\,\,\,\,\,\,\,\,\,\,\,\,\,\,\,\,\left(\!\bm{e}^{[j]}_{\bm{\Omega}}\!\!+\!c_R\bm{e}^{[j]}_{\bm{R}}\!\right)\!\bm{\mathrm{M}_d}^{[j]}\!\!\leq\! 0, \,\bm{\bar{J}}^{[j]}\!\!\geq\!\!\overset{\tiny \text{max}}{J}_{j}
\end{cases}
}
\label{Adaptive Law of Inertia Tensor}
\end{equation}    
where the adaptive parameter $\eta_{j}\in\mathbb{R}$ and $c_R\!\in\!\mathbb{R}$ are positive constants. The $1/\eta_{j}$ and $\mathfrak{s}_{j}\in\mathbb{R}$ are the adaptive rate and pull-back factor of the $j^{th}$ adaptive law \textit{``slice"}, respectively.  The constant $\overset{\tiny \text{max}}{J}_{j}\in\mathbb{R}$ denotes the adaptive upper bound imposed on the estimated moment of inertia along the $\bm{\vec{b}}_{j}$-axis. 

The estimation errors of inertia feature components
$\{\widetilde{J}_{j}\}_{j=1,2,3}$ are defined in a reciprocal form: 
\begin{align}
      \widetilde{J}_{j}\triangleq\frac{1}{\bm{J}^{[j]}}-\frac{1}{\bar{\bm{J}}^{[j]}}.
\label{Estimation errors of pseudo model}
\end{align}

\begin{algorithm}[!t]
\caption{SANM-augmented Geometric Attitude Control}
\label{alg:Attitude-Control}
\begin{algorithmic}[1]
\Require Initialized variables from \textbf{Algorithm}~\ref{alg:init}; incoming $\bm{R}_{{\bm{d}}},\bm{R},\bm{\Omega}$ at time step $t$ with step size $dt$
\While{attitude control is active (each $t \leftarrow t + dt$)}
    \State Update $\bm{e}_{\bm{R}}$, $\bm{e}_{\bm{\Omega}}$; compute $\dot{\bm{R}_{{\bm{d}}}},\bm{\Omega}_{{\bm{d}}},\bm{\dot{\Omega}}_{{\bm{d}}}$ \Comment{Eqs.~\eqref{errors},\eqref{Omegac}}
    \State Compute outputs of all adaptive law \textit{``slices"}: $\{\bm{\bar{J}}^{[j]}\}_{j=1,2,3}\leftarrow$ \textbf{Algorithm}~\ref{alg:J-update}
    \State Compute outputs of all neural network \textit{``slices"}: $\{\bm{\bar{\phi}}_{\bm{\textit{R}}}^{[j]}\}_{j=1,2,3}\leftarrow$ \textbf{Algorithm}~\ref{alg:nn-forward} $\leftarrow$ \textbf{Algorithm}~\ref{alg:proj}
    \State Compose desired control moment: $\{\bm{\mathrm{M}_d}^{[j]}\}_{j=1,2,3}$  \Comment{Eq.~\eqref{Md}}
    \State Send $\{\bm{\mathrm{M}_d}^{[j]}\}_{j=1,2,3}$ to actuators
\EndWhile
\end{algorithmic}
\vspace{-0.5em}
\rule{\linewidth}{0.4pt}
{\textbf{Note:} For \textbf{Algorithms}~\ref{alg:init}, \ref{alg:J-update}, \ref{alg:nn-forward}, \ref{alg:proj}, refer to Supplementary materials. When coupling with position control, the incoming desired attitude is computed by $\bm{R}_{{\bm{d}}}\leftarrow\bm{R}_{{\bm{c}}}:=[\bm{\vec{b}}_{1\bm{c}},\bm{\vec{b}}_{2\bm{c}},\bm{\vec{b}}_{3\bm{c}}]$, where $\bm{\vec{b}}_{1\bm{c}}:=\bm{\vec{b}}_{2\bm{c}}\times\bm{\vec{b}}_{3\bm{c}}$, $\bm{\vec{b}}_{2\bm{c}}:=(\bm{\vec{b}}_{3\bm{c}}\times\bm{\vec{b}}_{1\bm{d}})/(\|\bm{\vec{b}}_{3\bm{c}}\times\bm{\vec{b}}_{1\bm{d}}\|)$ and $\bm{\vec{b}}_{3\bm{c}}:= -\bm{\mathrm{F}_d}/\|\bm{\mathrm{F}_d}\|$. Here, $\bm{\mathrm{F}_d}\in\mathbb{R}^3$ denotes the desired resultant control force and $\bm{\vec{b}}_{1\bm{d}}(t)\in\mathbf{S}^2$ is given as the desired heading direction. For details, refer to Supp. \ref{supp:SE(3)}, and \cite{2010 Geometric tracking control of a quadrotor UAV on SE(3), 2013 Geometric nonlinear PID control of
a quadrotor UAV on SE(3)}.   }
\end{algorithm}

\subsubsection{\textbf{Design of the $j^{th}$ Neural Network \textit{``Slice"}}}
\label{Design of the Neural Network Slice}
Based on the universal approximation theorem \cite{1989 Multilayer feedforward networks are universal approximators}, the following submapping of SANM $\bm{\phi}_{\bm{\textit{R}}}^{[j]}\triangleq\bm{\mathcal{S}}^{N}_{j}(\bm{e}_{\bm{R}}^{[j]}, \bm{e}_{\bm{\Omega}}^{[j]}):\mathcal{C}_{j}\to\mathbb{R}$ can be approximated in a compact domain $\mathcal{C}_j\subset\mathbb{R}^{2}$ by multiple shallow neural networks (SNNs) with sufficient capacity. A Radial Basis Function (RBF) neural network (NN) with 2 inputs-$l$ hidden layer neurons-1 output (2-$l$-1) structure is deployed as follows:
\begin{equation}
\bm{\phi}_{\bm{\textit{R}}}^{[j]}=\bm{\mathcal{W}}_{\bm{\textit{R}} j}^{\top}\bm{\hbar}(\textbf{x}_{\bm{\textit{R}} j})+\epsilon_{\bm{\textit{R}} j},
\label{Phi}
\end{equation}
where $\textbf{x}_{\bm{\textit{R}} j}\in \mathbb{R}^2$ denotes the input vector of the $j^{th}$ neural network, and $\bm{\mathcal{W}}_{\bm{\textit{R}} j}\in\mathbb{W}_{\bm{\textit{R}} j}$ represents the corresponding weight vector bounded within a closed ball (compact set) $\mathbb{W}_{\bm{\textit{R}} j}=\{\bm{\mathcal{W}}_{\bm{\textit{R}} j}\in\mathbb{R}^l\,|\,\|\bm{\mathcal{W}}_{\bm{\textit{R}} j}\|\leq \overset{\tiny \text{max}}{W}_{j}\}$ with radius $\overset{\tiny \text{max}}{W}_{j}>0$. The $\bm{\hbar}(\cdot)\in\mathbb{R}^l$ denotes the Gaussian activation function and the $\epsilon_{\bm{\textit{R}} j}\in\mathbb{R}^+$ represents an arbitrarily small intrinsic approximation error, i.e.,  $\epsilon_{\bm{\textit{R}} j}\to 0^+$. 

The output of the $k^{th}$ hidden layer neuron in the $j^{th}$ neural network \textit{``slice"} is expressed as:
\begin{equation}
\begin{aligned}
&\bm{\hbar}^{[k]}(\textbf{x}_{\bm{\textit{R}} j})\in\mathbb{R}:=\mathrm{exp}\left(-\frac{\lVert\textbf{x}_{\bm{\textit{R}} j}-\textbf{c}_{kj}\rVert^2}{2b^{2}_{kj}}\right),
\end{aligned}
\label{Gaussian activation function}
\end{equation}
where $\textbf{c}_{kj}\in\mathbb{R}^2$ denotes the center vector of the $k^{th}$ neuron and $b_{kj}\in\mathbb{R}$ denotes the width of the $k^{th}$ Gaussian function for neuron index $k=1,2,\dots,l$. 

To approximate Eq.~\eqref{Phi}, the estimated rotational disturbance dynamics $\bm{\bar{\phi}}_{\bm{\textit{R}}}^{[j]}$ is modeled as the output of a neural network with time-varying estimated weights $\bm{\bar{\mathcal{W}}}_{\bm{\textit{R}} j}\in\mathbb{W}_{\bm{\textit{R}} j}$. This approximation is expressed as follows:
\begin{equation}
\begin{aligned}
\bm{\bar{\phi}}_{\bm{\textit{R}}}^{[j]}:=\bm{\bar{\mathcal{W}}}_{\bm{\textit{R}} j}^{\top}\bm{\hbar}(\textbf{x}_{\bm{\textit{R}} j}),
\end{aligned}
\label{Phi_hat}
\end{equation}
where input $\textbf{x}_{\bm{\textit{R}} j}:= \left(\bm{e}_{\bm{R}}^{[j]}, \bm{e}_{\bm{\Omega}}^{[j]}\right)^{\top}$ is the rotational error vector along the $\bm{\vec{b}}_j$-axis. From here, the $j^{th}$ neural network \textit{``slice"} is constructed and its structure is shown in Fig.~\ref{SANM_Structure}.  To guarantee boundedness of the weight estimates within the prescribed closed ball, a projection-based adaptive update is employed as in \textbf{Algorithm}~\ref{alg:proj}. According to the Lyapunov analysis in Supp.~\ref{Proofs of Proposition 1 and 2},  the nominal weight update laws $\{\bm{\dot{\bar{\mathcal{W}}}}_{\bm{\textit{R}} 
 j}^{\mathrm{nom}}\}_{j=1,2,3}$ are computed online via the following Lyapunov adaptation:
\begin{align}
\bm{\dot{\bar{\mathcal{W}}}}_{\bm{\textit{R}} 
 j}\leftarrow\bm{\dot{\bar{\mathcal{W}}}}_{\bm{\textit{R}} 
 j}^{\mathrm{nom}}:=&\gamma_{\bm{\textit{R}}j} \left(\bm{e}^{[j]}_{\bm{\Omega}}+c_R\bm{e}^{[j]}_{\bm{R}}\right)\bm{\hbar}(\textbf{x}_{\bm{\textit{R}} j}),
\label{Estimated Weights_R}
\end{align}
where the learning rates $\{\gamma_{\bm{\textit{R}}j}\}_{j=1,2,3}$ and $c_R$ are positive constants. 
The optimal weights that can be identified by these weight update laws are expressed as:
\begin{equation}
    \bm{\mathcal{W}}_{\bm{\textit{R}} j}^*\triangleq\mathrm{arg}\, \underset{\bm{\mathcal{W}}_{\bm{\textit{R}} j}\in\mathbb{W}_{\bm{\textit{R}} j}}{\mathrm{min}}\left(\mathrm{sup}\big{\vert}\bm{\phi}_{\bm{\textit{R}} }^{[j]}-\bm{\bar{\phi}}_{\bm{\textit{R}} }^{[j]}\big{\vert}\right), \label{W*}
\end{equation}
where $\mathrm{arg}\,\mathrm{min}$ denotes the value of $ \bm{\mathcal{W}}_{\bm{\textit{R}} j}$ that minimizes the supremum of the error between $\bm{\phi}_{\bm{\textit{R}}}^{[j]}$ and $\bm{\bar{\phi}}_{\bm{\textit{R}}}^{[j]}$.

The optimal approximation error is then defined as follows:
\begin{equation}
     \bm{\varpi}^{[j]}_{\bm{\textit{R}}}\triangleq\bm{\phi}_{\bm{\textit{R}}}^{[j]}-\bm{\bar{\phi}}_{\bm{\textit{R}}}^{[j]}(\textbf{x}_{\bm{\textit{R}} j}\vert\bm{\mathcal{W}}^*_{\bm{\textit{R}} j}),
     \label{varpi}
\end{equation}
where $\|\bm{\varpi}^{[j]}_{\bm{\textit{R}}}\|$ is bounded according to the universal approximation theorem \cite{1989 Multilayer feedforward networks are universal approximators} and \textbf{\textit{Proposition~3}}. Here, $\bm{\varpi}^{[j]}_{\bm{\textit{R}}}\in\mathbb{R}$ denotes the $j^{th}$ component of the optimal approximation error vector $\bm{\varpi}_{\bm{\textit{R}}}\in\mathbb{R}^{3}$.

From Eqs.~\eqref{Phi_hat}, \eqref{W*}, \eqref{varpi}, the problem of the approximation error $\bm{\phi}_{\bm{\textit{R}} }^{[j]}-\bm{\bar{\phi}}_{\bm{\textit{R}} }^{[j]}$ can be transformed into the problem of weight estimation error:
\begin{equation}
   \bm{\phi}_{\bm{\textit{R}} }^{[j]}-\bm{\bar{\phi}}_{\bm{\textit{R}} }^{[j]}=\bm{\tilde{\mathcal{W}}}_{\bm{\textit{R}} j}^{\top}\bm{\hbar}(\textbf{x}_{\bm{\textit{R}} j})+\bm{\varpi}^{[j]}_{\bm{\textit{R}}},
    \label{approximation error to weight error}
\end{equation}
where the weight estimation error is defined as:
\begin{equation}
    \bm{\tilde{\mathcal{W}}}_{\bm{\textit{R}} j}\triangleq\bm{\mathcal{W}}^*_{\bm{\textit{R}} j}-\bm{\bar{\mathcal{W}}}_{\bm{\textit{R}} j}.
    \label{weight error}
\end{equation}

\subsection{Propositions}
 First, we consider the following almost-global domain of attraction for the initial conditions of rotational dynamics:
\begin{equation}
{\small
   \begin{aligned}
    \mathcal{D}_{{\bm{\textit{R}}0}}\!\triangleq&\Big{\{}
\,\,\,0<\Psi_{\textit{R}}\big{(}\bm{R}(0),\bm{R}_{\bm{d}}(0)\big{)}<2,\\[-0pt]
&\,\,\,\,\,\,\|\bm{e_R}(0)\|=\sqrt{\Psi_{\textit{R}}(0)\big{(}2-\Psi_{\textit{R}}(0)\big{)}}<1,\\
&\,\,\,\,\,\,\|\bm{e}_{\bm{\Omega}}(0)\|^2< k_{R}\Big{(}2-\Psi_{\textit{R}}\big{(}\bm{R}(0),\bm{R}_{\bm{d}}(0)\big{)}\Big{)}-\frac{c_R^2}{2}\Big{\}},
\end{aligned} 
}
\label{D_R0}
\end{equation}
where $\Psi_{\textit{R}}\!:\!\mathbf{SO}(3)\times\mathbf{SO}(3)\!\to\!\mathbb{R}$ denotes an attitude configuration error scalar function:
\begin{equation}
    \Psi_{\textit{R}}(\bm{R},\bm{R}_{\bm{d}})\triangleq\frac{1}{2}\mathrm{tr}\Big{[}\mathrm{I}^{3\times3}-\bm{R}_{\bm{d}}^{\top}\bm{R}\Big{]}.
    \label{main:Psi_R}
\end{equation}
The $\Psi_{\textit{R}}:\mathbf{SO}(3)\times\mathbf{SO}(3)\to\mathbb{R}$ is positive definite and constrained by:
\begin{equation}
    \frac{1}{2}\|\bm{e_R}\|^{2}\leq\Psi_{\textit{R}}\leq\frac{1}{2-\psi_{\textit{R}}}\|\bm{e_R}\|^{2},
    \label{main:Psi_R_bound}
\end{equation}
with a positive scalar $0<\psi_{\textit{R}}<2$.
When $0\!<\!\Psi_{\textit{R}}\!<\!2$, it covers almost the entire $\mathbf{SO}(3)$, except for singular points corresponding to a rotation of exactly $180^\circ$. Within this domain of attraction, we present:

\textbf{\textit{Proposition 1: (Almost-Global Exponential Attractiveness)}}
Under the attitude control compensated by the SANM module and the initial condition that $\bm{z}_{\bm{\textit{R}}}(0)\!=\!(\|\bm{e_R}(0)\|,\|\bm{e_\Omega}(0)\|)^{\top}\!\!\in\!\mathcal{D}_{\bm{\textit{R}}0}$, the state solution of the rotational error dynamics $\bm{z}_{\bm{\textit{R}}}(t)=\left(\|\bm{e}_{\bm{R}}\|,\|\bm{e}_{\bm{\Omega}}\|\right)^{\top}\!\!\in\mathbb{R}^{2}$ is almost-globally exponentially attracted to a bounded residual set
\begin{equation}
    {
    \begin{aligned}
\mathcal{D}_{{\bm{\textit{R}}1}} \triangleq \Big{\{}\bm{z}_{\bm{\textit{R}}}\in\mathbb{R}^{2}:\|\bm{z}_{\bm{\textit{R}}}\|\le r_1\Big{\}},
    \end{aligned}
    }
\end{equation}
for a positive radius $r_{1}$. This result holds despite bounded disturbance effects, an unknown inertia tensor, and bounded approximation errors of the SANM module.

\textbf{\textit{Proof:}} Supp. \ref{Proofs of Proposition 1 and 2}. \textbf{\textit{Validation:}} Section \ref{Exp::Numerical-Simulation}

\textbf{\textit{Proposition 2: (Local Exponential Convergence under 
Time-varying Disturbances and Inertia Uncertainties within the Identification Region $\mathcal{D}_{\mathcal{C}}$ )}} Once the rotational error state enters the compact neural-network identification region 
\begin{equation}
    {
    \begin{aligned}
\mathcal{D}_{\mathcal{C}}
\;\triangleq\;
\Big\{
\bm{z}_{\bm{\textit{R}}}\in\mathbb{R}^2 
\;\big|\;
\|\bm{z}_{\bm{\textit{R}}}\|\le r_c
\Big\},
    \end{aligned}
    }
\end{equation}
for a positive radius $r_{c}$ (i.e., $\bm{z}_{\bm{\textit{R}}}(t_1)\in\mathcal{D}_{\mathcal{C}}$ for some $t_1\ge0$), the state solution of the rotational error dynamics $\bm{z}_{\bm{\textit{R}}}(t)$ exponentially converges into an arbitrarily small ball
\begin{equation}
    {
    \begin{aligned}
    \mathcal{B}_\epsilon 
    \triangleq 
    \Big\{
        \bm{z}_{\bm{\textit{R}}}\in\mathbb{R}^{2}
        \;\big|\;
        \|\bm{z}_{\bm{\textit{R}}}\|\le\epsilon
    \Big\},
    \end{aligned}
    }
\end{equation}
such that $\lim\limits_{t \to \infty}\bm{z}_{\bm{\textit{R}}}(t)\in\mathcal{B}_\epsilon$, where $\epsilon\to0^+$ denotes an arbitrarily small positive radius. The size of this radius depends on the mapping deviations of the control moment and the approximation errors of the neural network \textit{``slices"}. In addition, the estimation errors $\{ \widetilde{J}_{j}, \bm{\tilde{\mathcal{W}}}_{\bm{\textit{R}} j}\}_{j=1,2,3}$ remain uniformly bounded. The above result holds without the persistent excitation (PE) condition and despite the presence of bounded time-varying disturbance effects and an unknown inertia tensor.

\textbf{\textit{Proof:}} Supp. \ref{Proofs of Proposition 1 and 2}. \textbf{\textit{Validation:}} Sections \ref{Exp::Numerical-Simulation}, \ref{Exp::Impact-Disturbance}.

\textbf{\textit{Proposition 3: (Compact Set Constraint on Neural Network Inputs)}} Inside the neural-network identification region $\mathcal{D}_{\mathcal C}$, the rotational state error vector $\mathbf{E}_{\bm{\textit{R}}}$ is bounded by a compact set: $ \mathcal{C}\triangleq\Big{\{}\mathbf{E}_{\bm{\textit{R}}}\in\mathbb{R}^{6}| \,\|\mathbf{E}_{\bm{\textit{R}}}\|\leq\!\|\bm{e}_{\bm{R}}\| +\|\bm{e}_{\bm{\Omega}}\|\leq r_{c}\Big{\}}$. This implies that all inputs of the neural networks $\{\textbf{x}_{\bm{\textit{R}} j}\}_{j=1,2,3}$ are also bounded within their respective compact sets (denoted $\{\mathcal{C}_j\}_{j=1,2,3}$). 
This compactness condition satisfies the prerequisite of the universal approximation theorem \cite{1989 Multilayer feedforward networks are universal approximators}.

\textbf{\textit{Proof:}} Supp. \ref{supp:extended-proofs}. \textbf{\textit{Validation:}} Sections \ref{Exp::Numerical-Simulation} to \ref{Exp::Real-world-Flight}.

\textbf{\textit{Proposition 4: (ISpS of the Sampled-Data Closed Loop)}} 
Building upon the continuous-time exponential convergence result in \textbf{\textit{Proposition 2}}, the sampled-data implementation with zero-order hold (ZOH), sampling period $dt>0$, and bounded delay $0\le\tau<dt$, preserves a discrete-time practical exponential estimate of the form
\begin{equation}
\|\bm{z}_{\bm{\textit R}}(ndt)\|
\le \alpha_s e^{-\beta_s ndt}\|\bm{z}_{\bm{\textit R}}(0)\| + \epsilon_s,
\qquad \forall n\in\mathbb{N},
\end{equation}
for constants $\alpha_s>0$, $\beta_s\in(0,\beta)$ (with $\beta>0$), and $\epsilon_s>0$.   
The residual radius $\epsilon_s$ depends on the approximation errors and scales with $(dt,\tau)$.  
Hence, the sampled-data closed loop is \emph{Input-to-State Practically Stable (ISpS)} and exponentially decays toward a slightly enlarged residual ball.

\textbf{\textit{Proof:}} Supp. \ref{sec:ISpS_sampled}. \textbf{\textit{Validation:}} Section \ref{Exp::Impact-Disturbance}.

\section{Experimental Validation}
\label{Experiments Validation}

This section presents simulation and real-world experiments that provide comprehensive validation of the proposed method. In \textit{Experiment 1}, a numerical simulation was carried out in \textit{MATLAB Simulink}. In \textit{Experiment 4}, a Software-In-The-Loop (SITL) simulation was performed using the \textit{Gazebo Harmonic} physics engine. In \textit{Experiments 2, 3, 5}, testbed evaluations and flight tests were conducted in the real-world. The hardware configuration of the quadrotor is shown in Fig.~\ref{Quadrotor}. The flight control unit (FCU, \textit{brainfpv RADIX 2 HD}) integrated an \textit{STM32H750 ARM 480 MHz} processor with \textit{1 MB RAM} and a \textit{BMI270} inertial measurement unit (IMU). The propulsion system consisted of 1250 kV brushless DC motors combined with 8-inch tri-blade propellers. The power supply was provided by a 4-cell lithium polymer (LiPo) battery with a capacity of 3300 mAh. For communication with the \textit{ROS~2} and mo-cap systems, a \textit{Raspberry Pi~5 (8GB RAM)} was mounted onboard.
The SANM-augmented geometric attitude control \textbf{Algorithm}~\ref{alg:Attitude-Control} was written in \textit{C++} and integrated into the open-source \textit{ArduPilot 4.6} firmware. For calculating the Gaussian activation function in neural network \textit{``slices"}, the \texttt{AP\textunderscore Math.h} and \texttt{cmath} were included to support the computation of \texttt{expf()} function. 

\subsection{Experiment 1: (Numerical Simulation)}
\label{Exp::Numerical-Simulation}
\begin{figure}[t]
\centering
\begin{subfigure}{0.9\linewidth}
    \centering
    \includegraphics[scale=0.4]{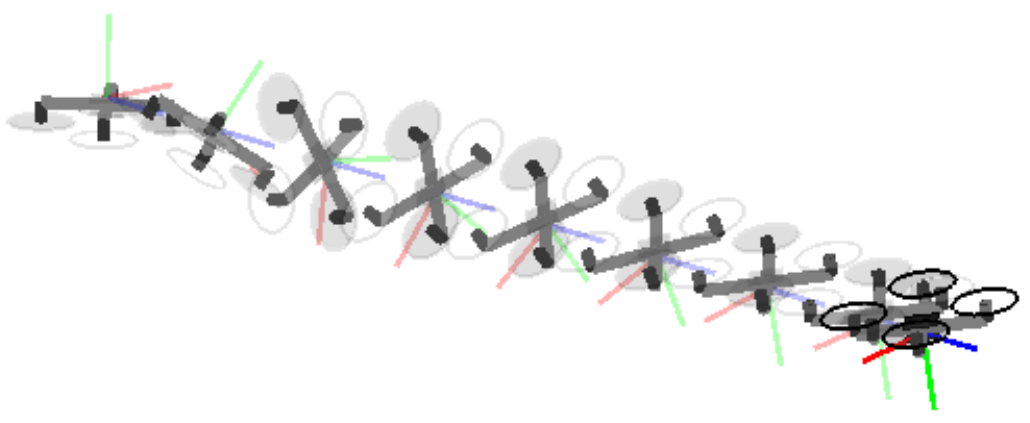}
    \caption{\footnotesize Snapshots of the attitude evolution showing that even with an almost $180^{\circ}$ flip, the SANM-augmented geometric attitude controller achieves convergence to the desired orientation, verifying the almost-global domain of attraction.}
  \label{fig:Quadrotor_almost_global}
\end{subfigure}
\vspace{0em}
\begin{subfigure}{0.45\linewidth}
    \centering
    \includegraphics[width=\linewidth]{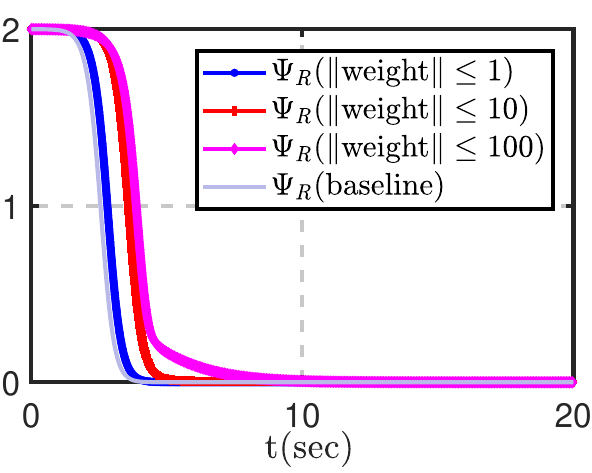}
    \vspace{-2em}
    \caption{\footnotesize Without external disturbances.}
    \label{fig:exp1_no_disturb}
\end{subfigure}
\hspace{0.5em}
\begin{subfigure}{0.45\linewidth}
    \centering
    \includegraphics[width=\linewidth]{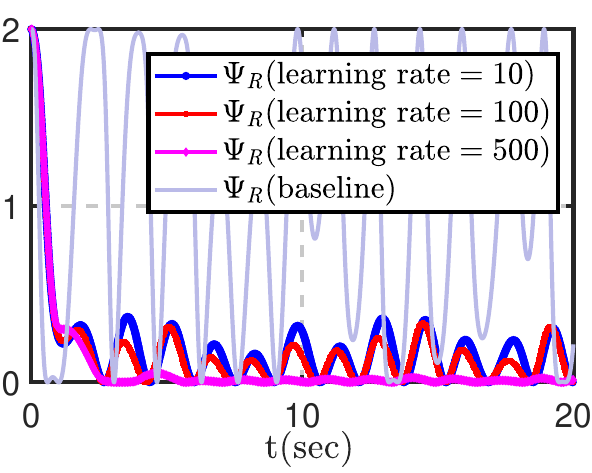}
    \vspace{-2em}
    \caption{\footnotesize With external disturbances.}
    \label{fig:exp1_with_disturb}
\end{subfigure}

\caption{\footnotesize \textit{Experiment 1-} Numerical validation of the almost-global attractiveness and local exponential convergence.  (b) During large-error convergence, increasing the bound of the network weights ($\overset{\tiny \text{max}}{W}_{j}$) tends to slow down the convergence process. (c)  Near the equilibrium, higher learning rates ($\gamma_{\bm{\textit{R}}j}$) yield superior disturbance rejection performance.}
\label{fig:combined}
\end{figure}

This experiment was conducted to numerically validate \textbf{\textit{Propositions 1 and 2}}.
The simulation was carried out in \textit{MATLAB Simulink} using a fixed-step \textit{ODE3 (Bogacki--Shampine)} solver with a step size of $dt=0.0025$ s. 

We first confirmed that the augmentation with the SANM module does not compromise the almost-global property of the baseline geometric attitude controller on $\mathbf{SO}(3)$ \cite{2011 Geometric tracking control of the attitude dynamics of a rigid body on SO(3)}. To isolate the effect of external disturbances, we conducted the experiment without any disturbance, as such disturbances can occasionally facilitate convergence. The initial attitude error was set to approximately $179^\circ$, corresponding to an attitude configuration error $\Psi_{\textit{R}}(0):=1.9998$. Then, Fig.~\ref{fig:exp1_no_disturb} illustrates that even with an initial attitude almost $180^{\circ}$ away from the desired orientation, corresponding to an almost-globally large configuration error, the attitude errors still converged. 

Then, we examined that the almost-global property holds even under \textbf{\textit{Scenario 2 ($\bm{J}$ is unknown)}} and time-varying disturbances. Except for model uncertainties, a bounded time-varying  disturbance effect 
\begin{equation}
{\small
\begin{aligned}
\bm{\phi}_{\bm{\textit{R}}}:=&\big{(}-0.5\sin(\sin(0.2t)t)-3\cos(2t),0,0\big{)}^{\top}
\end{aligned}
}
\notag
\end{equation}
was injected along the $\bm{\vec{b}}_{1}$-axis to emulate the external disturbance. As shown in Fig.~\ref{fig:exp1_with_disturb}, the SANM-augmented closed-loop behavior demonstrates:
(i) The rotational errors still converge from large initial conditions, confirming that the SANM augmentation preserves the almost-global attractiveness predicted in \textbf{\textit{Proposition 1}}.
(ii) Once the trajectory enters the identification region $\mathcal{D}_{\mathcal C}$, the error evolution exhibits the enhanced exponential convergence consistent with \textbf{\textit{Proposition 2}}.
This demonstrates that SANM not only maintains the intrinsic almost-global attractiveness property but also locally strengthens the convergence characteristics in the presence of time-varying disturbances and inertia uncertainties.
(iii) The disturbance-rejection performance improves with higher learning rates, which supports the positive correlation between the learning rate and the exponential convergence rate, as discussed in \textbf{\textit{Remark \ref{supp:extended-proofs}.2}}.
For extended information such as \textit{simulation settings} and \textit{control parameters}, refer to Supp. \ref{supp:Experiment-1}.

\subsection{Experiment 2: (Real-world Wind Disturbance)}
\label{Exp::Wind-Disturbance}

\begin{figure}[tp]
    \centering
    \captionsetup{labelformat=default, labelsep=colon}

    \begin{subfigure}{0.95\linewidth}
        \centering
        \includegraphics[width=\linewidth]{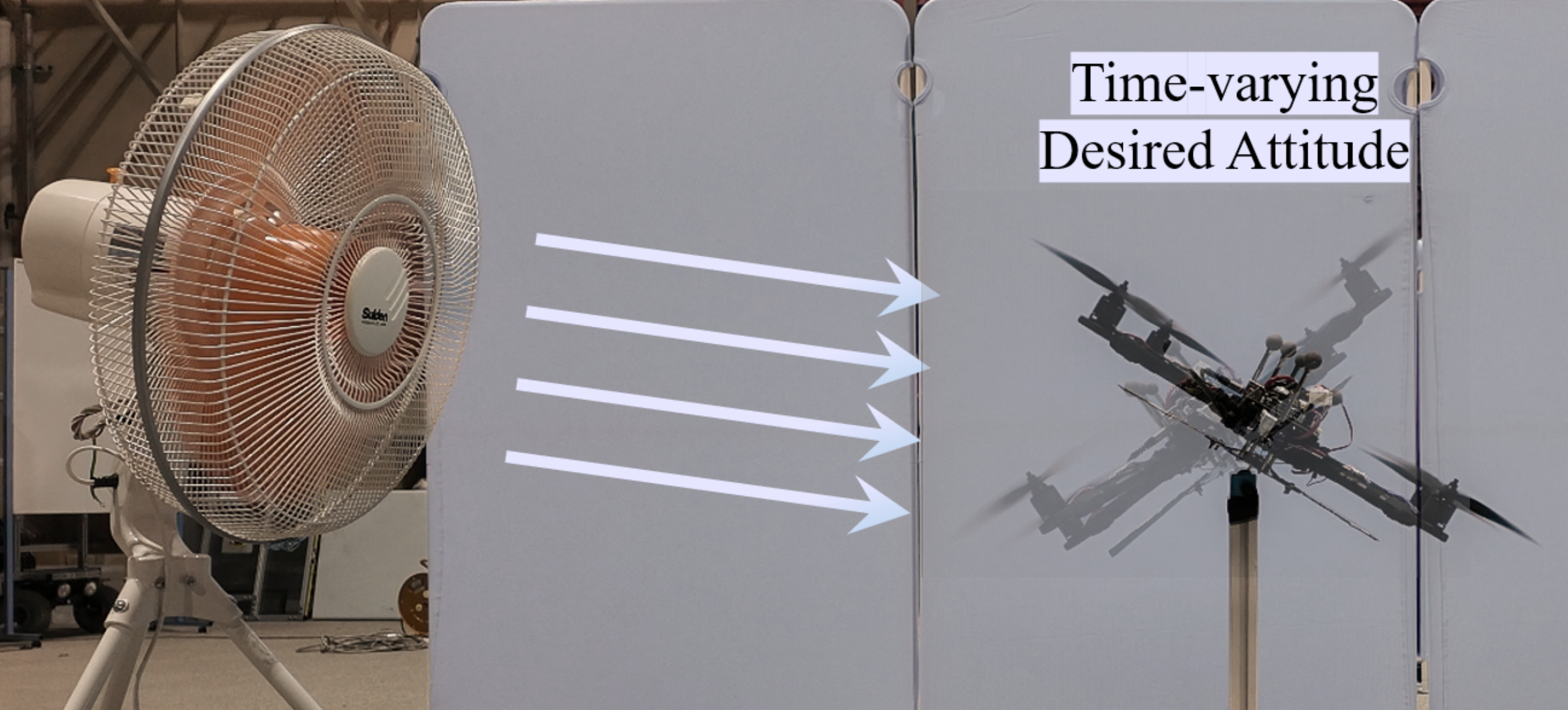}
        \caption*{} 
        \label{fig:testbed-wind-main}
    \end{subfigure}

    \vspace{-1em}  

    \begin{subfigure}{1\linewidth}
        \centering
        \includegraphics[width=\linewidth]{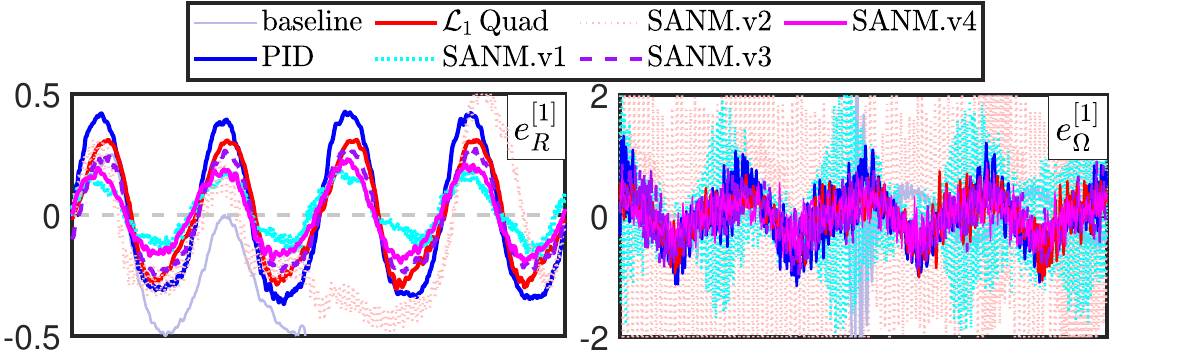}
        \caption*{} 
        \label{fig:testbed-wind-sub1}
    \end{subfigure}

    \vspace{-1.74em}

    \begin{subfigure}{1\linewidth}
        \centering
        \includegraphics[width=\linewidth]{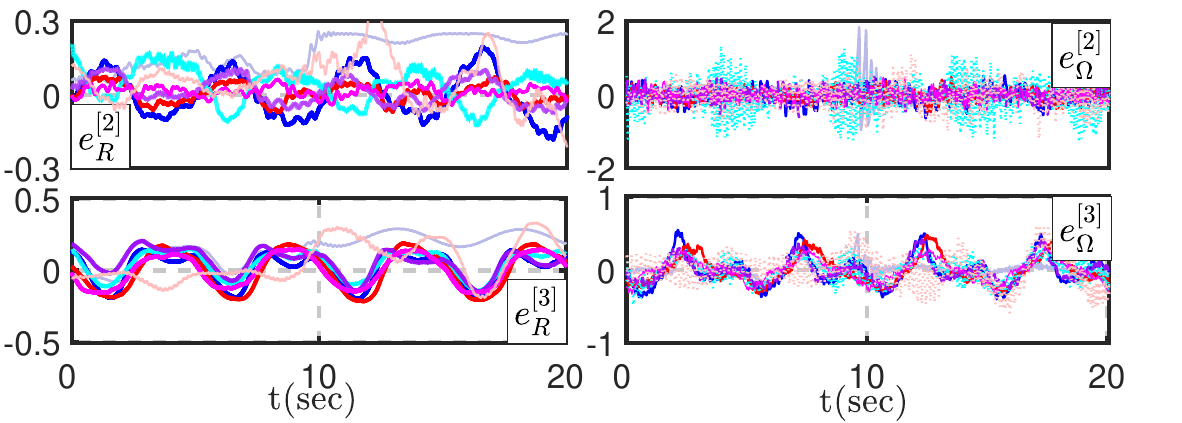}
        \caption*{} 
        \label{fig:testbed-wind-sub2}
    \end{subfigure}
    \vspace{-3em}
    \caption{\footnotesize \textit{Experiment 2-}Attitude tracking experiments under realistic wind disturbances. Three benchmark geometric controllers (baseline PD \cite{2011 Geometric tracking control of the attitude dynamics of a rigid body on SO(3)}, PID \cite{2013 Geometric nonlinear PID control of
a quadrotor UAV on SE(3)}, and $\mathcal{L}_1\,\mathrm{Quad}$ \cite{2025 L1Adaptive Augmentation of Geometric Control for Agile Quadrotors With Performance Guarantees}) are included for comparison. Four SANM variants v1, v2, v3, v4 are further included as ablation groups, where the adaptive law \textit{``slices"} are disabled to evaluate the neural-network-only (NN-only) mode. v1, v2 and v3 share the same learning rates $\{\gamma_{\bm{\textit{R}}j}\}_{j=1,2,3}\!:=\!\{35,35,10\}$  but use 3, 9, and 7 RBF neurons, respectively, yielding different RBF coverage densities under identical widths. v4 adopts the same 7-neuron structure as v3 but increases the learning rates to
$\{120,120,50\}$ to examine their effect on the residual error ball size. For video: \url{https://youtu.be/kDE5079TgCI}.}
\label{fig:testbed-wind}
\end{figure}

This experiment evaluated the performance of the proposed method by giving time-varying desired attitude to present the attitude tracking performance under realistic wind disturbances. The evaluation was conducted on a stationary testbed to isolate the attitude loop from the position loop. Three benchmark geometric controllers were included: a geometric PD \cite{2011 Geometric tracking control of the attitude dynamics of a rigid body on SO(3)} (baseline), a geometric PID \cite{2013 Geometric nonlinear PID control of
a quadrotor UAV on SE(3)}, and the current state-of-the-art geometric adaptive: $\mathcal{L}_1\,\mathrm{Quad}$ \cite{2025 L1Adaptive Augmentation of Geometric Control for Agile Quadrotors With Performance Guarantees}. These benchmarks provided a direct comparison for evaluating the robustness of SANM under wind disturbances.
In addition, four SANM variants were tested as ablation groups. In all four variants, the adaptive law \textit{``slices"} were intentionally disabled, such that disturbance identification relied solely on the  neural network \textit{``slices"}. This setting highlighted the intrinsic robustness of the neural network \textit{``slices"}. All benchmark controllers and all SANM variants were implemented with the same nominal PD gains ($k_{R}\!\!=\!\!100$, $ k_{\Omega}\!\!=\!\!80$).  The variants differed in their neural-network configurations and adaptation parameters, enabling a structured comparison of RBF coverage density and learning rates. The corresponding tracking errors are shown in Fig.~\ref{fig:testbed-wind}. From the benchmark comparison, it was clear that SANM achieved a smaller error ball under wind disturbances when properly configured. From the comparison among SANM.v1–v3, overly sparse RBFs became too localized, leading to large equivalent gains and inducing oscillations. In contrast, excessive overlap caused the RBFs to behave in a global manner, making it difficult to capture local disturbance features, thereby leading to closed-loop divergence. Therefore, a stability \textit{sweet spot} exists in the RBF coverage density. Furthermore, comparing v3 and v4 indicates that higher learning rates reduce the size of the residual error ball.  Additional information and guidance on selecting the RBF coverage density are provided in Supp. ~\ref{supp:sec:Experiment-2}.

\subsection{Experiment 3: (Real-world Impact Disturbance)}
\label{Exp::Impact-Disturbance}

\begin{figure}[tp]
    \centering
    \captionsetup{labelformat=default, labelsep=colon}

    \begin{subfigure}{0.95\linewidth}
        \centering
        \includegraphics[width=\linewidth]{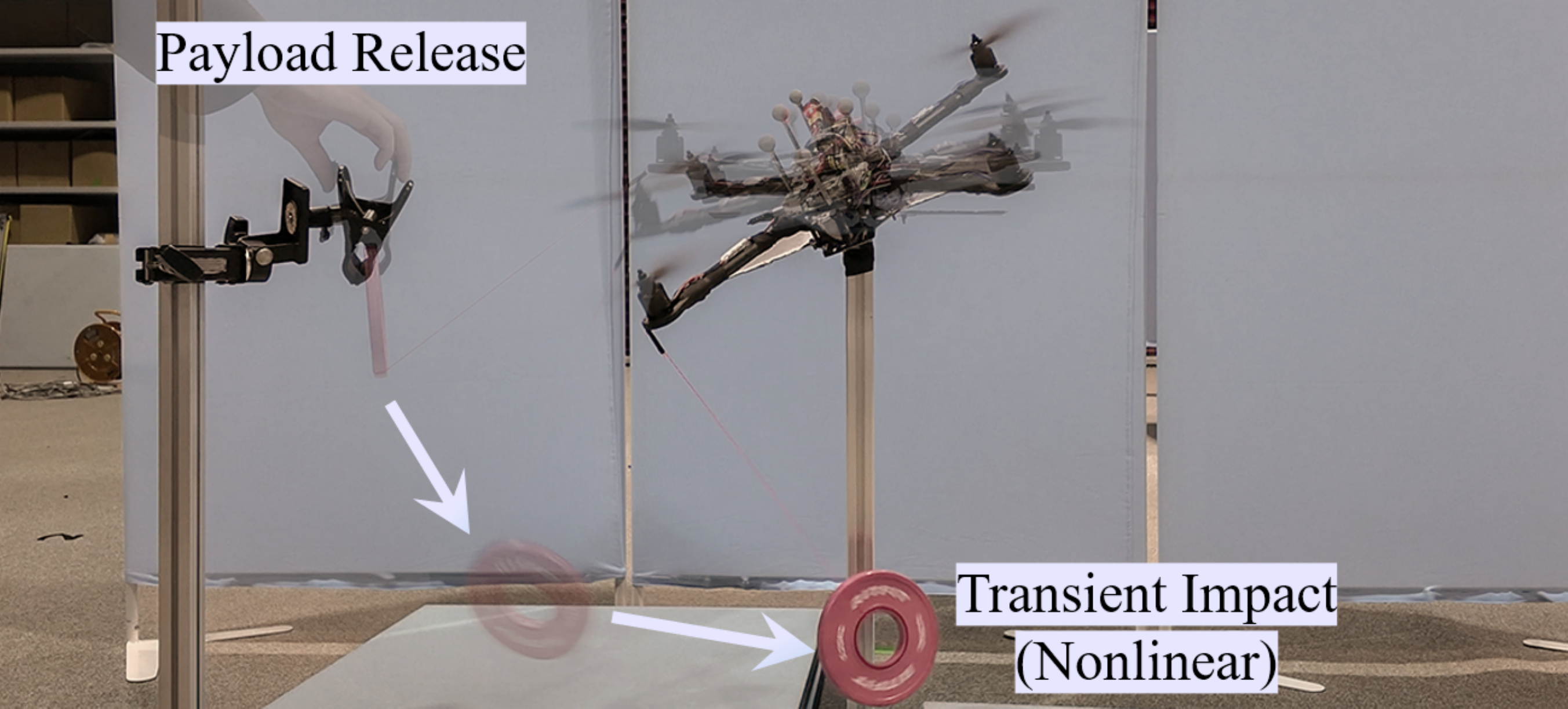}
        \caption*{} 
        \label{fig:testbed-impact-main}
    \end{subfigure}

    \vspace{-1em}  

\begin{subfigure}{0.48\linewidth}
    \centering
    \includegraphics[width=\linewidth]{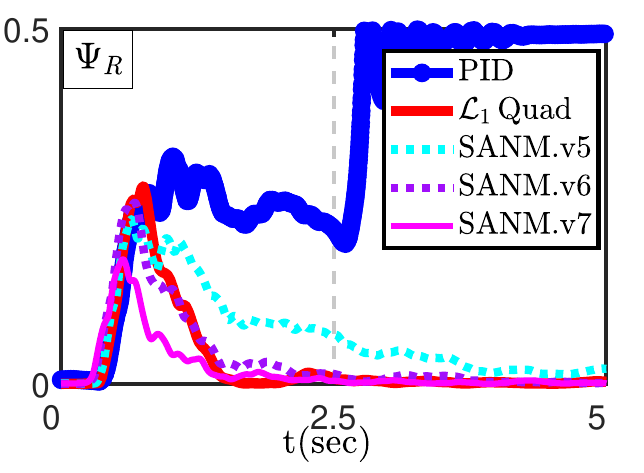}
    \vspace{-2em}
    \caption{\footnotesize Anti-impact convergence.}
    \label{fig:Anti-impact-convergence}
\end{subfigure}
\hspace{-1em}
\begin{subfigure}{0.49\linewidth}
    \centering
    \includegraphics[width=\linewidth]{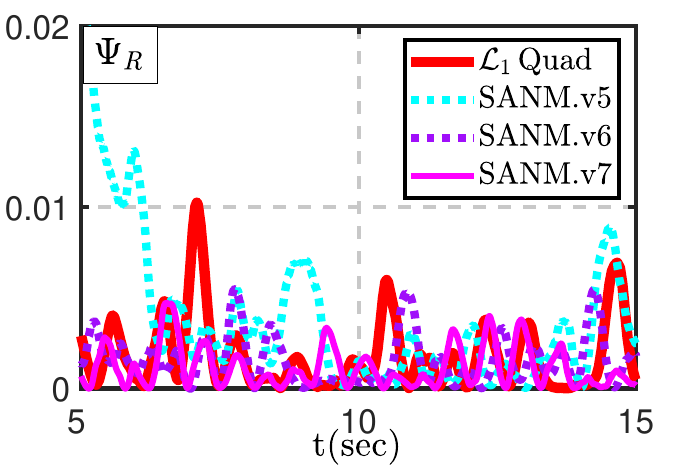}
    \vspace{-1.8em}
    \caption{\footnotesize Residual steady-state error.}
    \label{fig:Residual-steady-state-error}
\end{subfigure}
\begin{subfigure}{1\linewidth}
    \centering
    \includegraphics[width=1\linewidth]{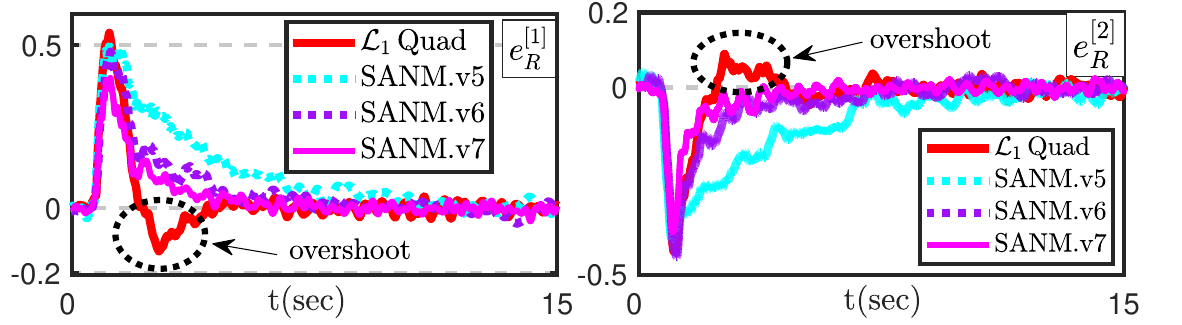}
    \vspace{-2em}
    \caption{\footnotesize Exponential convergence under impact disturbances.}
    \label{fig:Exponential-convergence-under-impact-disturbances}
\end{subfigure}

    \vspace{0em}
      \caption{\footnotesize \textit{Experiment 3-}Anti-impact experiments for attitude dynamics under transient disturbances. 
      Two benchmark geometric controllers (PID \cite{2013 Geometric nonlinear PID control of
a quadrotor UAV on SE(3)}, and $\mathcal{L}_1\,\mathrm{Quad}$ \cite{2025 L1Adaptive Augmentation of Geometric Control for Agile Quadrotors With Performance Guarantees}) are included for comparison. Three SANM variants v5, v6 and v7 are further included as variant groups, with the learning rates $\{\gamma_{\bm{\textit{R}}j}\}_{j=1,2,3}$ treated as the only varying variable (v5: $\{35,35,10\}$, v6: $\{80,80,30\}$, v7: $\{120,120,50\}$), while the adaptive law \textit{``slices"} are enabled and kept identical across all variants. (a) illustrates the anti-impact convergence behavior in the attitude  configuration error $\Psi_{\textit{R}}$. (b) presents the residual steady-state error behavior achieved after the system recovers from the impact. (c) compares SANM with the geometric $\mathcal{L}_1$-adaptive, demonstrating overshoot-free exponential convergence under transient (time-varying nonlinear) disturbances.  Overall, these results show that SANM achieves superior spike suppression and transient response compared with the benchmark controllers. The findings also validate the theoretical analysis (Supp.~\ref{supp:extended-proofs}), which establishes that the exponential convergence rate increases with the learning-rate magnitude. For video: \url{https://youtu.be/kDE5079TgCI}. }
      \label{fig:testbed-impact}
\end{figure}

This experiment evaluated the performance of the proposed method under a transient impact disturbance. The evaluation was performed on a testbed to isolate the attitude loop from the position loop. As illustrated in Fig.~\ref{fig:testbed-impact}, a 
$0.25$ kg payload attached to one arm of the quadrotor was released, generating a nonlinear impulsive disturbance moment on the quadrotor. Two geometric benchmark controllers were included for comparison: a geometric PID controller \cite{2013 Geometric nonlinear PID control of
a quadrotor UAV on SE(3)} and the current state-of-the-art geometric adaptive method: $\mathcal{L}_1\,\mathrm{Quad}$ \cite{2025 L1Adaptive Augmentation of Geometric Control for Agile Quadrotors With Performance Guarantees}. These benchmarks provided a direct reference for assessing the anti-impact robustness of SANM. 
In addition, three SANM variants (v5–v7) were tested as variant groups. All benchmarks and variants were also implemented with the same nominal PD gains ($k_{R}\!\!=\!\!100$, $ k_{\Omega}\!\!=\!\!80$). In all variants, the adaptive law \textit{``slices"} were enabled with identical configurations, and the neural network \textit{``slices"} consisted of 7 RBF neurons with the same coverage density as SANM.v4 in \textit{Experiment 2}. The variants differed only in learning rates ($\{35,35,10\}$, $\{80,80,30\}$ and $\{120,120,50\}$), allowing an evaluation of how the learning rate influenced the convergence rate under impact disturbances. The corresponding geometric attitude error measures are shown in Fig.~\ref{fig:testbed-impact}. These results confirmed that the transient impact disturbance did not violate the assumed continuity and boundedness of the acceleration-level disturbance term, thereby preserving the validity of the neural network approximation.

The comparison with the benchmark controllers shows that SANM exhibited markedly faster disturbance rejection and a smaller error ball after the impact, while also maintaining a high-damping, non-overshooting response, which reflects the theoretically proven \textbf{\textit{Exponential Convergence}} property. This contrasted with $\mathcal{L}_1\,\mathrm{Quad}$, which showed a convergence rate comparable to SANM but exhibited noticeable overshoot under impact. This behavior is expected, as the $\mathcal{L}_1$-adaptive controller does not provide exponential convergence guarantees under time-varying disturbances. The comparison of variants among SANM.v5–v7 further confirmed that higher learning rates yield faster convergence, consistent with the theoretical analysis (Supp.~\ref{supp:extended-proofs}) of the \textbf{\textit{Tunable Exponential Convergence Rate}}. In addition, higher learning rates and the resulting faster convergence also lead to a smaller steady-state error ball, as shown in Fig.~\ref{fig:Residual-steady-state-error}. 
Additional information about this experiment is provided in Supp.~\ref{supp:sec:Experiment-3}.

\subsection{Experiment 4: (Physics Simulation)}

\begin{figure*}[htbp] 
\vspace{-0em} 
  \centering
  \begin{subfigure}[b]{0.48\textwidth}
    \centering
    \includegraphics[width=\linewidth]{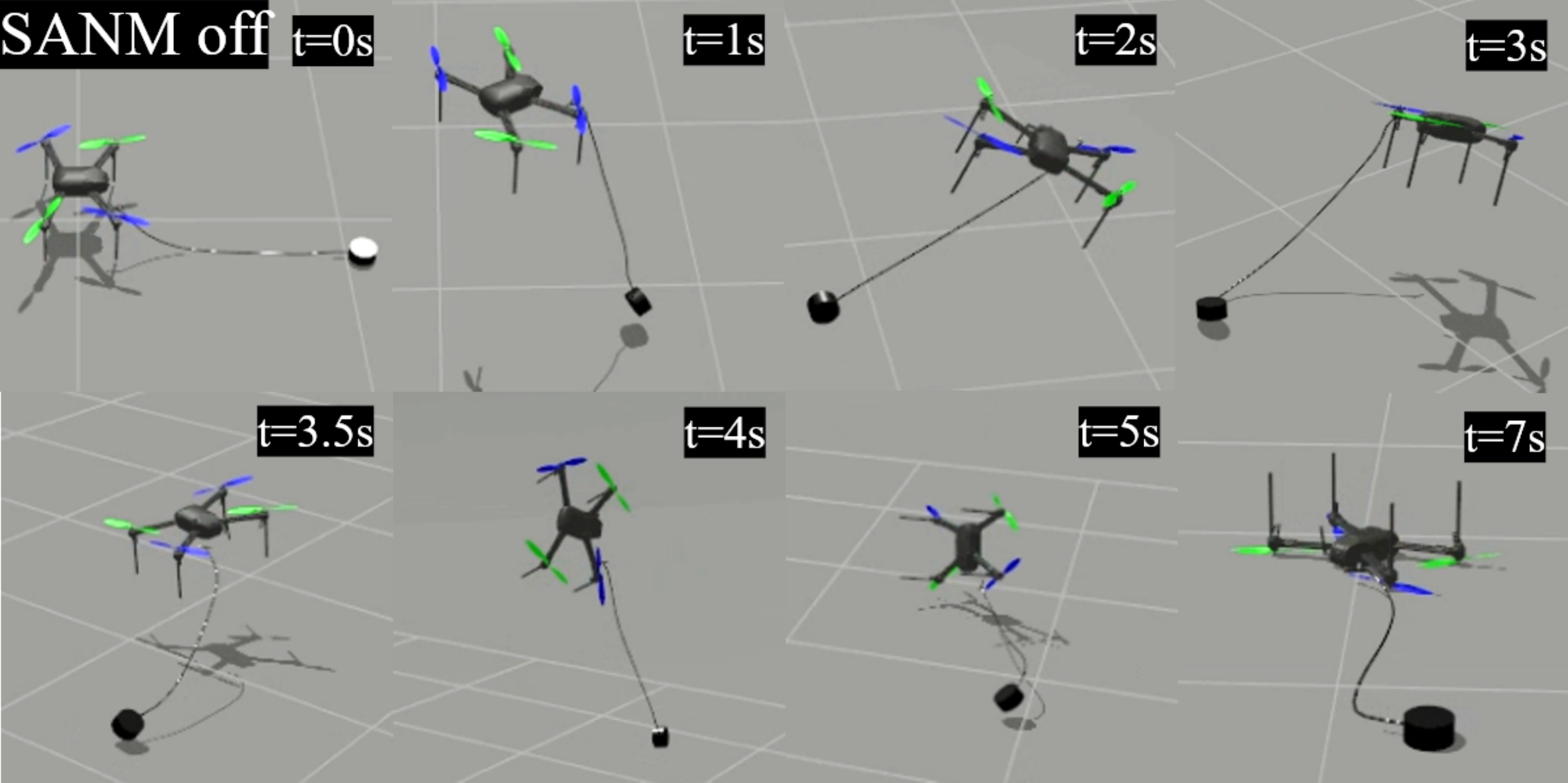}
  \end{subfigure}
  \vspace{-0em}
  \hspace{2mm}
  \begin{subfigure}[b]{0.48\textwidth}
    \centering
    \includegraphics[width=\linewidth]{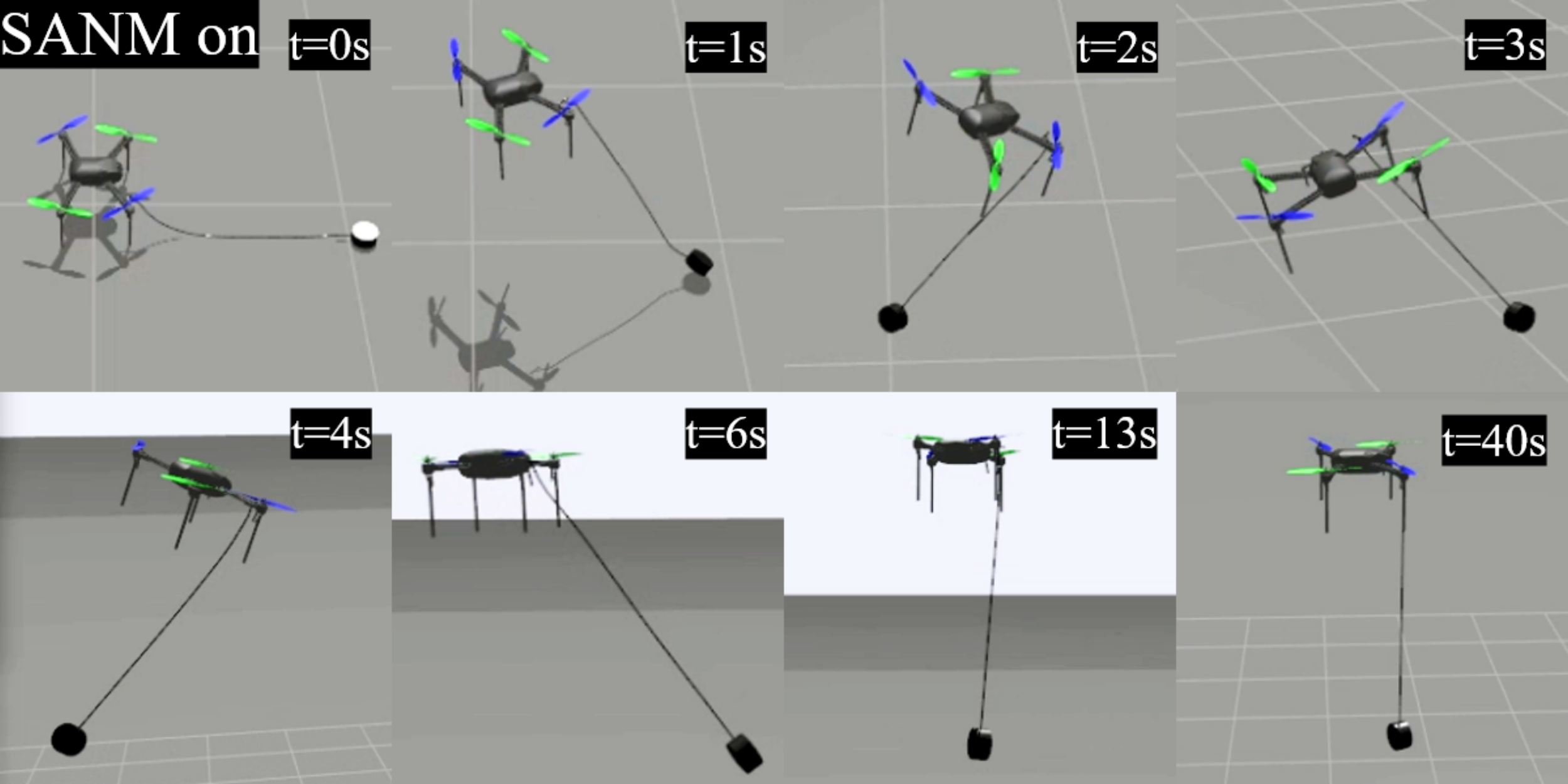}
  \end{subfigure}
  \hspace{0mm}
  \vspace{+0.5em}
  \captionsetup{labelformat=default, labelsep=colon}
  \begin{subfigure}[b]{0.49\textwidth}
    \centering
    \includegraphics[width=\linewidth]{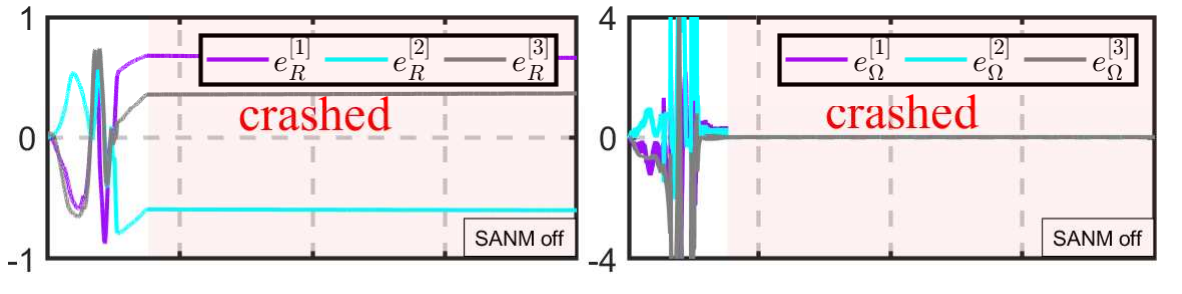}
    \label{}
  \end{subfigure}
  \hspace{1mm}
  \begin{subfigure}[b]{0.49\textwidth}
    \centering
    \includegraphics[width=\linewidth]{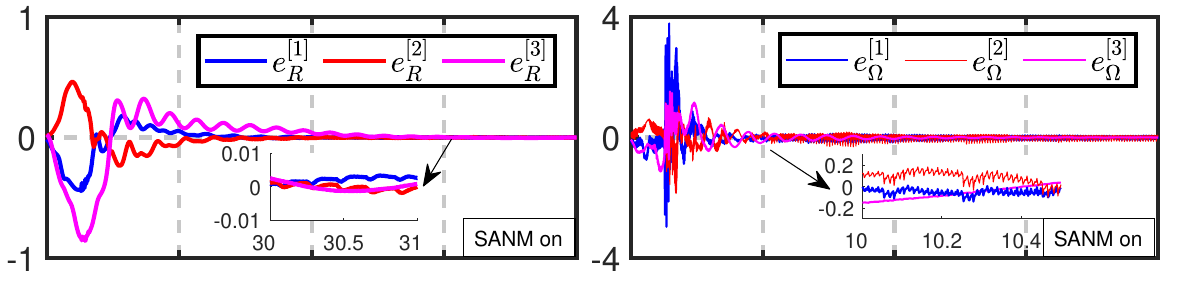}
    \label{}
  \end{subfigure}
  \hspace{0mm}
  \vspace{-1.5em} %
  \begin{subfigure}[b]{0.49\textwidth}
    \centering
    \includegraphics[width=\linewidth]{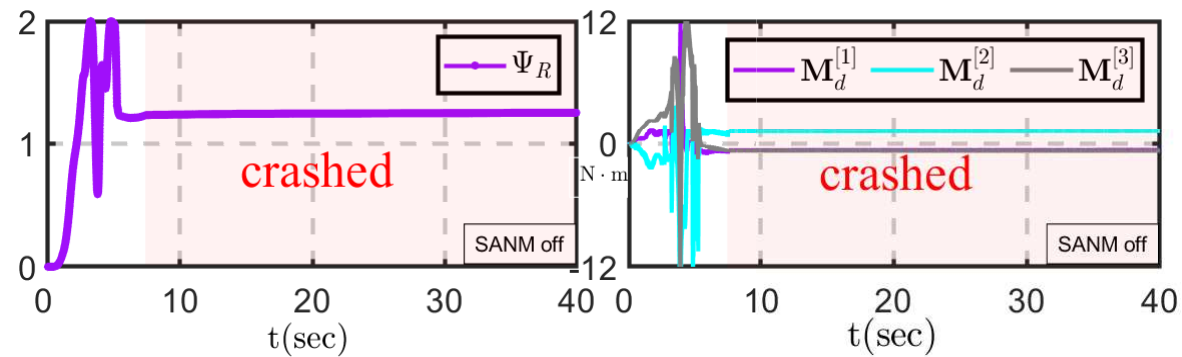}
     \caption{Performance of baseline geometric attitude control (SANM off)}
    \label{}
  \end{subfigure}
  \hspace{1mm}
  \begin{subfigure}[b]{0.49\textwidth}
    \centering
    \includegraphics[width=\linewidth]{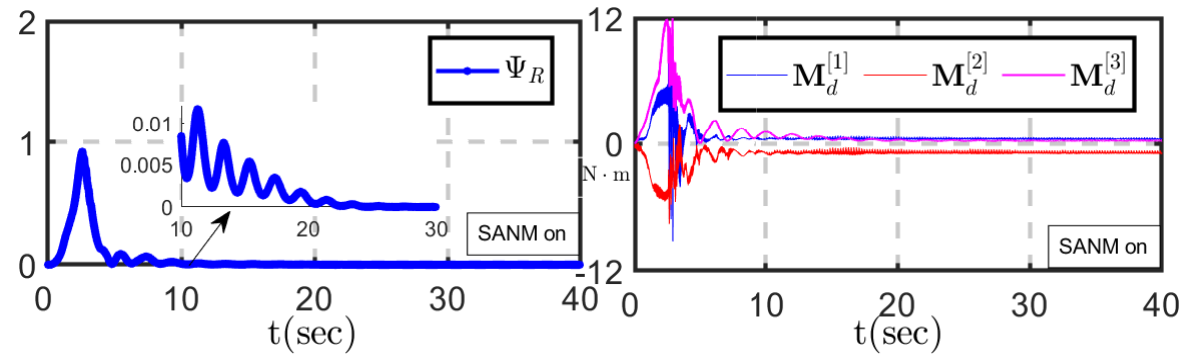}
     \caption{Performance of our geometric attitude control (SANM on)}
    \label{}
  \end{subfigure}
  \hspace{0mm}
    \vspace{-0em} %
    
  \caption{\textit{Experiment 4-}A high-fidelity physics simulation was used to validate 
the $\mathbf{SE}(3)$-\textbf{\textit{Compatibility}} of SANM. For video: \url{https://youtu.be/wadR-C_ZXIU}. }
   \vspace{-0em} %
  \label{Physics-simulation}
\end{figure*}

To validate the $\mathbf{SE}(3)$-\textbf{\textit{Compatibility}}, we further integrated the SANM-augmented attitude controller into the complete geometric control on $\mathbf{SE}(3)$ (Supp. \ref{supp:SE(3)}, and \cite{2010 Geometric tracking control of a quadrotor UAV on SE(3), 2013 Geometric nonlinear PID control of
a quadrotor UAV on SE(3)}). 
This experiment used a Software-In-The-Loop (SITL) simulation integrated with the \textit{Gazebo Harmonic} physics engine. This simulation environment replicated high-fidelity physical dynamics of quadrotor flight, including effects such as sensor noise, motor delay, and external disturbances. The geometric control loop was executed within the \textit{ArduPilot-SITL} environment at a frequency of 400~Hz. To introduce time-varying disturbance moments, the quadrotor model carried a payload suspended by a cable at an offset position from the center of mass, as shown in Fig.~\ref{Physics-simulation}.

The physical properties of the quadrotor-payload model were as follows: 
\begin{equation}
{\small
   \begin{array}{cc}
   &m = 1.6~\mathrm{kg}, \,m_p= 0.25~\mathrm{kg}, \,m_c= 0.02~\mathrm{kg}, \\
   &\bm{J}=10^{-2}\mathrm{diag}[1.1,2.0,2.3]~\mathrm{kg}\mathrm{m}^{2},
   \notag
\end{array} 
}
\label{Parameters of quadrotor-payload model}
\end{equation}
where $m$, $m_p$ and $m_c\in\mathbb{R}$ are the mass of quadrotor, payload and cable, respectively. 
The attitude controller was tested under \textbf{\textit{Scenario 2 ($\bm{J}$ is unknown)}} and the PD gains were chosen as $k_{R}=100$, $ k_{\Omega}=80$. For parameters of the \textit{``slices"}, refer to Supp. \ref{supp:sec:Experiment-4}.

 \begin{figure}[] %
  \vspace{-0em} %
  \centering
 \includegraphics[width=\linewidth]{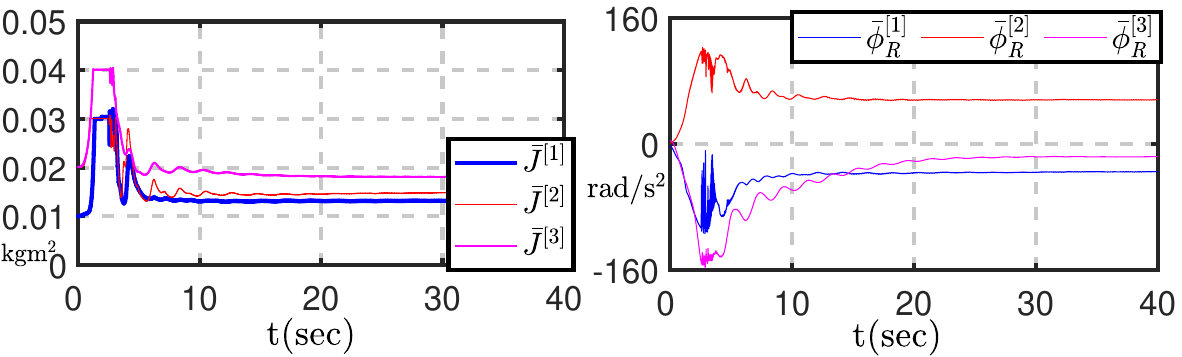}      
 \caption{ Real-time outputs of SANM in \textit{Experiment 4}.  } %
      \label{Physics-simulation_NN}
\end{figure}

During simulation, the desired heading direction was maintained $\bm{\vec{b}}_{1\bm{d}}(t):=(1,0,0)^{\top}$. We assigned a fixed target altitude for the quadrotor take-off, without setting any horizontal position commands. 
To assess the effect of the SANM module in isolation, we performed a controlled comparison between the proposed method (SANM on) and the baseline controller (SANM off), where all experimental settings and control parameters were kept identical except for the activation of SANM. The comparative results are presented in Fig.~\ref{Physics-simulation}.
The values of the estimated inertia and disturbance features identified online by SANM are shown in Fig.~\ref{Physics-simulation_NN}. 
These results empirically validate the $\mathbf{SE}(3)$-\textbf{\textit{Compatibility}}.

\begin{figure*}[htbp] %
\vspace{-0em} %
  \centering
  \begin{subfigure}[b]{0.48\textwidth}
    \centering
    \includegraphics[width=\linewidth]{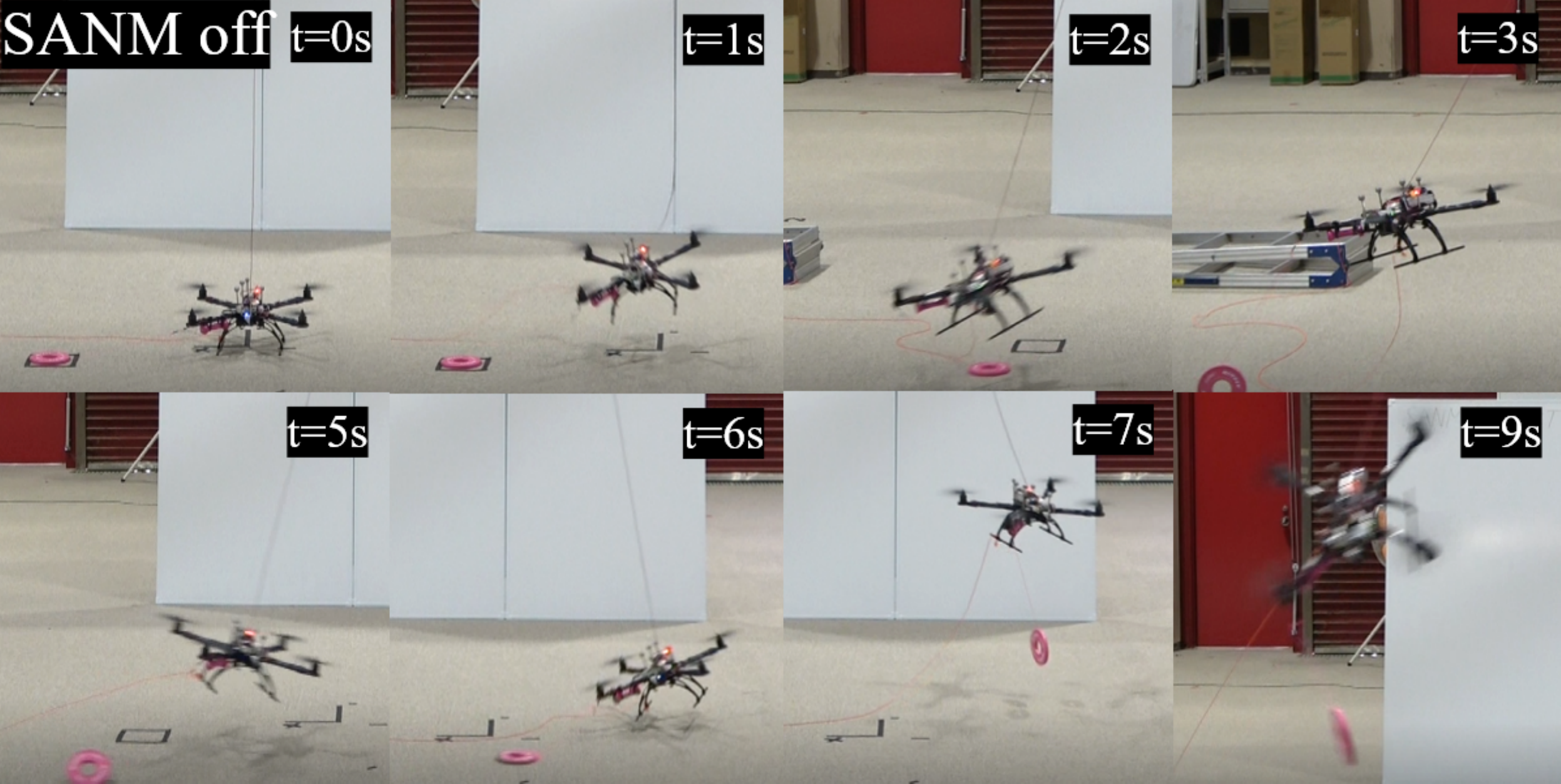}
    
  \end{subfigure}
  \vspace{-0em}
  \hspace{2mm}
  \begin{subfigure}[b]{0.48\textwidth}
    \centering
    \includegraphics[width=\linewidth]{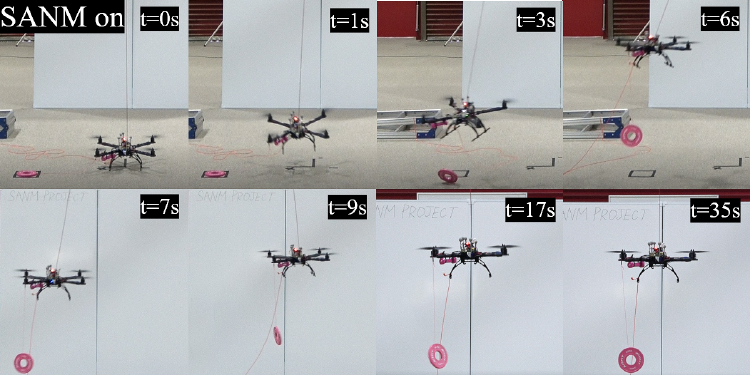}
  \end{subfigure}
  \hspace{0mm}
  \vspace{+0.5em}

  \captionsetup{labelformat=default, labelsep=colon}
  \begin{subfigure}[b]{0.49\textwidth}
    \centering
    \includegraphics[width=\linewidth]{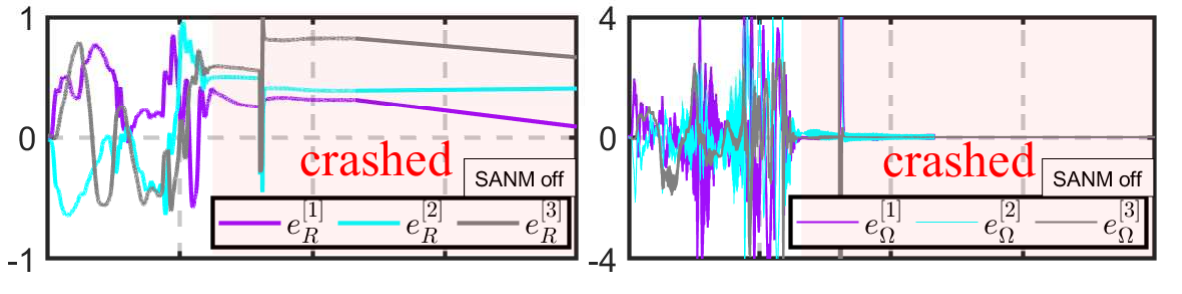}
    \label{}
  \end{subfigure}
  \hspace{1mm}
  \begin{subfigure}[b]{0.49\textwidth}
    \centering
    \includegraphics[width=\linewidth]{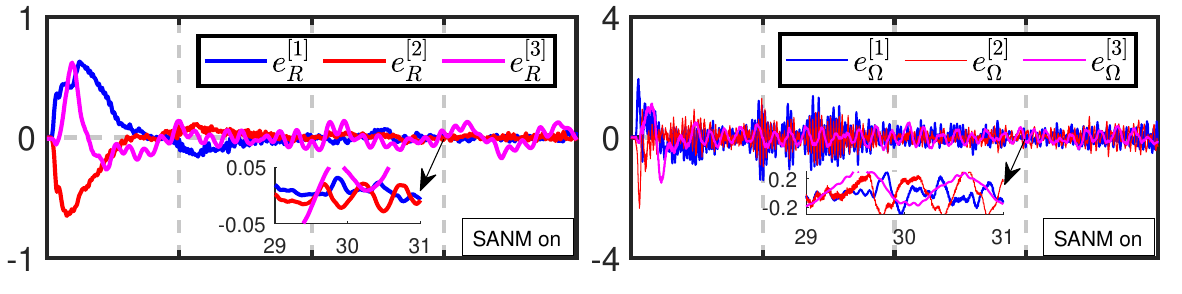}
    \label{}
  \end{subfigure}
  \hspace{0mm}
  \vspace{-1.5em} %
  \begin{subfigure}[b]{0.49\textwidth}
    \centering
    \includegraphics[width=\linewidth]{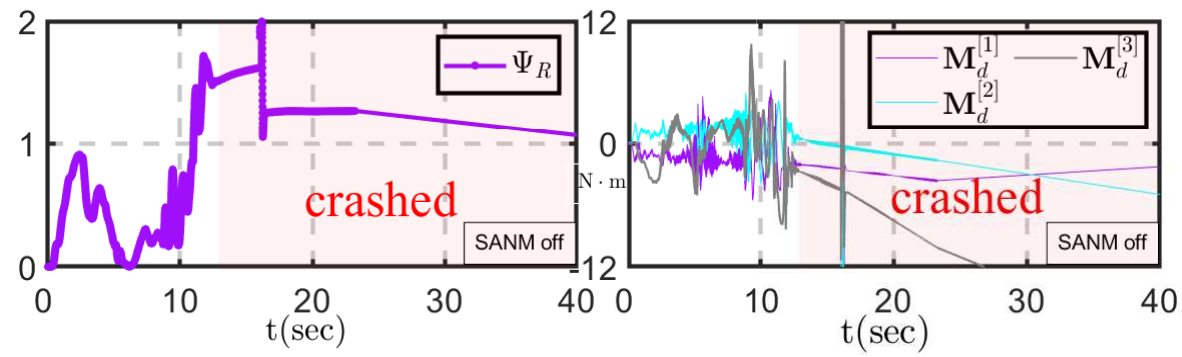}
    \caption{Performance of baseline geometric attitude control (SANM off)}
    \label{}
  \end{subfigure}
  \hspace{1mm}
  \begin{subfigure}[b]{0.49\textwidth}
    \centering
    \includegraphics[width=\linewidth]{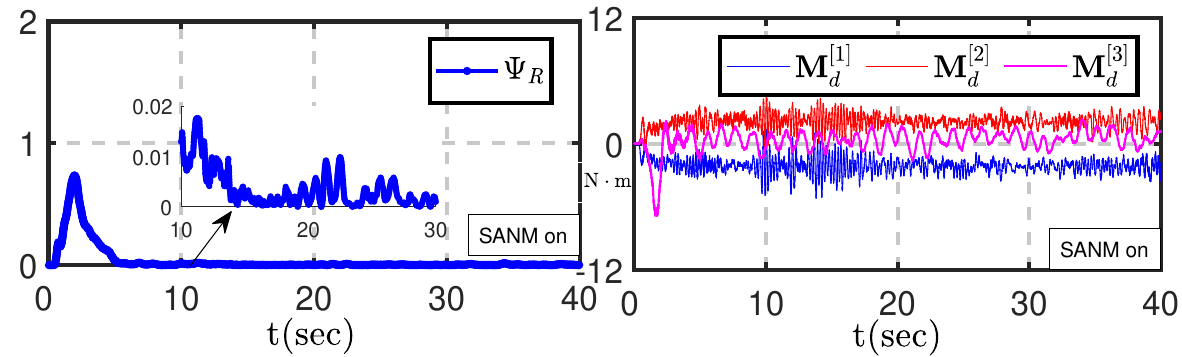}
    \caption{Performance of our geometric attitude control (SANM on)}
    \label{}
  \end{subfigure}
  \hspace{0mm}
    \vspace{-0em} %
    
  \caption{\textit{Experiment 5-}Real-world flight test to validate the lightweight nature and overall feasibility of SANM. For video: \url{https://youtu.be/wadR-C_ZXIU}. }
   \vspace{-0em} %
  \label{Real-flight}
\end{figure*}

\subsection{Experiment 5: (Real-world Flight)}
\label{Exp::Real-world-Flight}
This experiment was conducted in a motion capture (mo-cap) environment to further validate the real-world feasibility, building upon \textit{Experiment 4}. Here, three RBF neural networks with $l=5$ neurons were employed, such that each execution of SANM required only $3 \times 5 = 15$ evaluations of the \texttt{expf()} function for the hidden layer computation.
With this setup, the complete control loop, including the online learning process implemented by the SANM module, can run at 400~Hz on the \textit{STM32H750}-based FCU. During real-time flight experiments involving online estimation and compensation of an off-center suspended payload, the recorded memory usage was $255.2~\mathrm{kB}$, corresponding to $24.9\%$ RAM usage, while the CPU utilization averaged $77.2\%$ with a peak of $81.2\%$. For the deployment status of current learning-based methods \cite{2021 Geometric Adaptive Control With Neural Networks
for a Quadrotor in Wind Fields}, \cite{2024 Neural Moving Horizon Estimation for Robust Flight Control}, \cite{2022 Neural-Fly enables rapid learning for agile flight in strong winds}, \cite{2023 Quadrotor Neural Network Adaptive Control: Design and Experimental Validation}, \cite{2022 Neural-Swarm2: Planning and Control of Heterogeneous Multirotor Swarms Using Learned Interactions}, \cite{2024 Efficient Deep Learning of Robust Policies From MPC Using Imitation and Tube-Guided Data Augmentation}, see Table~\ref{TABLE_Deployment_Status}. 

 \begin{figure}[] %
  \vspace{-0em} %
  \centering
 \includegraphics[width=\linewidth]{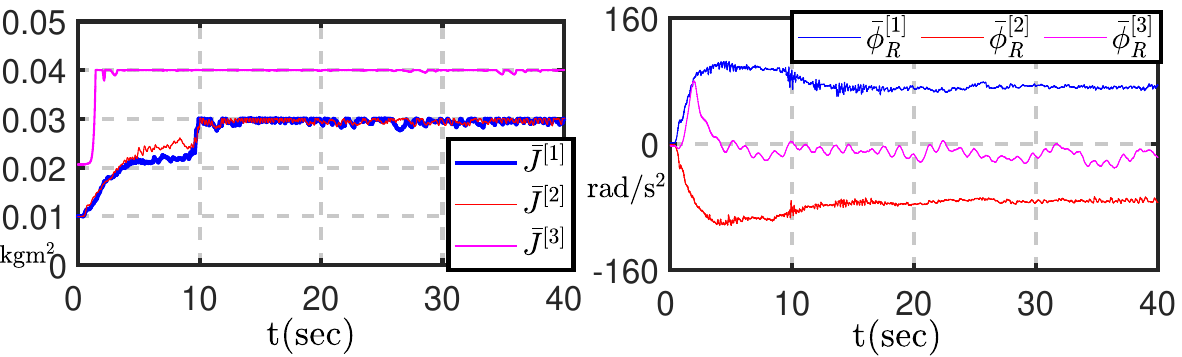}      
 \caption{Real-time outputs of SANM in \textit{Experiment 5}.  } %
      \label{Real-flight_NN}
\end{figure}
The main physical properties were maintained the same as in \textit{Experiment 4}. The mass of the quadrotor and payload were still $1.6\,\mathrm{kg}$ and $0.25\,\mathrm{kg}$. To establish  \textbf{\textit{Scenario 2 ($\bm{J}$ is unknown)}}, an additional $0.25\,\mathrm{kg}$ dumbbell was mounted at an offset position on the quadrotor arm, as shown in Figs.~\ref{Quadrotor} and \ref{Real-flight}. As a result, nearly $800\,\mathrm{g}$ of weight was loaded on a single rotor, which approached the maximum thrust capacity of our hardware. In this experiment, due to the sim-to-real gap, such as differences in motor thrust and response, some control parameters were reduced (see Supp. \ref{supp:sec:Experiment-5}). 
As shown in Figs.~\ref{Real-flight} and \ref{Real-flight_NN}, the real-world results exhibited strong agreement with the physics simulation. In addition, this experiment demonstrates the \textbf{\textit{Lightweight}} nature of SANM, showing that it remains computationally feasible even when integrated with the position-control loop. Furthermore, the close consistency between the physics simulation and the real-world experiments indicates that the SANM module achieves robust sim-to-real transfer performance.

\section{Conclusion}
\label{Conclusion}
This work fills a longstanding gap in embedded learning-based flight control by demonstrating a 400 Hz online learning 
controller running on \textit{STM32}-based microcontrollers. Prior to this, online learning in quadrotor control was typically considered computationally expensive and thus relied heavily on high-performance external platforms such as \textit{NVIDIA Jetson} boards or ground computers. In addition, SANM provides a theoretical guarantee of exponential convergence to an arbitrarily small ball under time-varying disturbances within the identification region, and the convergence rate is explicitly shown to be tunable. This represents a clear step beyond the conventional uniform ultimate boundedness (UUB)-type results commonly reported in state-of-the-art geometric control literature (e.g., \cite{2021 Geometric Adaptive Control With Neural Networks for a Quadrotor in Wind Fields}, \cite{2025 High Maneuverability and Efficiency Control for Hybrid Quadrotor With All-Moving Wings in SE(3) Based on Deep Reinforcement Learning}, \cite{2025 L1Adaptive Augmentation of Geometric Control for Agile Quadrotors With Performance Guarantees}). These stem from the \textit{$\mathbf{SO}(3)$-Preserving}, \textit{Geometry Consistency} properties and the universal approximation capability, revealing the significant potential of \textit{Sliced Learning} for  online disturbance identification in quadrotor geometric control. Moreover, grounded in the \textit{subspace sharing}, the framework offers  substantial extensibility for further development across the geometric control community.

\textbf{\textit{Future Work:}} Building upon the current formulation, a natural extension of the proposed SANM module is toward a 12-slice SANM design on $\mathbf{SE}(3)$. Specifically, the six-degree-of-freedom (6-DoF) rigid-body motion can be decomposed into three translational axes and three rotational axes, totaling six groups of dual slices, thereby enabling simultaneous compensation of both the position and attitude loops. A full-state version of SANM \cite{2025 Robustness Enhancement for Multi-Quadrotor Centralized Transportation System via Online Tuning and Learning}, \cite{2025 Online Identification using Adaptive Laws and Neural Networks for Multi-Quadrotor Centralized Transportation System} has also been preliminarily examined in centralized multi-quadrotor transportation, where it acts as a compensator for payload control.

\vspace{-33pt}
\begin{IEEEbiography}[{\includegraphics[width=1in,height=1.25in,clip,keepaspectratio]{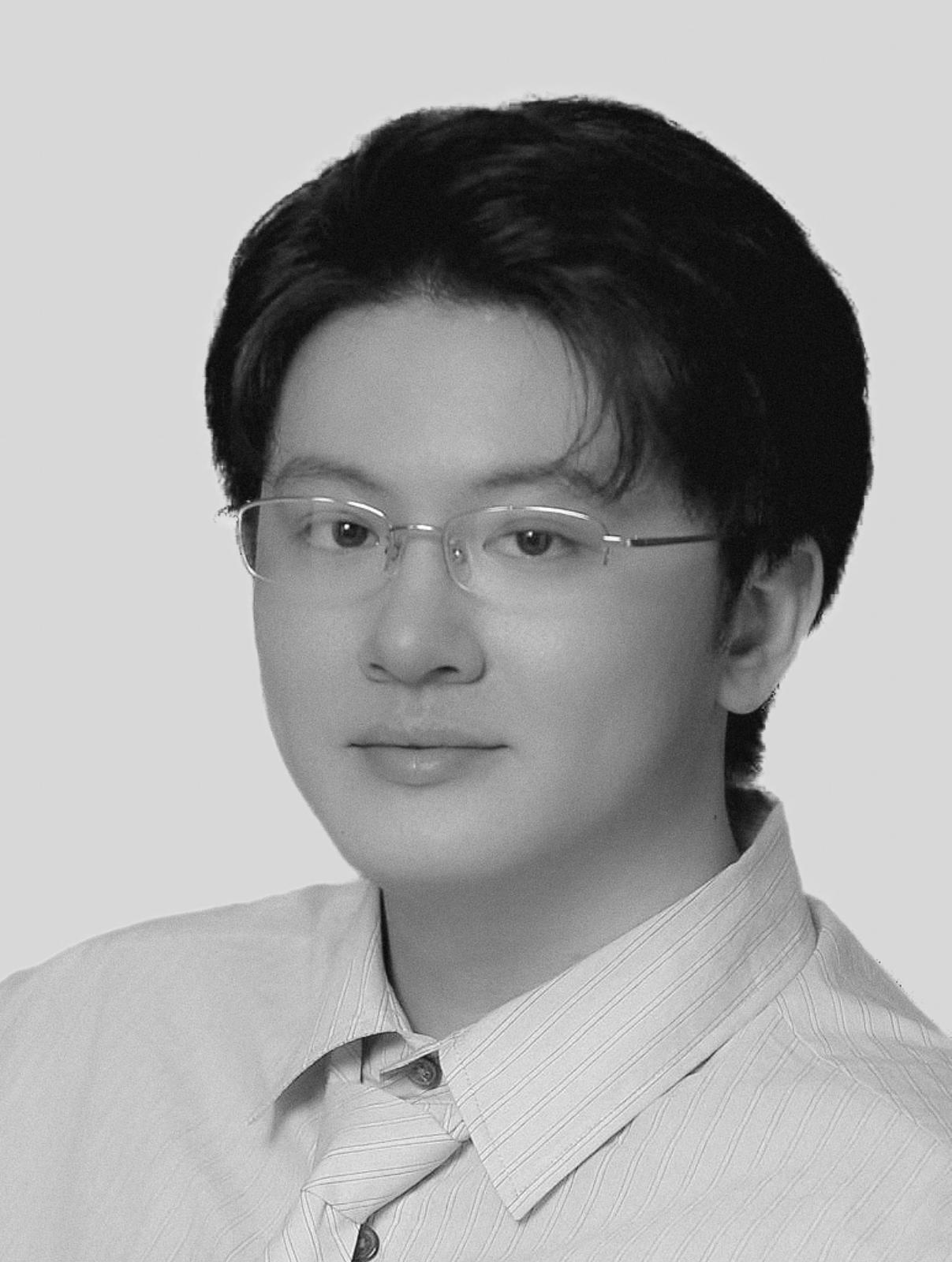}}]{Tianhua Gao} is currently working toward the Ph.D. degree with the Graduate School of Systems and Information Engineering, University of Tsukuba, Japan. During his doctoral study, he has been awarded the University of Tsukuba Fellowship and the JST SPRING Scholarship. He is also a Research Assistant at the Intelligent Systems Research Institute, National Institute of Advanced Industrial Science and Technology (AIST), Japan. His research interests focus on integrating control theory and neuroscience to advance the field of machine intelligence and control.
\end{IEEEbiography}
\vspace{-33pt}
\begin{IEEEbiography}
[{\includegraphics[width=1in,height=1.25in,clip,keepaspectratio]{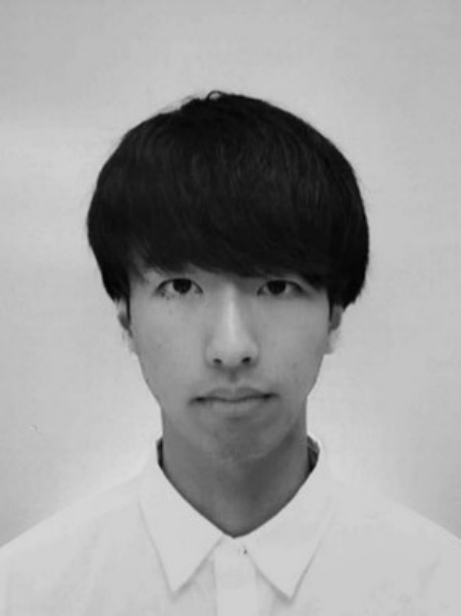}}]{Masashi Izumita}
received his B.E. degree from the Faculty of Advanced Engineering, Tokyo University of Science, Japan, in 2023, and his M.E. degree from the Graduate School of Science and Technology, University of Tsukuba, Japan, in 2025. He joined the Intelligent Systems Research Institute, National Institute of Advanced Industrial Science and Technology (AIST) as a research scientist in 2025. His research interests include autonomous aerial robotics, with a particular focus on multi-robot-systems, LiDAR–IMU–based SLAM, real-time perception and mapping using ROS frameworks, and path planning. He is a member of the Robotics Society of Japan (RSJ).
\end{IEEEbiography}
\vspace{-33pt}
\begin{IEEEbiography}
[{\includegraphics[width=1in,height=1.25in,clip,keepaspectratio]{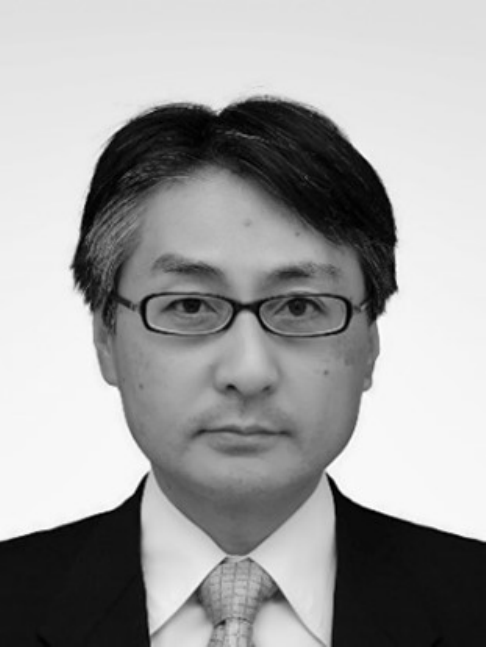}}]{Kohji Tomita}
(Member, IEEE) received his B.E., M.E. and Ph.D. degrees from the University of Tsukuba in 1988, 1990, and 1997, respectively. He joined the Mechanical Engineering Laboratory, AIST, MITI, in 1990, and has been conducting research at the National Institute of Advanced Industrial Science and Technology (AIST) as a senior research scientist since 2001. He was a visiting researcher at Dartmouth College from 2000 to 2001. His research interests span distributed systems from theory to application, including graph-based computational models, distributed software architectures, modular robotic systems, and autonomous unmanned aerial vehicles (UAVs).
\end{IEEEbiography}
\vspace{-33pt}
\begin{IEEEbiography}
[{\includegraphics[width=1in,height=1.25in,clip,keepaspectratio]{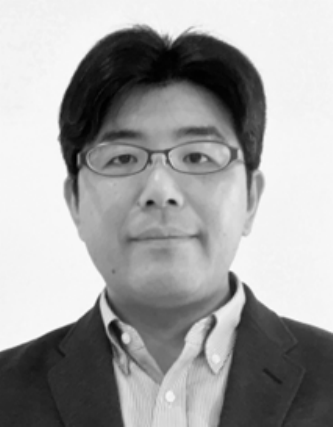}}]{Akiya Kamimura}
(Member, IEEE) received his M.E. and Ph.D. degrees from the Graduate School of Engineering, the University of Tokyo in 1997 and 2000, respectively. He joined the Mechanical Engineering Laboratory, AIST, MITI in 2000, and has been conducting research at the National Institute of Advanced Industrial Science and Technology (AIST) since 2001. From Mar. 2009 to Mar. 2010, he was a visiting researcher at the Information Sciences Institute, University of Southern California, USA. As of 2025, he serves as Deputy Director of the Intelligent Systems Research Institute at AIST. His research interests include modular robotics, field robotics, infrastructure inspection robots, and IoT communication systems. He has authored more than 90 articles and has acquired 17 patents in Japan, EPO, USA, Germany, and UK. He received the Best Paper Award at international conferences in 2002, 2003, 2006, and 2018. Since FY2017, he was an associate professor with the Partner Graduate School of Master’s and Doctoral Programs in Intelligent and Mechanical Interaction Systems, University of Tsukuba, and has been a professor since FY2021.
\end{IEEEbiography}

\vfill

\begin{supplementary}[Supplementary Material]

\setcounter{tocdepth}{2}   
\tableofcontents 

\clearpage
\section{Notation, Parameters, and Reference Tables}
\begin{table*}[!t]
  \centering
  \caption{List of symbol references.}
  \label{TABLE_Symbol_References}
  \begin{tabular}{>{\columncolor{gray!10}}c|l|>{\columncolor{gray!10}}c|l}
    \hline
    \hline
      $\bm{R}\in \mathbf{SO}(3)$ &Quadrotor attitude rotation matrix& $\bm{J}\!\in\!\mathbb{R}^{3\times3}$& \,\,Inertia tensor of the quadrotor \,\,\\
     $\bm{R}_{{\bm{d}}}\in \mathbf{SO}(3)$ &Desired quadrotor attitude rotation matrix& $\bm{\phi}_{\bm{\textit{R}}}\in \mathbb{R}^3$& \,\,Time-varying dynamics  term of rotational disturbance\,\\
    $\bm{\Omega}\in\mathbb{R}^3$& Angular velocity of the quadrotor& $\bm{e}_{\bm{R}}\in \mathbb{R}^3$& \,\,Attitude error of the quadrotor\\
     $\bm{\Omega}_{{\bm{d}}}\in\mathbb{R}^3$& Desired angular velocity of the quadrotor& $\bm{e}_{\bm{\Omega}}\in \mathbb{R}^3$& \,\,Angular velocity error of the quadrotor\\
   $\bm{\mathrm{M}}\in\mathbb{R}^3$ & Control moment& $\mathbf{E}_{\bm{\textit{R}}}\in\mathbb{R}^6$ & \,\,Rotational state error vector\,\,\\
   $\bm{\mathrm{M}_d}\in\mathbb{R}^3$ & Desired control moment& $\bm{J}^{\text{vec}}\in\mathbb{R}^{3}$ &\,\,Vectorized inertia tensor of the quadrotor\,\,\\
  $\textbf{x}_{\bm{\textit{R}} j}\in \mathbb{R}^2$ & Input vector of the $j^{th}$ neural network&$\bm{\mathcal{W}}_{\bm{\textit{R}} j}\in\mathbb{R}^{l}$ &\,\,Weight vector of the $j^{th}$ neural network\,\,\\
  $\bm{\hbar}(\cdot)\in\mathbb{R}^l$ & Gaussian activation function& $\bm{\varpi}_{\bm{\textit{R}}}\in\mathbb{R}^{3}$ &\,\,Vectorized inertia tensor of the quadrotor\,\,\\
    \hline
    \hline
  \end{tabular}
\end{table*}
\begin{table*}[!t]
  \centering
  \caption{List of parameter references.}
  \label{TABLE_Parameter_References}
  \begin{tabular}{>{\columncolor{gray!10}}c|l|>{\columncolor{gray!10}}c|l}
    \hline
    \hline
     $k_R,k_\Omega\in\mathbb{R}$ &PD controller gains& $c_R\in\mathbb{R}$& \,\,Positive constant for attitude error\,\,\\
     $1/\eta_{j}\in\mathbb{R}$ &Adaptive rate of $j^{th}$ adaptive law & $\mathfrak{s}_{j}\in\mathbb{R}$& \,\,Pull-back factor of $j^{th}$ adaptive law\,\\
    $\textbf{c}_{kj}\in\mathbb{R}^{2}$& Center vector of the $k^{th}$  RBF in $j^{th}$ neural network& $b_{kj}\in\mathbb{R}$& \,\,Width of the $k^{th}$ RBF in $j^{th}$ neural network\\
     $\gamma_{\bm{\textit{R}}j}\in\mathbb{R}$& Learning rate of $j^{th}$ neural network& $\zeta_j\in\mathbb{R}$& \,\,Dead-zone threshold\\
    \hline
    \hline
  \end{tabular}
\end{table*}
\begin{table*}[!t]
  \centering
  \caption{Deployment status of existing methods.}
  {\scriptsize
  \begin{tabular}{c|l|l|l}
    \hline
    \hline
    $\textbf{Method}$ & \textbf{Frequency}\!\! & \textbf{Platform}\!\!  & \textbf{Learning Type}\!\! \\
    \hline
  \text{Geometric-Adaptive \cite{2021 Geometric Adaptive Control With Neural Networks
for a Quadrotor in Wind Fields}} &\text{400 Hz} & \textit{NVIDIA Jetson}&\text{Online}\\
  \hline
  \text{Neural-Fly \cite{2022 Neural-Fly enables rapid learning for agile flight in strong winds}} &\text{50 Hz} & \textit{Raspberry Pi 4}&\text{Offline + Online}\\
  \hline
  \text{adaptive-NN \cite{2023 Quadrotor Neural Network Adaptive Control: Design and Experimental Validation}} &\text{45 Hz} & \textit{Ground PC}&\text{Online}\\
  \hline
  \text{NeuroMHE \cite{2024 Neural Moving Horizon Estimation for Robust Flight Control}} &\text{25 Hz} & \textit{Intel NUC}&\text{Offline + Online}\\
  \hline
   \text{Neural-Swarm2 \cite{2022 Neural-Swarm2: Planning and Control of Heterogeneous Multirotor Swarms Using Learned Interactions}} &\text{N/A} & \textit{STM32}&\!\!\text{Offline + Inference}\!\!\\
  \hline
  \text{DNN + RTMPC \cite{2024 Efficient Deep Learning of Robust Policies From MPC Using Imitation and Tube-Guided Data Augmentation}} &\text{500 Hz} & \textit{NVIDIA Jetson}&\!\!\text{Offline + Inference}\!\!\\
  \hline
  \rowcolor{gray!20}\text{SANM (our method)} &\text{400 Hz} & \textit{STM32H7}&\text{Online (\textit{Sliced})}\\
    \hline
    \hline
  \end{tabular}
  }
  \label{TABLE_Deployment_Status}
\end{table*}
\begin{table*}[!t]
\centering
\caption{Comparison among \textit{Robust Adaptive Attitude Control (Lee 2013 \cite{2013 Robust Adaptive Attitude Tracking on SO3 With an Application to a Quadrotor UAV})}, \textit{Geometric Neural-Network (NN) Adaptive Control (Bisheban 2021 \cite{2021 Geometric Adaptive Control With Neural Networks
for a Quadrotor in Wind Fields})}, \textit{$\mathcal{L}_1$ Quad (Wu 2025 \cite{2025 L1Adaptive Augmentation of Geometric Control for Agile Quadrotors With Performance Guarantees})}, and the Proposed \textit{SANM-augmented Geometric Attitude Control}}
{\footnotesize
\renewcommand{\arraystretch}{1.2}
\begin{tabular}{p{2.7cm} p{3.3cm} p{3.3cm} p{3.3cm} p{3.3cm} p{3.3cm}}
\toprule
\textbf{Criterion} 
& \textbf{ Robust Adaptive Attitude Control (Lee 2013)} 
& \textbf{Geometric NN Adaptive Control (Bisheban 2021)} 
& \textbf{$\mathcal{L}_1$ Quad (Wu 2025)} 
& \textbf{SANM (Our Method)} \\ 
\midrule
\textbf{Disturbance Model} 
& Bounded unstructured disturbance $\|\Delta\|\!\le\!\delta$; unknown inertia 
& Bounded time-varying wind-induced disturbance force $\Delta_1$ and moment $\Delta_2$ 
& Matched disturbance $\sigma_{\textit{m}}$ with bounded magnitude and bounded time derivative; 
unmatched disturbance $\sigma_{\textit{um}}$ bounded but not canceled; 
overall uncertainty $\sigma(t,x)$ is continuous and Lipschitz in state $x$. 
& Bounded time-varying continuous universal disturbance $\bm{\phi}_{\textit{R}}(\cdot)$ at the acceleration level, capable of absorbing unknown nonlinear acceleration effects; unknown inertia \\

\textbf{Adaptation Mechanism} 
& Estimates only inertia parameters $\hat{J}$ with $\sigma$-modification 
& Adapts NN weights $(\hat W_i,\hat V_i)$ (with projection and leakage); no inertia adaptation 
& Adaptation filtered by $C(s)$; separation of estimation and control loop via LPF 
& Co-adaptive: inertia adaptation + NN disturbance identification (with dead-zone and projection, without leakage); NN-only mode also supported\\ 

\textbf{Theoretical Guarantee (under disturbance)} 
& UUB + Exponential convergence to a residual set 
& \multicolumn{1}{c}{UUB}
& \multicolumn{1}{c}{UUB}
& Exponential convergence to an arbitrarily small ball + input–state practical stability (ISpS)  \\ 

\textbf{Convergence Rate Guarantee} 
& Exponential attractiveness to a residual set, but rate is not tunable
& No convergence rate guarantee 
& No convergence rate guarantee; LPF bandwidth affects transient behavior but does not constitute a formal rate bound
& Axis-wise Tunable exponential convergence rate \\ 

\textbf{Hardware Platform} 
& \multicolumn{1}{c}{OMAP 600 MHz} 
& \multicolumn{1}{c}{NVIDIA Jetson TX2}  
& \multicolumn{1}{c}{Pixhawk 4 mini}
& \multicolumn{1}{c}{STM32H750} \\ 

\textbf{Frequency} 
& \multicolumn{1}{c}{Unknwon} 
& \multicolumn{1}{c}{400~Hz}
& \multicolumn{1}{c}{400~Hz}
& \multicolumn{1}{c}{400~Hz} \\ 
\bottomrule
\end{tabular}
}
\label{tab:comparison_with_SOTA}
\end{table*}
This section provides supplementary reference tables used throughout the main text.
Table~\ref{TABLE_Symbol_References} lists the mathematical symbols and their corresponding physical meanings, units, and notations, 
while Table~\ref{TABLE_Parameter_References} summarizes the key experimental and simulation parameters, including their nominal values and sources of reference. 
These tables serve as a unified reference for readers to easily track the notation and parameter settings adopted in this work.
In addition, Table~\ref{TABLE_Deployment_Status} presents the deployment status of representative existing learning-based flight-control methods, detailing their operating frequency, onboard platform, and learning modality. A comprehensive comparison with representative geometric controllers is provided in Table~\ref{tab:comparison_with_SOTA}.
Together, these tables offer a consolidated reference to conveniently track the notation, parameter settings, and methodological context adopted in this work.

\section{Overview of This Work}
\label{supp:overview}
\begin{figure}[ht]
      \centering
      \includegraphics[scale=0.32]{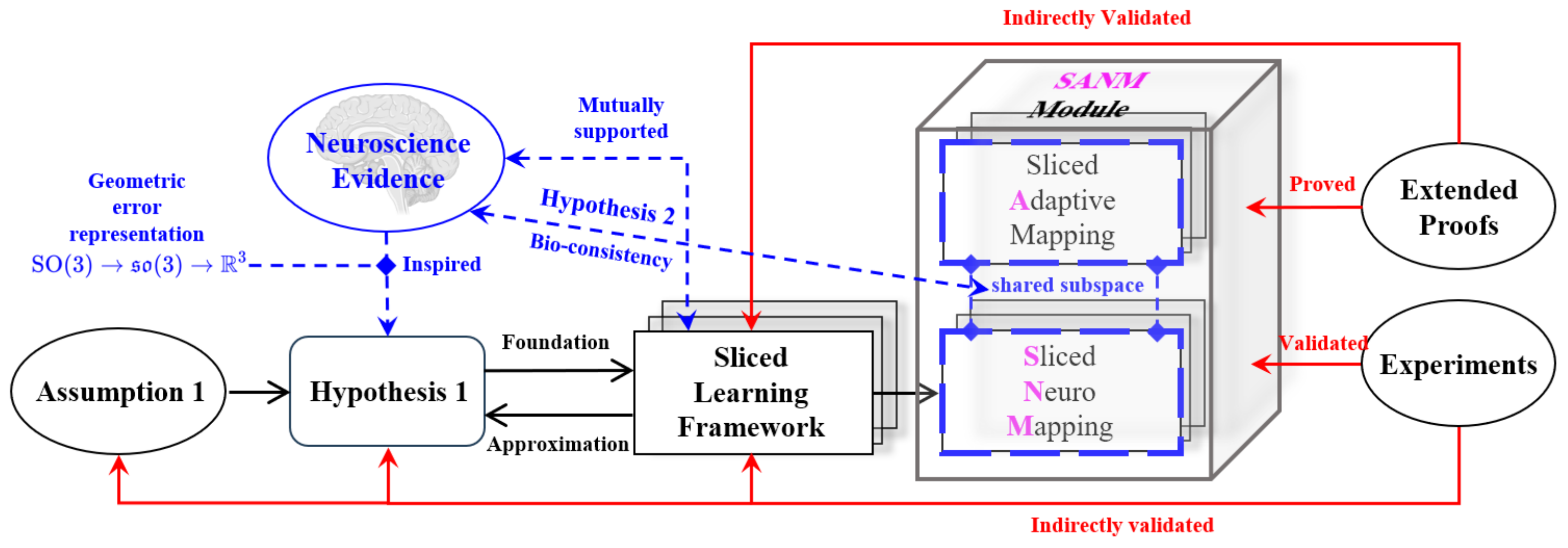}
      \caption{\footnotesize Overview of this work. The geometric error representation and neuroscience evidence motivate two key hypotheses that support the proposed \textit{Sliced Learning} framework. Based on this framework, the SANM module is constructed with axis-wise adaptive and neural mappings operating on shared subspaces. Theoretical proofs and experiments both directly validate the proposed SANM module, thereby indirectly supporting the underlying \textit{Sliced Learning} framework from which SANM is constructed. }
      \label{fig:overview}
\end{figure}

This work is motivated by the Lie-algebra–based error representation in geometric control \cite{2010 Geometric tracking control of a quadrotor UAV on SE(3)} and its consistency with the geometric mechanisms observed in biological neural systems \cite{2019 Cortical Areas Interact through a Communication Subspace}, \cite{2019 High-dimensional geometry of population responses in visual cortex}, \cite{2025 Multiplexed subspaces route neural activity across brain-wide networks}. Fig.~\ref{fig:overview} illustrates the conceptual flow from motivation to theory and validation.
The starting point of our framework is the Lie-algebraic geometric error representation, which maps attitude errors from $\mathbf{SO}(3)$ into $\mathfrak{so}(3)$ and subsequently into $\mathbb{R}^3$. This representation provides the \textit{Geometry Consistency} property for learning. Using \emph{errors}, rather than rotational \textit{states} (such as Euler angles), forms the basis of the proposed \textit{learning-from-error} strategy. Inspired by neuroscience findings on subspace-based information processing, we further introduce one modeling assumption and two central hypotheses:

\begin{itemize}
    \item \textbf{\textit{Assumption~1 (Local Pseudo-Inverse Mapping):}} In a compact, practically relevant operating region, there exists a local pseudo-inverse mapping from the rotational error vector to a set of disturbance-relevant quantities (desired moments, inertia-related terms, and disturbance dynamics).
    \item \textbf{\textit{Hypothesis~1 (Sliceability):}} The disturbance--error mapping relationship can be decomposed into axis-wise low-dimensional submappings (\textit{``slices"}).
    \item \textbf{\textit{Hypothesis~2 (Subspace Sharing):}} These axis-aligned geometric subspaces can be shared by different learning mechanisms, such as adaptive laws and neural networks, in a mutually supported manner.
\end{itemize}

Together, \textbf{\textit{Assumption~1}} and \textbf{\textit{Hypotheses~1 and 2}} establish the \textbf{\textit{Sliced Learning Framework}}. This framework reorganizes the otherwise high-dimensional disturbance identification problem into several parallel, interpretable, and geometry-consistent low-dimensional mappings. Its structure is supported by the intrinsic geometry of $\mathbf{SO}(3)$.

Based on this framework, we construct the \textbf{\textit{Sliced Adaptive-Neuro Mapping} (SANM) module}, consisting of:
\begin{enumerate}
    \item \textbf{\textit{Sliced Adaptive Mapping}} --- axis-wise adaptive laws estimating inertia-related effects.
    \item \textbf{\textit{Sliced Neuro Mapping}} --- shallow neural networks approximating time-varying disturbance components.
\end{enumerate}
Both components operate on a shared set of geometric subspaces and are updated from the same Lie-algebraic error inputs. The SANM output augments a standard geometric attitude controller \cite{2011 Geometric tracking control of the attitude dynamics of a rigid body on SO(3)}, preserving its geometric structure while enhancing robustness under disturbances and uncertainties.

The supplementary theoretical analysis provides formal guarantees for this construction. We establish almost-global exponential attractiveness of the rotational error dynamics, boundedness of all adaptive and neural parameters, and local exponential convergence to an arbitrarily small ball. These results validate the SANM module and indirectly validate the \textit{Sliced Learning} framework.

Finally, numerical simulations, testbed experiments, Gazebo simulations, and real flight tests collectively demonstrate the effectiveness of the proposed method. SANM consistently improves convergence, reduces residual errors, and enhances disturbance rejection under unknown inertia and strong time-varying disturbances, while remaining lightweight enough for embedded implementation on an STM32-class flight controller.

In summary, as illustrated in Fig.~\ref{fig:overview}, this work follows a structured progression:
\emph{geometric insight $\rightarrow$ neuroscience inspiration $\rightarrow$ Sliced Learning framework $\rightarrow$ SANM design $\rightarrow$ theoretical guarantees $\rightarrow$ experimental validation}.

\section{Quadrotor Geometric Control on $\mathbf{SE}(3)$}
\label{supp:SE(3)}

\begin{figure}[!t]
    \centering

    \begin{subfigure}[t]{0.35\linewidth}
        \centering
        \includegraphics[width=\linewidth]{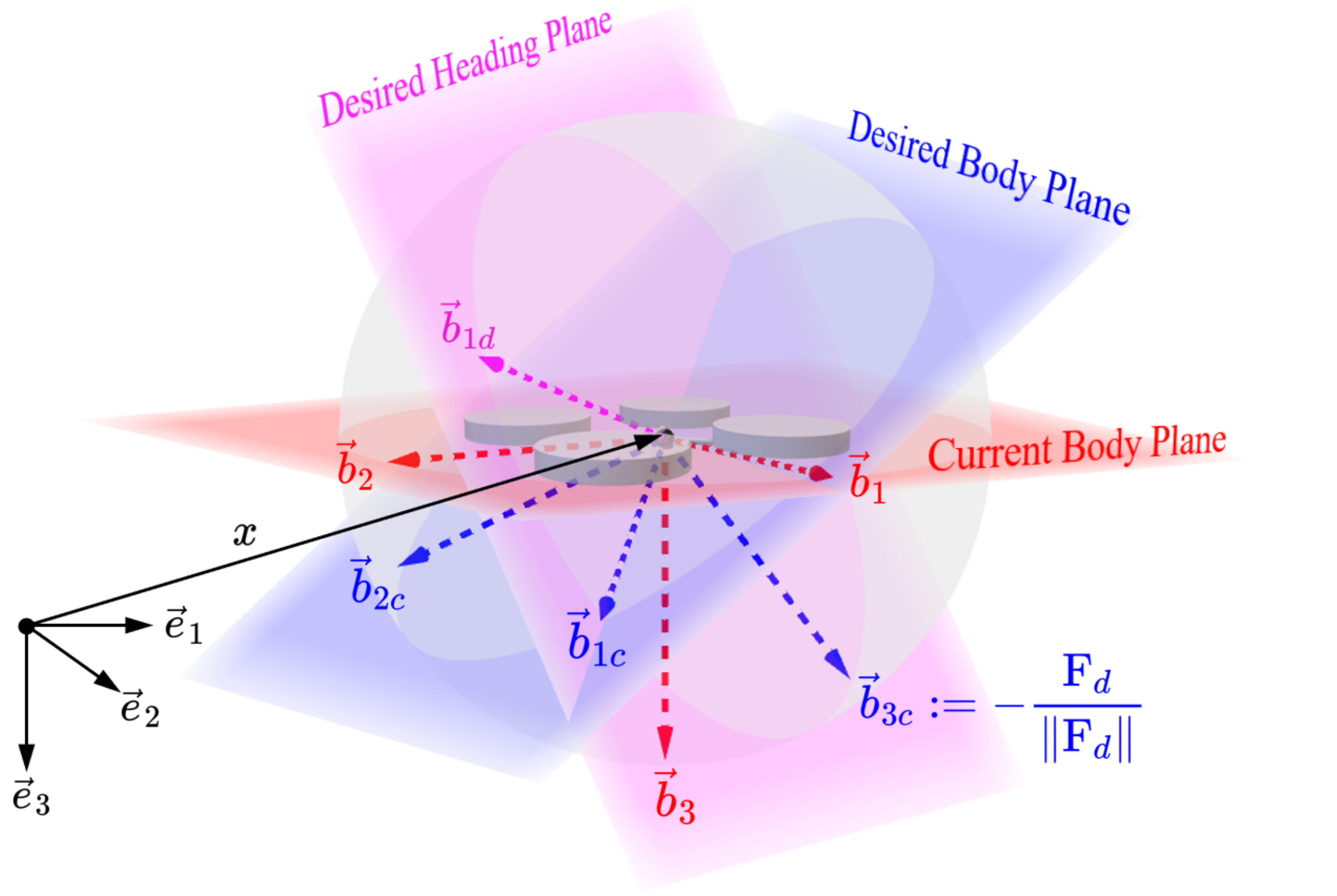}
        \caption{\footnotesize 
            Quadrotor modeling on $\mathbf{SE}(3)$.
        }
        \label{fig:SE(3)}
    \end{subfigure}
    \begin{subfigure}[t]{0.6\linewidth}
        \centering
        \includegraphics[width=\linewidth]{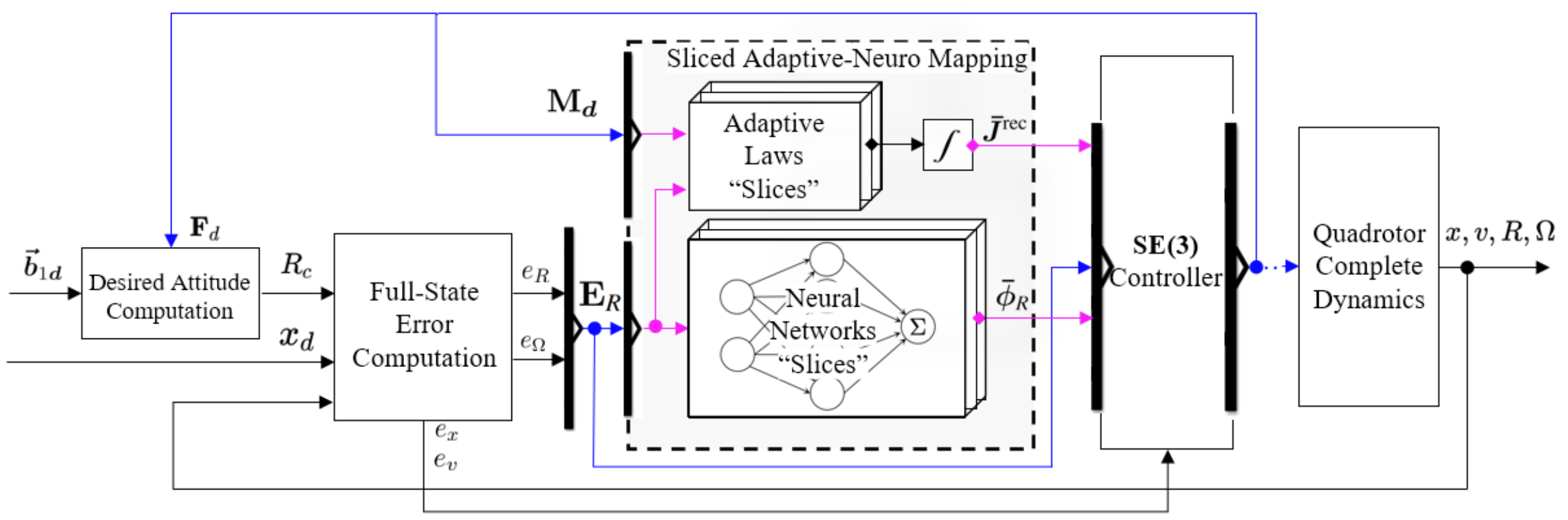} %
        \caption{\footnotesize 
            Coupling into $\mathbf{SE}(3)$.
        }
        \label{fig:Frame}
    \end{subfigure}
    \caption{\footnotesize
        Quadrotor modeling and geometric control on $\mathbf{SE}(3)$. The vectors $\bm{\vec{b}}_{1\bm{d}}$, $\bm{\vec{b}}_{1\bm{c}}$ and $\bm{\vec{b}}_{3\bm{c}}$ are coplanar and form the desired heading plane, while the vectors $\bm{\vec{b}}_{1\bm{c}}$ and  $\bm{\vec{b}}_{2\bm{c}}$ are coplanar and form the desired body plane.).
    }
    \label{fig:SE3_combined}
\end{figure}

A North-East-Down (NED) inertia frame $\mathcal{I}:=\{\bm{\vec{e}}_j\}_{1\leq j \leq 3}$ and an NED quadrotor body-fixed frame  $\mathcal{B}:=\{\bm{\vec{b}}_{j}\}_{1\leq j \leq 3}$ are defined as shown in Fig.~\ref{fig:SE(3)}. The quadrotor model is considered as a rigid body with its center of mass located at the geometric center of the structure, denoted as $\bm{x}\in\mathbb{R}^3$. The orientation of the quadrotor is described by a rotation matrix $\bm{R}\in \mathbf{SO}(3) = \{\bm{R}\in\mathbb{R}^{3\times3}\mid\bm{R}^{\top}\bm{R} = \bm{I}^{3\times3}, \mathrm{det}(\bm{R}) = 1\}$, which represents the rotation of $\mathcal{B}$ relative to $\mathcal{I}$.

The quadrotor geometric control pipeline adopts a multi-level control flow, involving hierarchical control signal transformations from high-level commands to low-level actuator inputs:
\begin{align}
\to\bm{\mathrm{w}_d}^{6\times1}\to{\renewcommand{\arraystretch}{1}\begin{pmatrix}f_d \\ \bm{\mathrm{M}_d}\\\end{pmatrix}}^{4\times1}\!\!\!\!\!\!\to \bm{T}_{\bm{d}}^{4\times1}\to\bm{\omega}^{4\times1}\!\!\to {\renewcommand{\arraystretch}{1}\begin{pmatrix}f \\ \bm{\mathrm{M}}\\\end{pmatrix}}^{4\times1}\!\!\!\!\!\!\!\!\to,
\end{align}
where $\bm{\mathrm{w}_d}=(\bm{\mathrm{F}_d}^{\top},\bm{\mathrm{M}_d}^{\top})^{\top}\in\mathbb{R}^{6}$ is the desired control wrench vector. $\bm{\mathrm{F}_d}\in\mathbb{R}^3$ denotes the desired resultant control force, and $\bm{\mathrm{M}_d}\in\mathbb{R}^3$ represents the desired resultant control moment.  $f_d\in\mathbb{R}$ is the desired total thrust projected from the desired resultant thrust $\bm{\mathrm{F}_d}$ onto the body-fixed frame $\bm{\vec{b}}_{3}$ axis: 
\begin{equation}
    f_d:=-\bm{\mathrm{F}_d} \cdot \bm{R}\bm{\vec{e}}_3,
\end{equation}
and the desired thrusts for each rotor $\bm{T}_{\bm{d}}\in\mathbb{R}^{4}=[T_{d1},T_{d2},T_{d3},T_{d4}]^{\top}$ are then computed by following the allocation mapping for the X-configuration: 
\begin{equation}
    \begin{aligned}
\bm{T}^{4\times1}_{\bm{d}}:=\frac{1}{4}{\renewcommand{\arraystretch}{1.2}\begin{bmatrix}1 & \frac{\sqrt{2}}{d}& \frac{\sqrt{2}}{d} & \frac{c'_M}{c'_T} \\ 1 & \scalebox{0.6}{$-$}\frac{\sqrt{2}}{d} & \frac{\sqrt{2}}{d} & \scalebox{0.6}{$-$}\frac{c'_M}{c'_T}\\1 & \scalebox{0.6}{$-$}\frac{\sqrt{2}}{d} & \scalebox{0.6}{$-$}\frac{\sqrt{2}}{d} & \frac{c'_M}{c'_T} \\1 & \frac{\sqrt{2}}{d} &\scalebox{0.6}{$-$}\frac{\sqrt{2}}{d}& \scalebox{0.6}{$-$}\frac{c'_M}{c'_T} \end{bmatrix}}{\renewcommand{\arraystretch}{1.2}\begin{pmatrix}f_d \\ \bm{\mathrm{M}_d}\\\end{pmatrix}}^{4\times1}\!\!\!\!\!\!\!\!,
\end{aligned}
\end{equation}
where $d\in\mathbb{R}$ represents the distance between the center of the body-fixed frame and the rotors. $c'_M\in\mathbb{R}$ and $c'_T\in\mathbb{R}$
are the constant thrust and moment reference coefficients. The rotor speeds for each motor $\bm{\omega}\in\mathbb{R}^4_{+}=[\omega_1,\omega_2,\omega_3,\omega_4]^{\top}$ are then derived as follows:
\begin{equation}
   \omega_i :=\sqrt{\frac{T_{di}}{c'_T}},
\end{equation}
where $\omega_i$ and $T_{di}$ are the rotor speed and desired thrust of the $i^{th}$ motor, respectively. Eventually, the actual resultant thrust $f\in\mathbb{R}$ and moment $\bm{\mathrm{M}}\in\mathbb{R}^3$ generated by the four rotors can be expressed through the following mapping:
\begin{align}
    {\renewcommand{\arraystretch}{1}\begin{pmatrix}f \\ \bm{\mathrm{M}}\\\end{pmatrix}}^{4\times1}\!\!\!\!\!=
    \setlength{\arraycolsep}{2pt}{{\renewcommand{\arraystretch}{1.2}\begin{bmatrix} c_{T} & c_{T}& c_{T}& c_{T} \\ \frac{\sqrt{2}}{2}dc_{T} & \scalebox{0.6}{$-$}\frac{\sqrt{2}}{2}dc_{T}& \scalebox{0.6}{$-$}\frac{\sqrt{2}}{2}dc_{T}& \frac{\sqrt{2}}{2}dc_{T}\\\frac{\sqrt{2}}{2}dc_{T} & \frac{\sqrt{2}}{2}dc_{T}& \scalebox{0.6}{$-$}\frac{\sqrt{2}}{2}dc_{T}& \scalebox{0.6}{$-$}\frac{\sqrt{2}}{2}dc_{T}\\c_{M} & \scalebox{0.6}{$-$}c_{M}& c_{M}& \scalebox{0.6}{$-$}c_{M} \end{bmatrix}}} {\renewcommand{\arraystretch}{1.2}\begin{pmatrix}\omega_{1}^{2}\\ \omega_{2}^{2}\\\omega_{3}^{2}\\\omega_{4}^{2}\\\end{pmatrix}},
\end{align}
where $c_{T}\in\mathbb{R}$ and $c_{M}\in\mathbb{R}$ are constant
thrust and moment physical coefficients of real rotor aerodynamics. The mapping deviations of resultant thrust $\Delta_f\in\mathbb{R}$ and moment $\bm{\Delta_{\mathrm{M}}}\in\mathbb{R}^{3}$ are defined as follows:
\begin{equation}
   \Delta_f\triangleq f-f_d,
   \label{mapping deviations_f}
\end{equation}
\begin{equation}
   \bm{\Delta_{\mathrm{M}}}\triangleq \bm{\mathrm{M}}-\bm{\mathrm{M}_d}.
   \label{mapping deviations_M}
\end{equation}
Given that $\|\bm{\omega}\|$, $\| c_T - c'_T\|$ and $\| c_M - c'_M\|$ are bounded, $\|\Delta_f\|$  and $\|\bm{\Delta_{\mathrm{M}}}\|$ are also bounded.  If $ c_T\to c'_T$, $ c_M\to c'_M$, it follows that $\|\Delta_f\|$ and $\|\bm{\Delta_{\mathrm{M}}}\|$ converge to zero.

\textbf{\textit{Remark \ref{supp:SE(3)}:}} Zero deviation is a common default assumption in multirotor control, but this work analyzes the impact of nonzero $\varepsilon_{\mathbf{M}}$ on stability (see~\ref{Proofs of Proposition 1 and 2}).

\section{Algorithm}

\begin{center}
\begin{minipage}{0.75\linewidth}
\begin{algorithm}[H]
\caption{Initialization (one-shot)}
\label{alg:init}
\begin{algorithmic}[1]
\Parameters PD controller gains $\{k_R,k_\Omega\!\in\!\mathbb{R}\}$; inertia tensor $\bm{J}\in\mathbb{R}^{3\times3}$ (if known); positive constant $c_R\!\in\!\mathbb{R}$; adaptive rates $\{1/\eta_{j}\!\in\!\mathbb{R}\}$, adaptive upper bounds $\{\overset{\tiny \text{max}}{J}_{j}\!\in\!\mathbb{R}\}$ and pull-back factors $\{\mathfrak{s}_{j}\!\in\!\mathbb{R}\}$; RBF center vectors $\{\textbf{c}_{kj}\in\mathbb{R}^{2}\}$, widths $\{b_{kj}\!\in\!\mathbb{R}\}$ and learning rates  $\{\gamma_{\bm{\textit{R}}j}\in\mathbb{R}\}$; weight bounds $\{\overset{\tiny \text{max}}{W}_{j}\!\in\!\mathbb{R}\}$, dead-zone thresholds $\{\zeta_j\in\mathbb{R}\}$.
\Require Initial desired attitude $\bm{R}_{{\bm{d}}}(0)$, current attitude $\bm{R}(0)$, current angular velocity $\bm{\Omega}(0)$.

\State Reset other desired states: $\dot{\bm{R}_{{\bm{d}}}}\!\leftarrow\!0$, $\bm{\Omega}_{{\bm{d}}}\!\leftarrow\!0$, $\bm{\dot{\Omega}}_{{\bm{d}}}\!\leftarrow\!0$
\State Compute initial errors: $\bm{e}_{\bm{R}}(0),\bm{e}_{\bm{\Omega}}(0)$ \Comment{Eqs.~\eqref{errors},\eqref{Omegac}}
\State Initialize adaptive law \textit{``slices"} (\textbf{Algorithm}~\ref{alg:J-update}): 
\Statex \hspace{1.2em} \textbf{for} $j = 1,2,3$ \textbf{do}
\Statex \hspace{1.2em} set initial estimated inertia feature $\bm{\bar{J}}(0)^{[j]}$
\Statex \hspace{1.2em} reset adaptive law $\bm{\dot{\bar{\mathit{J}}}}^{[j]}\!\leftarrow\!0$
\Statex \hspace{1.2em} \textbf{end for}
\State Initialize neural network \textit{``slices"} (\textbf{Algorithm}~\ref{alg:nn-forward}): 
\Statex \hspace{1.2em} \textbf{for} $j = 1,2,3$ \textbf{do}
\Statex \hspace{1.2em} set initial input $\textbf{x}_{\bm{\textit{R}} j}(0)\leftarrow \left(\bm{e}_{\bm{R}}(0)^{[j]}, \bm{e}_{\bm{\Omega}}(0)^{[j]}\right)^{\top}$
\Statex \hspace{1.2em} reset output of $k^{th}$ neuron $\{\bm{\hbar}^{[k]}(\textbf{x}_{\bm{\textit{R}} j})\!\leftarrow\!0\}_{k=1,2,\dots,l}$
\Statex \hspace{1.2em} reset estimated weight vector $\bm{\bar{\mathcal{W}}}_{\bm{\textit{R}} j}\in\mathbb{R}^l\!\leftarrow\!0$
\Statex \hspace{1.2em} reset weight update law $\bm{\dot{\bar{\mathcal{W}}}}_{\bm{\textit R} j}^{\mathrm{nom}} $, $\bm{\dot{\bar{\mathcal{W}}}}_{\bm{\textit{R}} 
j}\in\mathbb{R}^l\!\leftarrow\!0$
\Statex \hspace{1.2em} reset output of the neural network $\bm{\bar{\phi}}_{\bm{\textit{R}}}^{[j]}\!\leftarrow\!0$
\Statex \hspace{1.2em} \textbf{end for}
\State Reset desired control moment: $\bm{\mathrm{M}_d}\in\mathbb{R}^3\!\leftarrow\!0$

\State \Return all initialized parameters, states and errors
\end{algorithmic}
\vspace{-0.5em}
\rule{\linewidth}{0.4pt}
{\textbf{Note:} Index $j$ denotes the axis-wise subspace index ($j=1,2,3$), 
while $k$ denotes the neuron index ($k=1,2,\dots,l$). }
\end{algorithm}
\end{minipage}
\end{center}

\begin{center}
\begin{minipage}{0.75\linewidth}
\begin{algorithm}[H]
\caption{Adaptive Law \textit{``slices"}}
\label{alg:J-update}
\textbf{Require:} \textbf{Algorithm}~\ref{alg:init} has been executed 
\begin{algorithmic}[1]
\Parameters $c_R$, $\{1/\eta_{j},\overset{\tiny \text{max}}{J}_{j},\mathfrak{s}_{j}\}_{j=1,2,3}\leftarrow$ \textbf{Algorithm}~\ref{alg:init}
\Require $\bm{e}_{\bm{R}}$, $\bm{e}_{\bm{\Omega}}$, desired control moment $\{\bm{\mathrm{M}_d}^{[j]}\}_{j=1,2,3}$
\For{$j = 1,2,3$}
\If{$\left(\!\bm{e}^{[j]}_{\bm{\Omega}}\!\!+\!c_R\bm{e}^{[j]}_{\bm{R}}\!\right)\bm{\mathrm{M}_d}^{[j]}> 0$}
  \State $\bm{\dot{\bar{\mathit{J}}}}^{[j]} \leftarrow \frac{-\bm{\bar{J}}^{[j]^2}}{\eta_{j}}\!\left(\!\bm{e}^{[j]}_{\bm{\Omega}}\!\!+\!c_R\bm{e}^{[j]}_{\bm{R}}\!\right)\!\bm{\mathrm{M}_d}^{[j]}$
\Else
  \If{$\bm{\bar{J}}^{[j]}<\overset{\tiny \text{max}}{J}_{j}$}
    \State $\bm{\dot{\bar{\mathit{J}}}}^{[j]} \leftarrow \frac{-\bm{\bar{J}}^{[j]^2}}{\eta_{j}}\!\left(\!\bm{e}^{[j]}_{\bm{\Omega}}\!\!+\!c_R\bm{e}^{[j]}_{\bm{R}}\!\right)\!\bm{\mathrm{M}_d}^{[j]}$
  \Else
    \State $\bm{\dot{\bar{\mathit{J}}}}^{[j]} \leftarrow \mathfrak{s}_{j}\frac{-\bm{\bar{J}}^{[j]^2}}{\eta_{j}}$ \Comment{Eq.~\eqref{Adaptive Law of Inertia Tensor}}
  \EndIf
\EndIf
\State Integrate (Euler): $\bm{\bar{J}}^{[j]}\leftarrow \bm{\bar{J}}^{[j]} + \bm{\dot{\bar{\mathit{J}}}}^{[j]} dt$
\EndFor
\State \Return $\{\bm{\bar{J}}^{[j]}\}_{j=1,2,3}$
\end{algorithmic}
\end{algorithm}
\end{minipage}
\end{center}

\begin{center}
\begin{minipage}{0.75\linewidth}
\begin{algorithm}[H]
\caption{Neural Network \textit{``slices"}}
\label{alg:nn-forward}
\textbf{Require:} \textbf{Algorithm}~\ref{alg:init} has been executed 
\begin{algorithmic}[1]
\Parameters $c_R$, $\{\textbf{c}_{kj},b_{kj},\gamma_{\bm{\textit{R}}j}\}_{j=1,2,3}\leftarrow$ \textbf{Algorithm}~\ref{alg:init}
\Require $\bm{e}_{\bm{R}}$, $\bm{e}_{\bm{\Omega}}$
\For{$j = 1,2,3$}
 \State Set neural network input $\textbf{x}_{\bm{\textit{R}} j}\leftarrow \left(\bm{e}_{\bm{R}}^{[j]}, \bm{e}_{\bm{\Omega}}^{[j]}\right)^{\top}$
 \For{$k=1,2,\dots,l$}
 \State $\bm{\hbar}^{[k]}(\textbf{x}_{\bm{\textit{R}} j})\leftarrow\mathrm{exp}\left(-\frac{\lVert\textbf{x}_{\bm{\textit{R}} j}-\textbf{c}_{kj}\rVert^2}{2b^{2}_{kj}}\right)$ \Comment{Eq.~\eqref{Gaussian activation function}}
 \EndFor
 \State $\bm{\dot{\bar{\mathcal{W}}}}_{\bm{\textit{R}} 
 j}^{\mathrm{nom}}\leftarrow\gamma_{\bm{\textit{R}}j} \left(\bm{e}^{[j]}_{\bm{\Omega}}+c_R\bm{e}^{[j]}_{\bm{R}}\right)\bm{\hbar}(\textbf{x}_{\bm{\textit{R}} j})$  \Comment{Eq.~\eqref{Estimated Weights_R}}
 \State $\bm{\dot{\bar{\mathcal{W}}}}_{\bm{\textit{R}} 
j}\leftarrow$\textbf{Algorithm}~\ref{alg:proj}$\leftarrow\bm{\dot{\bar{\mathcal{W}}}}_{\bm{\textit{R}} 
 j}^{\mathrm{nom}}$
 \State $\bm{\bar{\phi}}_{\bm{\textit{R}}}^{[j]}\leftarrow\bm{\bar{\mathcal{W}}}_{\bm{\textit{R}} j}^{\top}\bm{\hbar}(\textbf{x}_{\bm{\textit{R}} j})$ \Comment{Eq.~\eqref{Phi_hat}}
 \State Integrate (Euler): $\bm{\bar{\mathcal{W}}}_{\bm{\textit{R}} 
 j}\leftarrow \bm{\bar{\mathcal{W}}}_{\bm{\textit{R}} 
 j} + \bm{\dot{\bar{\mathcal{W}}}}_{\bm{\textit{R}} 
 j} dt$
\EndFor
\State \Return $\{\bm{\bar{\phi}}_{\bm{\textit{R}}}^{[j]}\}_{j=1,2,3}$
\end{algorithmic}
\vspace{-0.5em}
\rule{\linewidth}{0.4pt}
{\textbf{Note:} $l$ denotes the number of neurons. }
\end{algorithm}
\end{minipage}
\end{center}

\begin{center}
\begin{minipage}{0.75\linewidth}
\begin{algorithm}[H]
\caption{Dead-zone and Projection}
\label{alg:proj}
\textbf{Require:} \textbf{Algorithm}~\ref{alg:init} has been executed 
\begin{algorithmic}[19]
\Parameters $\{\overset{\tiny \text{max}}{W}_{j}, \zeta_j\}_{j=1,2,3}\leftarrow$ \textbf{Algorithm}~\ref{alg:init}
\Require $\{\textbf{x}_{\bm{\textit{R}} j},\bm{\dot{\bar{\mathcal{W}}}}_{\bm{\textit{R}} 
 j}^{\mathrm{nom}}\}_{j=1,2,3}\leftarrow$ \textbf{Algorithm}~\ref{alg:nn-forward}
\If{$\|\textbf{x}_{\bm{\textit{R}} j}\| \le \zeta_j$}
    \State $\bm{\dot{\bar{\mathcal{W}}}}_{\bm{\textit R} j} \leftarrow \mathbf{0}$  \Comment{enter dead-zone}
\Else \Comment{exit dead-zone}
 \If{$\|\bm{\bar{\mathcal W}}_{\bm{\textit R} j}\|<\overset{\tiny \text{max}}{W}_{j} \,\,\text{or}\,
     \Big{(}\|\bm{\bar{\mathcal W}}_{\bm{\textit R} j}\|= \overset{\tiny \text{max}}{W}_{j} \,\text{and}\,
     (\bm{\dot{\bar{\mathcal{W}}}}_{\bm{\textit R} j}^{\mathrm{nom}})^{\!\top}\bm{\bar{\mathcal W}}_{\bm{\textit R} j}\le 0\Big{)}$}
    \State $\bm{\dot{\bar{\mathcal{W}}}}_{\bm{\textit R} j}\leftarrow \bm{\dot{\bar{\mathcal{W}}}}_{\bm{\textit R} j}^{\mathrm{nom}}$\Comment{non-projected}
\Else
    \State $\bm{\dot{\bar{\mathcal{W}}}}_{\bm{\textit R} j}\leftarrow
    \Big(\mathbf{I}^{l\times l}-\dfrac{\bm{\bar{\mathcal W}}_{\bm{\textit R} j}\bm{\bar{\mathcal W}}_{\bm{\textit R} j}^{\top}}
    {\bm{\bar{\mathcal W}}_{\bm{\textit R} j}^{\top}\bm{\bar{\mathcal W}}_{\bm{\textit R} j}}\Big)
    \bm{\dot{\bar{\mathcal{W}}}}_{\bm{\textit R} j}^{\mathrm{nom}}$
    \Comment{projected}
\EndIf
\EndIf
\State \Return Weight update law $\bm{\dot{\bar{\mathcal{W}}}}_{\bm{\textit R} j}$
\end{algorithmic}
\rule{\linewidth}{0.4pt}
{\textbf{Note:} $\mathbf{I}^{l\times l}$ denotes the $l\times l$ identity matrix.}
\end{algorithm}
\end{minipage}
\end{center}

\section{Extended Proofs}
\label{supp:extended-proofs}

The subsequent Lyapunov analysis is conducted in the following open domain:
\begin{equation}
   \begin{aligned}
    \mathcal{D}\!=&\Big{\{}\Big{(}\bm{e_R}, \bm{e_\Omega},(\widetilde{J}_{j},\bm{\tilde{\mathcal{W}}}_{\bm{\textit{R}} j})_{j=1,2,3}\Big{)}\!\in\mathbb{R}^3\!\times\!\mathbb{R}^3\!\times\!\prod^{3}_{j=1}(\mathbb{R}\!\times\!\mathbb{R}^{l})\big{|}\\[-5pt]
& \,\,\,\,\,\,\|\bm{e_R}\|\!+\!\|\bm{e_\Omega}\|\!+\!\sum_{j=1}^3(\|\widetilde{J}_{j}\|+\|\bm{\tilde{\mathcal{W}}}_{\bm{\textit{R}} j}\|) < r_d\Big{\}},
\end{aligned} 
\label{D}
\end{equation}
for a positive constant $r_d$. In this domain, $\|\bm{e_R}\|$ is also bounded by $\|\bm{e_R}\|=\sqrt{\Psi_{\textit{R}}(2-\Psi_{\textit{R}})}\leq\sqrt{\psi_{\textit{R}}(2-\psi_{\textit{R}})}<1$ with a positive scalar $0<\psi_{\textit{R}}<2$ and an attitude configuration error scalar function proposed in \cite{2010 Geometric tracking control of a quadrotor UAV on SE(3)}:

\begin{equation}
    \Psi_{\textit{R}}(\bm{R},\bm{R}_{\bm{d}})\triangleq\frac{1}{2}\mathrm{tr}\Big{[}\mathrm{I}^{3\times3}-\bm{R}_{\bm{d}}^{\top}\bm{R}\Big{]},
    \label{Psi_R}
\end{equation}
where $\Psi_{\textit{R}}:\mathbf{SO}(3)\times\mathbf{SO}(3)\to\mathbb{R}$ is positive definite and constrained by:
\begin{equation}
    \frac{1}{2}\|\bm{e_R}\|^{2}\leq\Psi_{\textit{R}}\leq\frac{1}{2-\psi_{\textit{R}}}\|\bm{e_R}\|^{2}.
    \label{Psi_R_bound}
\end{equation}
\subsection{Rotational Error Dynamics}
\label{Rotational Error Dynamics}
Following the formulation in \cite{2010 Geometric tracking control of a quadrotor UAV on SE(3)} and \cite{2011 Geometric tracking control of the attitude dynamics of a rigid body on SO(3)}, the rotational error dynamics are given by:
\begin{align}
\bm{\dot{e}_R}&=\frac{1}{2}\left(\bm{R_d}^{\top}\bm{R}[\bm{e_{\Omega}}]_{\times}+[\bm{e_{\Omega}}]_{\times}\bm{R}^{\top}\bm{R_d}\right)^{\vee}\notag\\
&=\frac{1}{2}\left(\mathrm{tr}\big{[}\bm{R}^{\top}\bm{R_d}\big{]}\mathrm{I}^{3\times3}-\bm{R}^{\top}\bm{R_d}\right)\bm{e_{\Omega}}\notag\\
&\equiv Y(\bm{R_d}^{\top}\bm{R})\bm{e_{\Omega}},\label{rotational error dynamics1}\\
\bm{\dot{e}}_{\bm{\Omega}}&=\bm{\dot{\Omega}}+[\bm{\Omega}]_{\times}\bm{R}^{\top}\bm{R_d}\bm{\Omega_d}-\bm{R}^{\top}\bm{R_d}\bm{\dot{\Omega}_d},
\label{rotational error dynamics2}
\end{align}
where $\|Y(\bm{R_d}^{\top}\bm{R})\|\leq1$ for any $\bm{R_d}^{\top}\bm{R}\in\mathbf{SO}(3)$.

In \textbf{\textit{Scenario 1 ($\bm{J}$ is known)}}, substituting Eqs.~\eqref{Dynamics with Augmented Disturbance} and \eqref{mapping deviations} into Eq.~\eqref{rotational error dynamics2}, we have:
\begin{align}
\bm{\dot{e}_{\Omega}}=&\bm{J}^{-1}\left(\bm{\mathrm{M}_d}-[\bm{\Omega}]_{\times}\bm{J}\bm{\Omega}+\bm{\Delta_{\mathrm{M}}}\right)+\bm{\phi}_{\bm{\textit{R}}}\notag\\
    &+[\bm{\Omega}]_{\times}\bm{R}^{\top}\bm{R_d}\bm{\Omega_d}-\bm{R}^{\top}\bm{R_d}\bm{\dot{\Omega}_d}.
    \label{rotational error dynamics3}
\end{align}
Further substituting Eqs.~\eqref{Md} and \eqref{Estimation errors of pseudo model}, the angular velocity error dynamics along the $\bm{\vec{b}}_{j}$-axis is rearranged as:
\begin{equation}
{
    \begin{aligned}
    \bm{\dot{e}_{\Omega}}^{[j]}=&-k_{R}\bm{e}^{[j]}_{\bm{R}}-k_{\Omega}\bm{e}^{[j]}_{\bm{\Omega}
}+\widetilde{J}_{j}\bm{\mathrm{M}_d}^{[j]}+\left(\bm{\phi}_{\bm{\textit{R}}}^{[j]}-\bm{\bar{\phi}}_{\bm{\textit{R}}}^{[j]}\right)\\
    &\!\!\!\!+\underbrace{(\bm{J}^{-1}[\bm{\Omega}]_{\times}\bm{J}\bm{\Omega})^{[j]}}_{\text{if $\bm{J}$ is known}}-\left(\bm{J}^{-1}[\bm{\Omega}]_{\times}\bm{J}\bm{\Omega}\right)^{[j]}+\left(\bm{J}^{-1}\bm{\Delta_{\mathrm{M}}}\right)^{[j]},\\[-5pt]
\end{aligned}
}
\label{rotational error dynamics4}
\end{equation}
with the fact that $\bm{J}$ is always diagonalizable such that $\bm{J}^{-1[j]}=1/{\bm{J}^{[j]}}$. From Eq.~\eqref{approximation error to weight error}, the problem of the approximation error $\bm{\phi}_{\bm{\textit{R}}}^{[j]}-\bm{\bar{\phi}}_{\bm{\textit{R}}}^{[j]}$ is transformed into the problem of the weight estimation error $\bm{\tilde{\mathcal{W}}}_{\bm{\textit{R}} j}$. Therefore, the rotational error dynamics can ultimately be expressed as:
\begin{equation}
{
    \begin{aligned}
    \bm{\dot{e}_{\Omega}}^{[j]}=&-k_{R}\bm{e}^{[j]}_{\bm{R}}-k_{\Omega}\bm{e}^{[j]}_{\bm{\Omega}
}+\widetilde{J}_{j}\bm{\mathrm{M}_d}^{[j]}+\bm{\tilde{\mathcal{W}}}_{\bm{\textit{R}} j}^{\top}\bm{\hbar}(\textbf{x}_{\bm{\textit{R}} j})+\bm{\varpi}^{[j]}_{\bm{\textit{R}}}\\
    &\!\!\!\!+\underbrace{(\bm{J}^{-1}[\bm{\Omega}]_{\times}\bm{J}\bm{\Omega})^{[j]}}_{\text{if $\bm{J}$ is known}}-\left(\bm{J}^{-1}[\bm{\Omega}]_{\times}\bm{J}\bm{\Omega}\right)^{[j]}+\left(\bm{J}^{-1}\bm{\Delta_{\mathrm{M}}}\right)^{[j]},\\[-5pt]
\end{aligned}
}
\label{rotational error dynamics5}
\end{equation}
where, if the knowledge of the inertia tensor $\bm{J}$ is augmented in Eq.~\eqref{Md}, the sixth term appears to cancel out the seventh term.

In \textbf{\textit{Scenario 2 ($\bm{J}$ is unknown)}}, since the neural networks intervene to learn and compensate for the internal disturbance term $\bm{J}^{-1}[\bm{\Omega}]_{\times}\bm{J}\bm{\Omega}$, by substituting
Eq.~\eqref{Dynamics with Augmented Disturbance2} into Eq.~\eqref{rotational error dynamics2} instead of Eq.~\eqref{Dynamics with Augmented Disturbance}, all the terms associated with $\bm{J}^{-1}[\bm{\Omega}]_{\times}\bm{J}\bm{\Omega}$ in Eqs.~\eqref{rotational error dynamics3}, \eqref{rotational error dynamics4} and \eqref{rotational error dynamics5} vanish. 
\subsection{Lyapunov Candidate}
\subsubsection{Candidate for Rotational State Errors }
Define a Lyapunov candidate function for the rotational state errors as:
\begin{equation}
{
\begin{aligned}
\bm{\mathcal{V}_{\bm{\textit{R},s}}}=k_{R}\Psi_{\textit{R}}+\bm{\sum}_{j=1}^3\Big{(}\frac{1}{2}\|\bm{e}^{[j]}_{\bm{\Omega}}\|^2+c_R\bm{e}^{[j]}_{\bm{R}}\bm{e}^{[j]}_{\bm{\Omega}}\Big{)},
\end{aligned}
}
\label{V_R,s}
\end{equation}
where $k_R$ and $c_R$ are positive constants. The compact form of Eq.~\eqref{V_R,s} can be expressed as:
\begin{equation}
{
\begin{aligned}
\bm{\mathcal{V}_{\bm{\textit{R},s}}}=k_{R}\Psi_{\textit{R}}+\frac{1}{2}\|\bm{e}_{\bm{\Omega}}\|^2+c_R\bm{e}_{\bm{R}}\cdot\bm{e}_{\bm{\Omega}}.
\end{aligned}
}
\label{V_R,s2}
\end{equation}
From Eq.~\eqref{Psi_R_bound} and Cauchy–Schwarz inequality, the lower and upper bounds of $\bm{\mathcal{V}_{\bm{\textit{R},s}}}$ are given by:
\begin{equation}
    \begin{aligned}    
\bm{z}^{\top}_{\bm{\textit{R}}}\bm{\mathfrak{M}}_{\bm{\textit{R}}1}\,\bm{z}_{\bm{\textit{R}}}\leq\bm{\mathcal{V}_{\bm{\textit{R},s}}}\leq&\bm{z}^{\top}_{\bm{\textit{R}}}\bm{\mathfrak{M}}_{\bm{\textit{R}}2}\,\bm{z}_{\bm{\textit{R}}},
    \end{aligned}
    \label{V_R,s_quadratic1}
\end{equation}
where $\bm{z}_{\bm{\textit{R}}}=\left(\|\bm{e}_{\bm{R}}\|,\|\bm{e}_{\bm{\Omega}}\|\right)^{\top}\!\!\in\mathbb{R}^{2}$ and
\begin{equation}
    \begin{aligned}    
      \bm{\mathfrak{M}}_{\bm{\textit{R}}1}={\renewcommand{\arraycolsep}{5pt}\renewcommand{\arraystretch}{1.5}\begin{bmatrix}
\frac{k_{R}}{2}&-\frac{c_R}{2}\\
       -\frac{c_R}{2} &\frac{1}{2}\\
    \end{bmatrix}}
    \end{aligned},
    \begin{aligned}    
      \bm{\mathfrak{M}}_{\bm{\textit{R}}2}={\renewcommand{\arraycolsep}{5pt}\renewcommand{\arraystretch}{1.5}\begin{bmatrix}
\frac{k_{R}}{2-\psi_{\textit{R}}}&\frac{c_R}{2}\\
       \frac{c_R}{2} &\frac{1}{2}\\
    \end{bmatrix}}
    \end{aligned}.
    \label{Matrix_R_1and2}
\end{equation}
If positive constant $c_R$ is chosen sufficiently small to satisfy
\begin{equation}
{
    \begin{aligned}
    c_R\!< \min\left\{\sqrt{k_R},\sqrt{\frac{2k_{R}}{2-\psi_{\textit{R}}}}\right\},
\end{aligned}
}\label{cR_condition}
\end{equation}
matrices $ \bm{\mathfrak{M}}_{\bm{\textit{R}}1}$ and $ \bm{\mathfrak{M}}_{\bm{\textit{R}}2}$ become positive definite, which implies $\bm{\mathcal{V}_{\bm{\textit{R},s}}}$ is positive definite and bounded by:
\begin{equation}
    \begin{aligned}    
\lambda_{\min}( \bm{\mathfrak{M}}_{\bm{\textit{R}}1})\|\bm{z}_{\bm{\textit{R}}}\|^2\leq\bm{\mathcal{V}_{\bm{\textit{R},s}}}\leq\lambda_{\max}(\bm{\mathfrak{M}}_{\bm{\textit{R}}2})\|\bm{z}_{\bm{\textit{R}}}\|^2,
    \end{aligned}
    \label{V_R,s_quadratic2}
\end{equation}
where $\lambda_{\min}(\bullet)$ and $\lambda_{\max}(\bullet)$ denote the minimum and maximum eigenvalues of a matrix, respectively.

\subsubsection{Candidate for Rotational Estimation Errors}
Next, we define the Lyapunov candidate function for rotational estimation errors:
\begin{equation}
{
\begin{aligned}
\bm{\mathcal{V}}_{\bm{\textit{R}},e}\!=\!\bm{\sum}_{j=1}^3\Big{(}\frac{1}{2}\eta_{j}\widetilde{J}_j^2\!+\!\frac{1}{2\gamma_{\bm{\textit{R}}j}}\bm{\tilde{\mathcal{W}}}_{\bm{\textit{R}} j}^{\top}\bm{\tilde{\mathcal{W}}}_{\bm{\textit{R}} j}\Big{)},
\end{aligned}
}
\label{V_e}
\end{equation}
where the  $\eta_{j}$ and $\gamma_{\bm{\textit{R}}j}$ are positive constants. 

\subsubsection{Complete Candidate}
\label{}
Combining Eq.~\eqref{V_R,s} and Eq.~\eqref{V_e}, the Lyapunov candidate function for the complete rotational error dynamics is rearranged and given as follows: 
\begin{equation}
{\small
\begin{aligned}
\bm{\mathcal{V}}_{\bm{\textit{R}}}\!=\!k_{R}\Psi_{\textit{R}}\!+\!\!\bm{\sum}_{j=1}^3\!\!\Big{(}\frac{1}{2}\|\bm{e}^{[j]}_{\bm{\Omega}}\|^2\!\!+\!c_R\bm{e}^{[j]}_{\bm{R}}\bm{e}^{[j]}_{\bm{\Omega}}\!\!+\!\frac{1}{2}\eta_{j}\widetilde{J}_j^2\!+\!\frac{1}{2\gamma_{\bm{\textit{R}}j}}\bm{\tilde{\mathcal{W}}}_{\bm{\textit{R}} j}^{\top}\bm{\tilde{\mathcal{W}}}_{\bm{\textit{R}} j}\!\Big{)}.
\end{aligned}
}
\label{V}
\end{equation}
From Eqs.~\eqref{V_R,s}, \eqref{V_R,s_quadratic2}, \eqref{V_e} and \eqref{V}, it holds that:
\begin{equation}
{
\begin{aligned}
     \lambda_{\min}( \bm{\mathfrak{M}}_{\bm{\textit{R}}1})\|\bm{z}_{\bm{\textit{R}}}\|^2\!+\!\bm{\mathcal{V}}_{\bm{\textit{R},e}}\leq\bm{\mathcal{V}}_{\bm{\textit{R}}}\leq\lambda_{\max}(\bm{\mathfrak{M}}_{\bm{\textit{R}}2})\|\bm{z}_{\bm{\textit{R}}}\|^2\!+\!\bm{\mathcal{V}}_{\bm{\textit{R},e}}.
\end{aligned}
}
\label{V_R_bound1}
\end{equation}

\textbf{\textit{Lemma \ref{supp:extended-proofs}.1:}} Given that $\bm{\mathcal{V}}_{\bm{\textit{R},e}}$ is positive definite and bounded, it holds that there always exist positive constants $\mathfrak{p}_1$ and $\mathfrak{p}_2$ outside the ball with arbitrary bounded radius $\epsilon$ around $\bm{z}_{\bm{\textit{R}}}=0$  such that:
\begin{equation}
{
\begin{aligned}
     \mathfrak{p}_1\lambda_{\min}( \bm{\mathfrak{M}}_{\bm{\textit{R}}1})\|\bm{z}_{\bm{\textit{R}}}\|^2\!\leq\bm{\mathcal{V}}_{\bm{\textit{R}}}\leq\mathfrak{p}_2\lambda_{\max}(\bm{\mathfrak{M}}_{\bm{\textit{R}}2})\|\bm{z}_{\bm{\textit{R}}}\|^2.
\end{aligned}
}
\label{VR_p1p2bound}
\end{equation}

\textbf{\textit{Proof:}}  With the fact that $\bm{\mathcal{V}}_{\bm{\textit{R},e}}$ is positive definite and bounded, if $\|\bm{z}_{\bm{\textit{R}}}\|\geq\epsilon$ and the radius $\epsilon$ is positive and bounded, there always exist bounded positive constants $\mathfrak{p}_1$ and $\mathfrak{p}_2$ to satisfy:
\begin{equation}
{
\begin{aligned}
    \mathfrak{p}_1&\leq1+\frac{\bm{\mathcal{V}}_{\bm{\textit{R},e}}}{\lambda_{\min}( \bm{\mathfrak{M}}_{\bm{\textit{R}}1})\|\bm{z}_{\bm{\textit{R}}}\|^2}, \\
     \mathfrak{p}_2&\geq1+\frac{\bm{\mathcal{V}}_{\bm{\textit{R},e}}}{\lambda_{\max}( \bm{\mathfrak{M}}_{\bm{\textit{R}}2})\epsilon^2}\geq1+\frac{\bm{\mathcal{V}}_{\bm{\textit{R},e}}}{\lambda_{\max}( \bm{\mathfrak{M}}_{\bm{\textit{R}}2})\|\bm{z}_{\bm{\textit{R}}}\|^2}.\\ 
\end{aligned}
}
\label{Proof_of_Corollary1}
\end{equation}

Substituting these into Eq.~\eqref{V_R_bound1} yields Eq.~\eqref{VR_p1p2bound}. Therefore, \textbf{\textit{Lemma \ref{supp:extended-proofs}.1}} is established.

\textbf{\textit{Remark \ref{supp:extended-proofs}.2 (Tunable Exponential Convergence Rate):}} \textbf{\textit{Lemma \ref{supp:extended-proofs}.1}} essentially reveals how the update rates of the neural networks and adaptive laws determine the exponential convergence rate of the state solution $\bm{z}_{\bm{\textit{R}}}(t)$. The $\mathfrak{p}_1$ and $\mathfrak{p}_2$
characterize the exponential convergence rate of the state solution $\bm{z}_{\bm{\textit{R}}}(t)$ (see Eq.~\eqref{NES condition1}). Once the state approaches the error ball $\|\bm{z}_{\bm{\textit{R}}}\|=\epsilon$ in Eq.~\eqref{small-ball}, the $\eta_{j}$ and $1/\gamma_{\bm{\textit{R}}j}$ dominate the $\bm{\mathcal{V}}_{\bm{\textit{R}},e}$. Then we can choose smaller constants $\eta_{j}$ and $1/\gamma_{\bm{\textit{R}}j}$ to obtain smaller $\mathfrak{p}_2$. Hence, choosing smaller constants $\eta_{j}$ and $1/\gamma_{\bm{\textit{R}}j}$ yields smaller $\mathfrak{p}_2$ and consequently a larger $\beta$ (see Eq.~\eqref{dot_V_R_decay}), which results in a faster theoretical exponential convergence rate.  However, it should be noted that a larger $\beta$ also tightens the allowable range of the discrete sampling period $dt^\star$ (analyzed in Eq.~\eqref{eq:h_tau_bounds}). In practice, an excessively large $\beta$ may therefore cause sensitivity to sampling and computation delays, implying a trade-off between analytical convergence speed and digital implementation robustness.

\textbf{\textit{Remark \ref{supp:extended-proofs}.3 (Dependence of the Guaranteed Rate on the Residual Ball Size):}}
A subtle theoretical point concerns the global constant $\mathfrak{p}_2$. Its existence as constructed in the proof  requires the ratio $\frac{\bm{\mathcal{V}}_{\bm{\textit{R},e}}}{\| \bm{z}_{\bm{\textit{R}}} \|^2}$ to be uniformly bounded for all $\| \bm{z}_{\bm{\textit{R}}} \| \geq \epsilon$. This requirement is formally captured by the condition:
\begin{equation}
    \mathfrak{p}_2 \geq 1 + \sup_{\| \bm{z}_{\bm{\textit{R}}} \| \geq \epsilon} \left( \frac{\bm{\mathcal{V}}_{\bm{\textit{R},e}}}{\lambda_{\max}(\bm{\mathfrak{M}_{R2}}) \| \bm{z}_{\bm{\textit{R}}} \|^2} \right).
\end{equation}
In practice, this condition holds since the system dynamics and adaptive learning laws prevent the parameter errors from growing disproportionately. Nevertheless, this construction implies that the guaranteed exponential convergence rate $\beta$ depends on the ball size $\epsilon$. Specifically, if $\epsilon$ is extremely small, the worst-case convergence rate may become very slow. Fortunately, as shown in Supp.~\ref{supp::para::Local-Exponential-Convergence}, high-precision tracking—corresponding to a small effective $\epsilon$—can be achieved by reducing the residual errors $\varepsilon_R$ and $\varepsilon_M$. This approach ensures a practically satisfactory convergence rate without requiring an excessively small $\epsilon$ in the stability analysis.

\subsection{Exponential Attractiveness and Convergence}
\label{Proofs of Proposition 1 and 2}
The time derivative of the rotational candidate function is derived by the fact that $\dot{\Psi}_{\textit{R}}=\bm{e}_{\bm{R}}\cdot\bm{e}_{\bm{\Omega}}$ \cite{2010 Geometric tracking control of a quadrotor UAV on SE(3)}:
\begin{equation}
    {
    \begin{aligned}    \bm{\dot{\mathcal{V}}}_{\bm{\textit{R}}}\!=&k_{R}\bm{e}_{\bm{R}}\cdot\!\bm{e}_{\bm{\Omega}}+\bm{\sum}_{j=1}^3\Bigg{\{}\bm{e}_{\bm{\Omega}}^{[j]}\bm{\dot{e}}_{\bm{\Omega}}^{[j]}+c_R\bm{e}^{[j]}_{\bm{\Omega}}\bm{\dot{e}}^{[j]}_{\bm{R}}+c_R\bm{e}^{[j]}_{\bm{R}}\bm{\dot{e}}^{[j]}_{\bm{\Omega}}\\
&+\eta_{j}\widetilde{J}_j\frac{\bm{\dot{\bar{\mathit{J}}}}^{[j]}}{\bm{\bar{J}}^{[j]^2}}-\frac{1}{\gamma_{\bm{\textit{R}}j}}\big{(}\bm{\mathcal{W}}^*_{\bm{\textit{R}} j}\!-\!\bm{\bar{\mathcal{W}}}_{\bm{\textit{R}} j}\big{)}\!^{\top}\bm{\dot{\bar{\mathcal{W}}}}_{\bm{\textit{R}} j}
    \Bigg{\}}\\
    =&\bm{\sum}_{j=1}^3\Bigg{\{}k_{R}\bm{e}_{\bm{R}}^{[j]}\bm{e}_{\bm{\Omega}}^{[j]}+(\bm{e}_{\bm{\Omega}}^{[j]}+c_R\bm{e}^{[j]}_{\bm{R}})\bm{\dot{e}}_{\bm{\Omega}}^{[j]}+c_R\bm{e}^{[j]}_{\bm{\Omega}}\bm{\dot{e}}^{[j]}_{\bm{R}}\\
    &+\eta_{j}\widetilde{J}_j\frac{\bm{\dot{\bar{\mathit{J}}}}^{[j]}}{\bm{\bar{J}}^{[j]^2}}-\frac{1}{\gamma_{\bm{\textit{R}}j}}\big{(}\bm{\mathcal{W}}^*_{\bm{\textit{R}} j}-\bm{\bar{\mathcal{W}}}_{\bm{\textit{R}} j}\big{)}^{\top}\bm{\dot{\bar{\mathcal{W}}}}_{\bm{\textit{R}} j}
    \Bigg{\}}.
    \end{aligned}
    }   
\end{equation}
Substituting Eq.~\eqref{rotational error dynamics5}, we derive:
\begin{equation}
    {
    \begin{aligned}    \bm{\dot{\mathcal{V}}}_{\bm{\textit{R}}}\!=&\bm{\sum}_{j=1}^3\!\Bigg{\{}\!\!-k_{R}c_{R}\|\bm{e}_{\bm{R}}^{[j]}\|^{2}\!-\!k_{\Omega}\|\bm{e}_{\bm{\Omega}}^{[j]}\|^{2}\!-\!k_{\Omega}c_{R}\bm{e}_{\bm{\Omega}}^{[j]}\bm{e}_{\bm{R}}^{[j]}+c_R\bm{e}_{\bm{\Omega}}^{[j]}\bm{\dot{e}}_{\bm{R}}^{[j]}\\[-5pt]
&+\!\!\left(\!\bm{e}^{[j]}_{\bm{\Omega}}\!+\!c_R\bm{e}^{[j]}_{\bm{R}}\!\right)\!\!\bigg{\{}\bm{\varpi}^{[j]}_{\bm{\textit{R}}}\!+\!\left(\bm{J}^{-1}\!\!\bm{\Delta_{\mathrm{M}}}\right)^{[j]}\!\!\bigg{\}}\\[-5pt]
    &+\widetilde{J}_j
\bigg{\{}\left(\bm{e}^{[j]}_{\bm{\Omega}}+c_R\bm{e}^{[j]}_{\bm{R}}\right)\bm{\mathrm{M}_d}^{[j]}+\eta_{j}\frac{\bm{\dot{\bar{\mathit{J}}}}^{[j]}}{\bm{\bar{J}}^{[j]^2}}\bigg{\}}\\
    &\!\!\!\!\!\!\!+\frac{1}{\gamma_{\textit{R}j}}\left(\bm{\mathcal{W}}^*_{\bm{\textit{R}} j}-\bm{\bar{\mathcal{W}}}_{\bm{\textit{R}} j}\right)^{\top}\!\bigg{\{}\gamma_{\bm{\textit{R}}j}\left(\bm{e}^{[j]}_{\bm{\Omega}}+c_R\bm{e}^{[j]}_{\bm{R}}\right)\bm{\hbar}(\textbf{x}_{\textit{R}j})\!-\!\bm{\dot{\bar{\mathcal{W}}}}_{\textit{R}j}\bigg{\}}\!\Bigg{\}}\\
    &+(\bm{e}_{\bm{\Omega}}+c_R\bm{e}_{\bm{R}})(\underbrace{\bm{J}^{-1}[\bm{\Omega}]_{\times}\bm{J}\bm{\Omega}}_{\text{if $\bm{J}$ is known}}-\bm{J}^{-1}[\bm{\Omega}]_{\times}\bm{J}\bm{\Omega}).\\[-0pt]
    \end{aligned}
    }
    \label{dot_V_R}
\end{equation}
In both \textbf{\textit{Scenario 1 ($\bm{J}$ is known)}} and \textbf{\textit{Scenario 2 ($\bm{J}$ is unknown)}}, the last line of Eq.~\eqref{dot_V_R} can be canceled (see Supp.~\ref{Rotational Error Dynamics}).
  In addition, if $\bm{\dot{\bar{\mathit{J}}}}^{[j]}$ is chosen as in Eq.~\eqref{Adaptive Law of Inertia Tensor} and $\bm{\dot{\bar{\mathcal{W}}}}_{\bm{\textit{R}} 
 j}$ follows \textbf{Algorithm~\ref{alg:proj}} and Eq.~\eqref{Estimated Weights_R}, the last three lines vanish.
 Given that $\|\bm{\Delta_{\mathrm{M}}}\|$ converges to zero if the aerodynamic coefficients are precisely chosen (see Supp. \ref{supp:SE(3)}), we consider the lower and upper bounds of  $\left(\bm{J}^{-1}\!\bm{\Delta_{\mathrm{M}}}\right)\!^{[j]}$ as follows:
 \begin{equation}
{
\begin{aligned}
0\leq\|\left(\bm{J}^{-1}\!\bm{\Delta_{\mathrm{M}}}\right)\!^{[j]}\|\leq\|\frac{\bm{\Delta_{\mathrm{M}}}}{\lambda_{\min}(\bm{J})}\|\leq\frac{\varepsilon_{\mathbf{M}}}{\lambda_{\min}(\bm{J})},
\end{aligned}
}
\label{DeltaM_bound}
\end{equation}
where $\varepsilon_{\mathbf{M}}$ is defined as the upper bound of  $\|\bm{\Delta_{\mathrm{M}}}\|$. 

Then, since $\|\bm{e_R}\|<1$ and $\|\bm{\dot{e}}_{\bm{R}}\|\leq\|\bm{e}_{\bm{\Omega}}\|$ from Eq.~\eqref{rotational error dynamics1}, we can apply the foregoing bounds from Eq.~\eqref{DeltaM_bound} to obtain the upper bound of $\bm{\dot{\mathcal{V}}}_{\bm{\textit{R}}}$:
\begin{equation}
    {
    \begin{aligned}    
    \bm{\dot{\mathcal{V}}}_{\bm{\textit{R}}}\leq
    &\!-\!k_{R}c_{R}\|\bm{e}_{\bm{R}}\|^{2}\!-\!\left(k_{\Omega}\!-\!c_R\right)\|\bm{e}_{\bm{\Omega}}\|^{2}\!+\!k_{\Omega}c_{R}\|\bm{e}_{\bm{\Omega}}\|\|\bm{e}_{\bm{R}}\|\!\\
    &+c_R\bigg{(}\varepsilon_{\bm{\textit{R}}}+\frac{\varepsilon_{\mathbf{M}}}{\lambda_{\min}(\bm{J})}\bigg{)}\|\bm{e}_{\bm{R}}\|+\bigg{(}\varepsilon_{\bm{\textit{R}}}+\frac{\varepsilon_{\mathbf{M}}}{\lambda_{\min}(\bm{J})}\bigg{)}\|\bm{e}_{\bm{\Omega}}\|,
    \end{aligned}
    } 
    \label{dot_V_R_bound1}
\end{equation}
where $\varepsilon_{{\bm{\textit{R}}}}\!\in\!\mathbb{R}$ is defined as an upper bound for the optimal approximation error $\bm{\varpi}_{\textit{R}}$ of the neural networks:
\begin{equation}
{\small
\begin{aligned}
\|\bm{\varpi}^{[j]}_{\textit{R}}\|\leq\|\bm{\varpi}_{\textit{R}}\|
\leq\varepsilon_{{\bm{\textit{R}}}}.
\end{aligned}
}
\label{varpix_bound}
\end{equation}
By choosing $k_\Omega > c_R$, we can apply Young’s inequality to the last line, yielding:
\begin{equation}
    {
    \begin{aligned}    
&c_R\bigg{(}\varepsilon_{\bm{\textit{R}}}+\frac{\varepsilon_{\mathbf{M}}}{\lambda_{\min}(\bm{J})}\bigg{)}\|\bm{e}_{\bm{R}}\|\leq\frac{c_R^2\bigg{(}\varepsilon_{\bm{\textit{R}}}+\frac{\varepsilon_{\mathbf{M}}}{\lambda_{\min}(\bm{J})}\bigg{)}^2}{2k_{R}c_{R}}+\frac{k_{R}c_{R}}{2}\|\bm{e}_{\bm{R}}\|^{2},\\
    &\bigg{(}\!\varepsilon_{\bm{\textit{R}}}\!+\!\frac{\varepsilon_{\mathbf{M}}}{\lambda_{\min}(\bm{J})}\!\bigg{)}\|\bm{e}_{\bm{\Omega}}\|\!\leq\!\frac{\bigg{(}\varepsilon_{\bm{\textit{R}}}\!+\!\frac{\varepsilon_{\mathbf{M}}}{\lambda_{\min}(\bm{J})}\bigg{)}^2}{2\left(k_{\Omega}\!\!-\!c_R\right)}\!+\!\frac{k_{\Omega}\!-\!c_R\!}{2}\|\bm{e}_{\bm{\Omega}}\|^2.
    \end{aligned}
    } 
    \label{dot_V_R_bound2}
\end{equation}
From here, we can reformulate Eq.~\eqref{dot_V_R_bound1} into:
\begin{equation}
    \begin{aligned}    
    \bm{\dot{\mathcal{V}}}_{\bm{\textit{R}}}\leq&-\bm{z}^{\top}_{\bm{\textit{R}}}\!\bm{\mathcal{M}}_{\bm{\textit{R}}}\,\bm{z}_{\bm{\textit{R}}}+\mathbf{C}_{\bm{\textit{R}}},
    \end{aligned}
    \label{dot_V_R_quadratic}
\end{equation}
where $\bm{z}_{\bm{\textit{R}}}=\left(\|\bm{e}_{\bm{R}}\|,\|\bm{e}_{\bm{\Omega}}\|\right)^{\top}\!\!\in\mathbb{R}^{2}$. The matrix $\bm{\mathcal{M}}_{\bm{\textit{R}}}\in\mathbb{R}^{2\times2}$ is given by:
\begin{equation}
    \begin{aligned}    
       \bm{\mathcal{M}}_{\bm{\textit{R}}}={\renewcommand{\arraycolsep}{5pt}\renewcommand{\arraystretch}{1.5}\begin{bmatrix}
\frac{k_Rc_R}{2}&\frac{-k_{\Omega}c_R}{2}\\
       \frac{-k_{\Omega}c_R}{2} &\frac{k_{\Omega}-c_R}{2}\\
    \end{bmatrix}}
    \end{aligned},
    \label{Matrix_R}
\end{equation}
and the constant term is expressed as:
\begin{equation}
{
    \begin{aligned}    
\mathbf{C}_{\bm{\textit{R}}}=\frac{c_R\bigg{(}\varepsilon_{\bm{\textit{R}}}+\frac{\varepsilon_{\mathbf{M}}}{\lambda_{\min}(\bm{J})}\bigg{)}^2}{2k_{R}}+\frac{\bigg{(}\varepsilon_{\bm{\textit{R}}}+\frac{\varepsilon_{\mathbf{M}}}{\lambda_{\min}(\bm{J})}\bigg{)}^2}{2\left(k_{\Omega}\!-\!c_R\!\right)}.
    \end{aligned}
    }
    \label{C_R}
\end{equation}
Combining with Eq.~\eqref{cR_condition}, if the positive constant $c_R$ is sufficiently small to satisfy:
\begin{equation}
{
    \begin{aligned}
    c_R\!< \min\left\{\frac{k_Rk_{\Omega}}{k_{\Omega}^2+k_R},\sqrt{k_R},\sqrt{\frac{2k_{R}}{2-\psi_{\textit{R}}}},k_\Omega\right\},\label{cR bound}
\end{aligned}
}
\end{equation}
it follows that matrix $\bm{\mathcal{M}}_{\bm{\textit{R}}}$ is positive definite. Therefore, Eq.~\eqref{dot_V_R_quadratic} can be further expressed as:
\begin{equation}
    \begin{aligned}    
    \bm{\dot{\mathcal{V}}}_{\bm{\textit{R}}}\leq-\lambda_{\min}(\bm{\mathcal{M}}_{\bm{\textit{R}}})\|\bm{z}_{\bm{\textit{R}}}\|^2+\mathbf{C}_{\bm{\textit{R}}},
    \end{aligned}
    \label{dot_V_R_quadratic2}
\end{equation}
where $\mathbf{C}_{\bm{\textit{R}}}>0$. To proceed, substituting Eq.~\eqref{VR_p1p2bound}, it holds that:
\begin{equation}
    \begin{aligned}    
    \bm{\dot{\mathcal{V}}}_{\bm{\textit{R}}}\leq-2\beta\bm{\mathcal{V}_{\bm{\textit{R}}}}+\mathbf{C}_{\bm{\textit{R}}},
    \end{aligned}
    \label{dot_V_R_decay}
\end{equation}
with $\beta=\frac{\lambda_{\min}(\bm{\mathcal{M}}_{\bm{\textit{R}}})}{2\mathfrak{p}_2\lambda_{\max}(\bm{\mathfrak{M}}_{\bm{\textit{R}}2})}$.
Inequality~\eqref{dot_V_R_decay} establishes a standard 
exponential-decay form for the Lyapunov function, except for the 
presence of the bounded term $\mathbf{C}_{\bm{\textit{R}}}$. 
This implies that the closed-loop error trajectories decay 
exponentially whenever the Lyapunov function is sufficiently large, 
and converge toward a bounded neighborhood whose size is determined 
by $\mathbf{C}_{\bm{\textit{R}}}$. 
This leads to the following characterization of the 
\textit{almost-global exponential attractiveness} of the closed-loop 
rotational error dynamics.

\paragraph{Almost-Global Exponential Attractiveness}

The exponential decay form in~\eqref{dot_V_R_decay} indicates that the
Lyapunov function decreases exponentially whenever its value is
sufficiently large compared to the bounded term $\mathbf{C}_{\bm{\textit{R}}}$.
This bounded term aggregates both the disturbance effects and the
neural-network approximation errors.
Since the universal approximation theorem guarantees small
approximation error only within the neural identification region
$\mathcal{D}_{\mathcal{C}}$, the approximation error outside this region
may reach its worst-case bound $\varepsilon_{{\bm{\textit{R}}}}$.
Hence, the smallest attainable value of $\bm{\dot{\mathcal{V}}}_{\bm{\textit{R}}}$ may
remain strictly above zero, preventing convergence to the origin and
leading instead to convergence toward a forward-invariant residual set
whose radius increases with the maximal approximation error.

Even in this scenario, the projection and dead-zone mechanisms ensure
that all adaptive weights remain uniformly bounded.
Therefore, although the approximation error may be large outside
$\mathcal{D}_{\mathcal{C}}$, inequality~\eqref{dot_V_R_decay} still
enforces exponential decay of the Lyapunov function until the trajectory
reaches the corresponding residual set.

Together, these observations imply that the error trajectories must approach a bounded residual region whose radius is determined by the constant $\mathbf{C}_{\bm{\textit{R}}}$.
To make this dependence explicit, we integrate the differential inequality~\eqref{dot_V_R_decay} and obtain the following exponential bound:
\begin{equation}
    \begin{aligned}    
    \|\bm{z}_{\bm{\textit{R}}}(t)\|\leq\alpha\|\bm{z}_{\bm{\textit{R}}}(0)\|e^{-\beta t}+r_1,
    \end{aligned}
    \label{NES condition1}
\end{equation}
where $\alpha\!=\!\!\sqrt{\frac{\mathfrak{p}_2\lambda_{\max}(\bm{\mathfrak{M}}_{\bm{\textit{R}}2})}{\mathfrak{p}_1\lambda_{\min}(\bm{\mathfrak{M}}_{\bm{\textit{R}}1})}}$, $r_1\!=\!\!\sqrt{\frac{\mathbf{C}_{\bm{\textit{R}}}}{2\beta\mathfrak{p}_1\lambda_{\min}(\bm{\mathfrak{M}}_{\bm{\textit{R}}1})}}$. 
Given the explicit bound in~\eqref{NES condition1}, it becomes clear that the closed-loop trajectories do not converge to the origin but instead approach a bounded neighborhood whose radius is determined by the radius $r_1$.
This motivates the definition of the following residual set:
\begin{equation}
    {
    \begin{aligned}
\mathcal{D}_{{\bm{\textit{R}}1}} \triangleq \Big{\{}\bm{z}_{\bm{\textit{R}}}\in\mathbb{R}^{2}:\|\bm{z}_{\bm{\textit{R}}}\|\le r_1\Big{\}},
    \end{aligned}
    }
\end{equation}
The inequality~\eqref{NES condition1} further implies that the distance from 
$\bm{z}_{\bm{\textit{R}}}(t)$ to this set decays exponentially:
\[
\operatorname{dist}\!\left(\bm{z}_{\bm{\textit{R}}}(t),\,\mathcal{D}_{{\bm{\textit{R}}1}}\right)
\;\le\;
\alpha\,
\operatorname{dist}\!\left(\bm{z}_{\bm{\textit{R}}}(0),\,\mathcal{D}_{{\bm{\textit{R}}1}}\right)
e^{-\beta t},
\qquad \forall\, t\ge 0,
\]
which shows that $\mathcal{D}_{{\bm{\textit{R}}1}}$ is exponentially attractive from 
almost all initial conditions (excluding the $180^\circ$ attitude 
ambiguity).  
This establishes the \textit{almost-global exponential attractiveness}, as stated in \textbf{\textit{Proposition 1}}.

\paragraph{Local Exponential Convergence within the Identification Region $\mathcal{D}_{\mathcal{C}}$}
\label{supp::para::Local-Exponential-Convergence}
We now refine the almost-global exponential attractiveness established
above by analyzing the closed-loop behavior \emph{inside a neural-network identification region $\mathcal{D}_{\mathcal{C}}$}.  
In particular, we show that once the rotational error state enters a
compact domain on which each neural network \textit{``slice"} admits a uniformly
small approximation error, the closed-loop dynamics exhibit
\emph{local exponential convergence} toward an arbitrarily small ball $\mathcal{B}_\epsilon$ around the equilibrium point $\bm{z}_{\bm{\textit{R}}}=\mathbf{0}$.  
By the universal approximation theorem, each neural \textit{``slice"} of the 
SANM module can approximate its associated disturbance feature with 
arbitrarily small error on any prescribed compact set of neural-network inputs.  
Therefore, there exists a compact region 
$\mathcal{D}_{\mathcal C}\subset\mathbb{R}^2$ of rotational error states 
on which the approximation error of each \textit{``slice"} is uniformly bounded by a 
smaller upper bound $\varepsilon_{{\bm{\textit{R}}}}$.  
We define this compact neural-network identification region as
\begin{equation}
    {
    \begin{aligned}
\mathcal{D}_{\mathcal{C}}
\;\triangleq\;
\Big\{
\bm{z}_{\bm{\textit{R}}}\in\mathbb{R}^2 
\;\big|\;
\|\bm{z}_{\bm{\textit{R}}}\|\le r_c
\Big\}.
    \end{aligned}
    }
\end{equation}
Once the rotational error enters the neural-identification region $\mathcal{D}_{\mathcal C}$, 
i.e., $\bm{z}_{\bm{\textit{R}}}(t_1)\in\mathcal{D}_{\mathcal C}$ for some $t_1\ge 0$, the \textit{``slice"} inputs of the SANM module remain within a 
prescribed compact domain. By the universal approximation theorem, there 
exist ideal weights such that the approximation error of each neural 
\textit{``slice"} is uniformly bounded by a small upper bound $\varepsilon_{{\bm{\textit{R}}}}$.  
Consequently, the radius $r_1$ in the inequality~\eqref{NES condition1} can be replaced by a tighter radius $\epsilon$, with $\epsilon<r_1<r_c$:
\begin{equation}
    \begin{aligned}    
\|\bm{z}_{\bm{\textit{R}}}(t)\|
\le \alpha\,\|\bm{z}_{\bm{\textit{R}}}(t_1)\|
e^{-\beta (t-t_1)} + \epsilon,
\qquad \forall\, t\ge t_1.
    \end{aligned}
    \label{NES condition2}
\end{equation}
By properly selecting the parameters of the neural network, including the number of neurons $l$ in the hidden layer, the center vectors $\textbf{c}_{kj}$ and the width $b_{kj}$, the universal approximation theorem \cite{1989 Multilayer feedforward networks are universal approximators} ensures that the upper bound of the approximation error can be made arbitrarily small, i.e., $\varepsilon_{\bm{\textit{R}}}\to0^+$. Furthermore, the precise identification of aerodynamic coefficients drives the upper bound $\varepsilon_{\mathbf{M}}$ to converge to zero. 
As $\varepsilon_{\bm{\textit{R}}}\to0^+$, $\varepsilon_{\mathbf{M}}\to0$, it follows that $\mathbf{C}_{\bm{\textit{R}}}\to0^+$ and $\epsilon\to0^+$.
This shows that once the trajectory enters $\mathcal{D}_{\mathcal C}$, the state solution of the rotational error dynamics $\bm{z}_{\bm{\textit{R}}}(t)$ exponentially converges into an arbitrarily small ball
\begin{equation}
    {
    \begin{aligned}
    \mathcal{B}_\epsilon 
    \triangleq 
    \Big\{
        \bm{z}_{\bm{\textit{R}}}\in\mathbb{R}^{2}
        \;\big|\;
        \|\bm{z}_{\bm{\textit{R}}}\|\le\epsilon
    \Big\}.
    \end{aligned}
    }
\label{small-ball}
\end{equation}
Therefore, \textbf{\textit{Proposition 2}} is established. This implies that the rotational state error vector $\|\mathbf{E}_{\bm{\textit{R}}}(t)\|=\|\big{(} \bm{e}^{\top}_{\bm{R}}(t),\bm{e}^{\top}_{\bm{\Omega}}(t)\big{)}^{\top}\|$ also converges to an arbitrarily small ball, and hence there exists a compact set $\mathcal{C}$ such that $\mathbf{E}_{\bm{\textit{R}}}(t) \in \mathcal{C}$ for all $t \geq t_1$. 
Consequently, once the trajectory enters the neural-network identification region $\mathcal{D}_{\mathcal{C}}$ at some time $t_1>0$,
\textbf{\textit{Proposition 3}} holds $\forall t \geq t_1$.

With \textbf{\textit{Propositions 2 and 3}} established, the overall behavior of the SANM-compensated rotational error dynamics can be interpreted through a hierarchy of nested invariant regions. Specifically, the error trajectory evolves almost-globally within $\mathcal{D}_{{\bm{\textit{R}}0}}$, is first exponentially attracted into the residual set $\mathcal{D}_{{\bm{\textit{R}}1}}$, and—once inside the neural-network identification region $\mathcal{D}_{\mathcal C}$—exhibits local exponential convergence towards an arbitrarily small ball $\mathcal{B}_\epsilon$.
This layered stability behavior is summarized schematically in Fig.~\ref{fig:nested_a}, which illustrates the nested set relations and the associated convergence mechanisms of the SANM-augmented closed loop.

\textbf{\textit{Remark \ref{supp:extended-proofs}.4:}} The above analysis established that $\bm{\mathcal{V}_{\bm{\textit{R},s}}}(t)$ is decreasing outside the ball $\mathcal{B}_\epsilon$.
From Cauchy-Schwarz and Young's inequalities, the initial value $\bm{\mathcal{V}_{\bm{\textit{R},s}}}(0)$ of Eq.~\eqref{V_R,s2} satisfies the following:
\begin{equation}
    {
    \begin{aligned}
        \bm{\mathcal{V}_{\bm{\textit{R},s}}}(0)\leq k_{R}\Psi_{\textit{R}}\big{(}\bm{R}(0),\bm{R}_{\bm{d}}(0)\big{)}+\|\bm{e}_{\bm{\Omega}}(0)\|^2+\frac{c_R^2}{2}\|\bm{e}_{\bm{R}}(0)\|^2.
    \end{aligned}
    }
\end{equation}
If $ \bm{z}_{\bm{\textit{R}}}(0)=(\|\bm{e}_{\bm{R}}(0)\|, \|\bm{e}_{\bm{\Omega}}(0)\|)^{\top}$ satisfies Eq.~\eqref{D_R0} and 
\begin{equation}
    {
    \begin{aligned}
       c_R<\sqrt{2k_{R}\Big{(}2-\Psi_{\textit{R}}\big{(}\bm{R}(0),\bm{R}_{\bm{d}}(0)\big{)}\Big{)}},
    \end{aligned}
    }
\end{equation}
the following bound can be deduced:
\begin{equation}
    {
    \begin{aligned}
        k_{R}\Psi_{\textit{R}}\big{(}\bm{R}(t),\bm{R}_{\bm{d}}(t)\big{)}\leq\bm{\mathcal{V}_{\bm{\textit{R},s}}}(t) \leq\bm{\mathcal{V}_{\bm{\textit{R},s}}}(0) <2k_{R}.
    \end{aligned}
    }
\end{equation}
As a result, $\Psi_{\textit{R}}(\bm{R}(t),\bm{R}_{\bm{d}}(t))\!<\!2$ holds $\forall t \geq 0$.

\begin{figure}[t]
    \centering
    \begin{subfigure}[t]{0.47\linewidth}
        \centering
        \includegraphics[width=\linewidth]{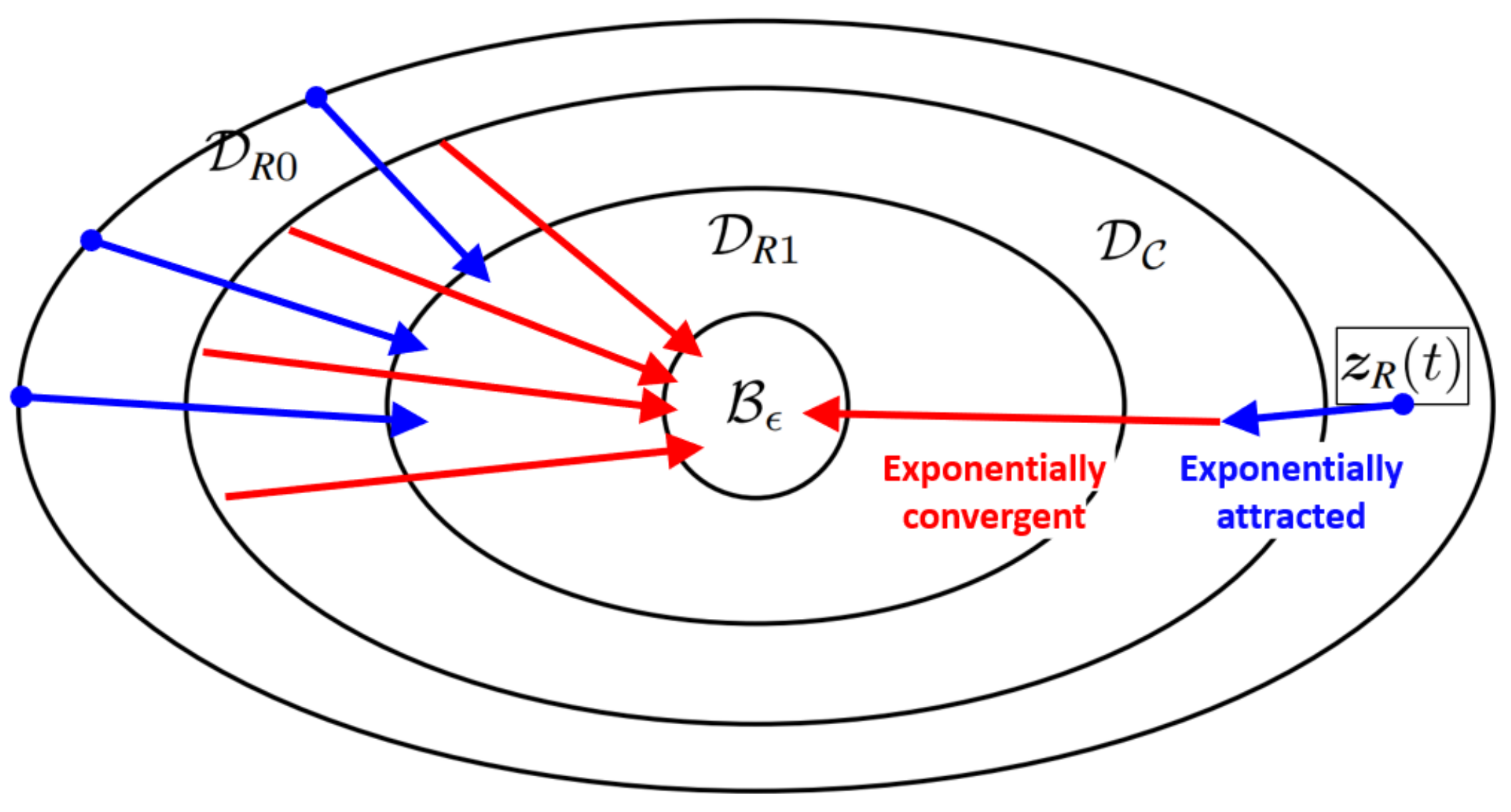}
        \caption{Nominal exponential attractiveness and convergence.}
        \label{fig:nested_a}
    \end{subfigure}
    \hfill
    \begin{subfigure}[t]{0.47\linewidth}
        \centering
        \includegraphics[width=\linewidth]{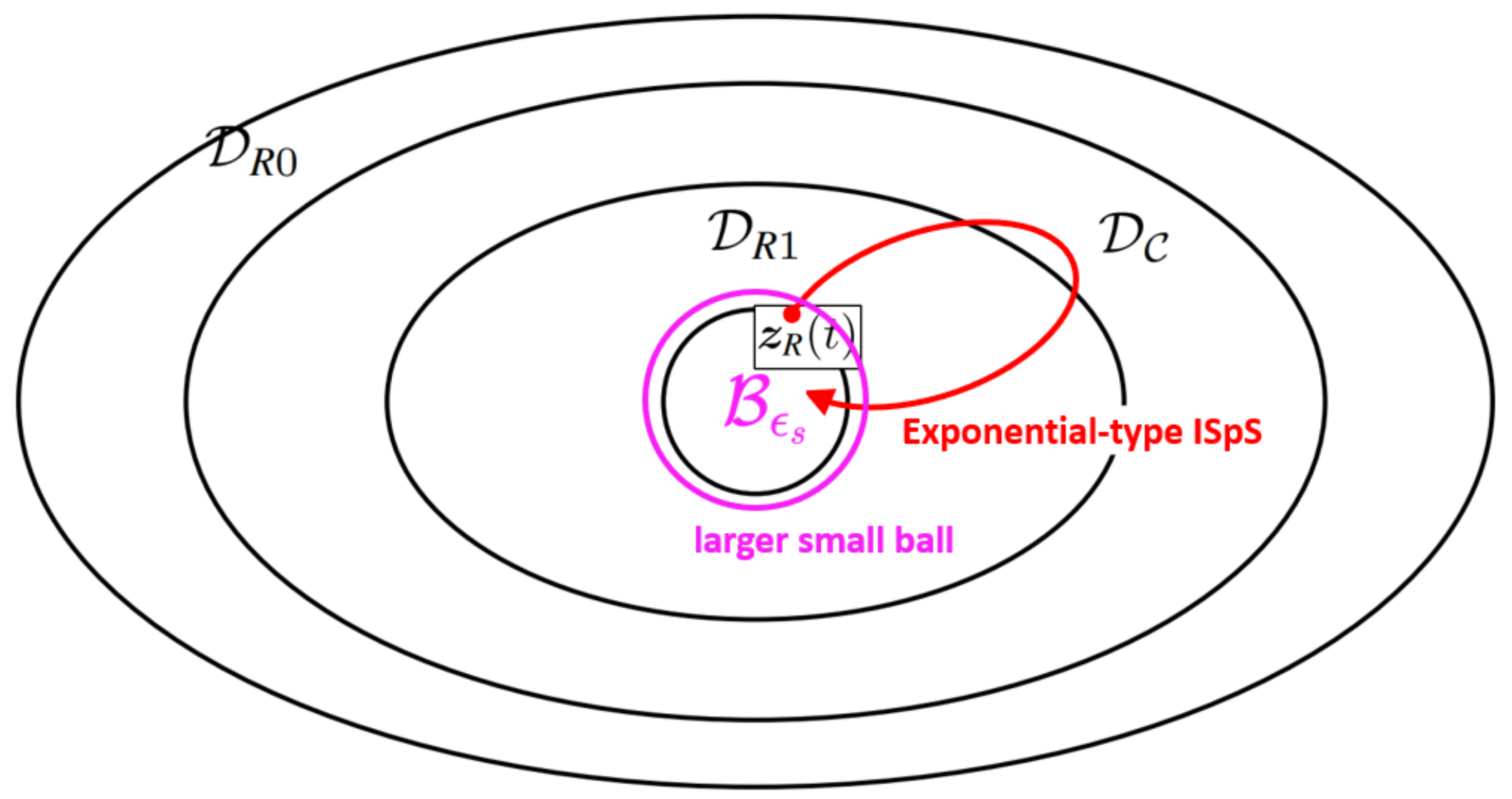}
        \caption{Sampled-data exponential-type ISpS.}
        \label{fig:nested_b}
    \end{subfigure}

    \caption{
    Nested stability structure of the SANM-compensated rotational error dynamics.
    The error-state vector $\bm{z}_{\bm{\textit{R}}}(t)=(\|\bm{e}_{\bm{R}}\|,\|\bm{e}_{\bm{\Omega}}\|)^{\top}$ evolves inside the almost-global domain 
    $\mathcal{D}_{\bm{\textit{R}}0}$.  
    (a) In the nominal case, the trajectory is first \emph{exponentially attracted} into the residual set 
    $\mathcal{D}_{\bm{\textit{R}}1}$ and, once inside the neural-network identification region $\mathcal{D}_{\mathcal{C}}$, it becomes 
    \emph{exponentially convergent} to an arbitrarily small ball $\mathcal{B}_{\epsilon}$.  
    (b) Under sampled-data implementation with bounded time-varying disturbances, the system exhibits 
transient deviations within $\mathcal{D}_{\bm{\textit{R}}1}$ but retains 
\emph{Input-to-State Practical Stability (ISpS)}, preserving the 
exponential decay toward a slightly enlarged residual ball $\mathcal{B}_{\epsilon_s}$.  \textit{Nested set relation:} 
$\mathcal{B}_\epsilon \subset \mathcal{B}_{\epsilon_s} \subset \mathcal{D}_{\bm{\textit{R}}1} \subset 
\mathcal{D}_{\mathcal{C}} \subset \mathcal{D}_{\bm{\textit{R}}0}$.
    }
    \label{fig:nested_sets}
\end{figure}

\subsection{ISpS under Sampled-Data Implementation}\label{sec:ISpS_sampled}

In this subsection, we show that the continuous-time result established in \eqref{dot_V_R_decay} implies an \emph{Input-to-State Practical Stability (ISpS)} property for the sampled-data realization with zero-order hold (ZOH). Let $dt>0$ denote the sampling period (the controller runs at $1/dt$~Hz) and assume that the reference signals $(\bm{R}_{\bm d},\bm{\Omega}_{\bm d},\dot{\bm{\Omega}}_{\bm d})$ are piecewise $C^1$ in each sampling interval $[n\,dt,(n+1)\,dt)$, while the implementation is ZOH within each interval. We also allow for a bounded computation/sensor delay $\tau\in[0,\overset{\tiny \text{max}}{\tau})$ with $\overset{\tiny \text{max}}{\tau}<dt$. 

\subsubsection{Continuous-time differential inequality}
From \eqref{dot_V_R_decay} and the bounds \eqref{C_R}, \eqref{cR bound}, there exist strictly positive constants $\beta$ and $\mathbf{C}_{\bm{\textit R}}$ such that the Lyapunov function $\bm{\mathcal{V}}_{\bm{\textit R}}$ satisfies, for all $t\ge0$,
\begin{equation}
\dot{\bm{\mathcal{V}}}_{\bm{\textit R}}(t) \le -2\beta\,\bm{\mathcal{V}}_{\bm{\textit R}}(t) + \mathbf{C}_{\bm{\textit R}}.
\label{eq:cts_diff_ineq}
\end{equation}
Here $\mathbf{C}_{\bm{\textit R}}$ depends quadratically on the bounded residuals $\varepsilon_{\bm{\textit R}}$ and $\varepsilon_{\mathbf M}$ as in \eqref{C_R}, hence it is an \emph{input-magnitude} term. If $\varepsilon_{\bm{\textit R}}\to0^+$ and $\varepsilon_{\mathbf M}\to0$, we have $\mathbf{C}_{\bm{\textit R}}\to0^+$.

The differential inequality~\eqref{eq:cts_diff_ineq} ensures the exponential decay of 
$\bm{\mathcal{V}}_{\bm{\textit R}}(t)$ in continuous time. 
To investigate whether this property persists under sampled-data implementation, 
we first characterize the deviation between the ideal continuous-time dynamics and its sampled realization.

\subsubsection{Implementation residual}
Let $dt>0$ and $T_n\triangleq[n\,dt,(n+1)\,dt)$. Under ZOH with delay $\tau\in[0,\overset{\tiny \text{max}}{\tau})$, the implemented closed-loop vector field is $\mathfrak{f}^{\mathrm{impl}}(\cdot,t)$, while the ideal one is $\mathfrak{f}^{\mathrm{ideal}}(\cdot,t)$. Along the implemented trajectory,
\begin{equation}
\dot{\bm{\mathcal V}}_{\bm{\textit R}}(t)
\le -2\beta\,\bm{\mathcal V}_{\bm{\textit R}}(t)+\mathbf C_{\bm{\textit R}}
+\delta(t),\quad t\in T_n,
\end{equation}
with the residual defined by
\begin{equation}
\delta(t)\triangleq\nabla \bm{\mathcal V}_{\bm{\textit R}}(\bm{z}_{\bm{\textit{R}}}(t))^\top\!\Big(
\mathfrak{f}^{\mathrm{impl}}(\bm{z}_{\bm{\textit{R}}}(t),t)-\mathfrak{f}^{\mathrm{ideal}}(\bm{z}_{\bm{\textit{R}}}(t),t)\Big).
\label{eq:delta_min}
\end{equation}

Then, we analyze the Lyapunov function over each sampling interval as follows.

\paragraph{Step 1: integration over one sampling interval}
Fix $n\in\mathbb{N}$ and integrate \eqref{eq:cts_diff_ineq} over $[n\,dt,(n+1)\,dt)$. By applying the standard comparison lemma and Grönwall’s inequality within each interval, we obtain
\begin{equation}
\bm{\mathcal{V}}_{\bm{\textit R}}\big((n\!+\!1)dt\big)\ \le\ e^{-2\beta dt}\,\bm{\mathcal{V}}_{\bm{\textit R}}(n\,dt) + \frac{\mathbf{C}_{\bm{\textit R}}}{2\beta}\Big(1-e^{-2\beta dt}\Big) + \underbrace{\int_{n\,dt}^{(n+1)\,dt} e^{-2\beta\big((n+1)dt-s\big)}\,\delta(s)\,ds}_{\Delta_n},
\label{eq:one_step_bound_exact}
\end{equation}
where $\Delta_n$ is the \emph{discretization residual} in $[n\,dt,(n+1)\,dt)$ induced by the ZOH implementation and the bounded delay $\tau<\overset{\tiny \text{max}}{\tau}$. Because the closed-loop vector field and the update laws are locally Lipschitz on a compact set (see Section~\ref{Proofs of Proposition 1 and 2}), there exists a constant $L>0$ such that
\begin{equation}
0\ \le\ \Delta_n\ \le\ L\,\big(dt^2+dt\,\tau\big)\qquad \text{for all }n\in\mathbb{N}.
\label{eq:Delta_bound}
\end{equation}
Thus $\Delta_n$ is of higher order in $dt$ (and linear in a small delay~$\tau$).

\paragraph{Step 2: discrete ISpS inequality}
Using $e^{-2\beta dt}\le 1-2\beta dt+\beta^2 dt^2$ (Taylor expansion) and \eqref{eq:Delta_bound}, the one-step bound \eqref{eq:one_step_bound_exact} implies the following \emph{linear} difference inequality for sufficiently small $dt$:
\begin{equation}
\bm{\mathcal{V}}_{\bm{\textit R}}(n\!+\!1) \le \big(1-\beta dt\big)\,\bm{\mathcal{V}}_{\bm{\textit R}}(n) + dt\,\mathbf{C}_{\bm{\textit R}}\ +\ \overset{\tiny \text{max}}{L}\,(dt^2+dt\,\tau),
\label{eq:discrete_ISpS}
\end{equation}
with a constant $\overset{\tiny \text{max}}{L}>0$ (absorbing numerical coefficients). Inequality \eqref{eq:discrete_ISpS} is the standard \emph{discrete ISpS form}: the ``decay term” $(1-\beta dt)\bm{\mathcal{V}}_{\bm{\textit R}}(n)$ is strictly contractive for $dt<\beta^{-1}$, while the ``input terms” are $dt\,\mathbf{C}_{\bm{\textit R}}$ (from bounded approximation residuals) and a small affine perturbation $\overset{\tiny \text{max}}{L}(dt^2+dt\tau)$ (from sampling and delay).

\paragraph{Step 3: uniform bound and persistence of exponential estimate}
By unfolding \eqref{eq:discrete_ISpS} over $n$ and summing the geometric series,
\begin{equation}
\bm{\mathcal{V}}_{\bm{\textit R}}(n)\ \le\ (1-\beta dt)^n\,\bm{\mathcal{V}}_{\bm{\textit R}}(0)\ +\ \frac{dt\,\mathbf{C}_{\bm{\textit R}}+\overset{\tiny \text{max}}{L}(dt^2+dt\tau)}{\beta dt}.
\label{eq:V_uniform_bound}
\end{equation}
Equivalently, there exist \emph{slightly inflated} constants $\alpha_s\!>\!0$ and $\beta_s\!\in\!(0,\beta)$ such that
\begin{equation}
\|\bm{z}_{\bm{\textit R}}(ndt)\|\ \le\ \alpha_s\,e^{-\beta_s ndt}\,\|\bm{z}_{\bm{\textit R}}(0)\|\ +\ \epsilon_s,
\label{eq:ISpS_exp_estimate}
\end{equation}
where the residual radius satisfies
\begin{equation}
\epsilon_s\ =\ \sqrt{\frac{1}{\mathfrak{p}_1\lambda_{\min}(\bm{\mathfrak{M}}_{\bm{\textit R}1})}\cdot \frac{\mathbf{C}_{\bm{\textit R}}}{2\beta}\ + \mathbf{C}_{\delta}\,(dt+\tau)}\ \xrightarrow[dt,\tau\to 0]{}\ \sqrt{\frac{\mathbf{C}_{\bm{\textit R}}}{2\beta\,\mathfrak{p}_1\lambda_{\min}(\bm{\mathfrak{M}}_{\bm{\textit R}1})}},
\label{eq:epsilon_d}
\end{equation}
for $\mathbf{C}_{\delta}>0$. Hence the discrete-time exponential estimate \eqref{eq:ISpS_exp_estimate} \emph{persists} with slightly relaxed constants $(\alpha_s,\beta_s)$ compared to the continuous-time ones $(\alpha,\beta)$ in \eqref{NES condition1} and \eqref{NES condition2}.
Consequently, the sampled-data trajectory exponentially approaches a 
slightly enlarged residual ball
\begin{equation}
\mathcal{B}_{\epsilon_s}
\;\triangleq\;
\Big\{
    \bm{z}_{\bm{\textit R}}\in\mathbb{R}^{2}
    \;\big|\;
    \|\bm{z}_{\bm{\textit R}}\|\le \epsilon_s
\Big\},
\end{equation}
with a slightly enlarged radius $\epsilon_s$, as shown in Fig.~\ref{fig:nested_b}.

\subsubsection{Allowable sampling period and delay}
Note that the constant $\beta>0$ is determined first by the continuous-time Lyapunov analysis in \eqref{dot_V_R_decay}, and is therefore independent of the sampled-data implementation parameters $dt$ and $\tau$. 
From \eqref{eq:discrete_ISpS}, the contraction factor remains valid if
\begin{equation}
0<1-\beta dt<1,
\end{equation}
which yields $dt<\frac{1}{\beta}$.
Moreover, since $0\le\tau<dt$, we have
\begin{equation}
dt^2+dt\tau \le 2dt^2.
\end{equation}
Therefore, a conservative sufficient condition to ensure that the sampled-data perturbation term $\overset{\tiny \text{max}}{L}(dt^2+dt\tau)$ remains dominated by the nominal decay term is
\begin{equation}
2\overset{\tiny \text{max}}{L}dt^2 \le \beta dt,
\end{equation}
which implies $dt\le\frac{\beta}{2\overset{\tiny \text{max}}{L}}$.
Combining the above sufficient conditions and \eqref{eq:discrete_ISpS}-\eqref{eq:V_uniform_bound}, it suffices to choose
\begin{equation}
0\ <\ dt\ <\ dt^\star := \min\left\{\frac{1}{\beta},\ \frac{\beta}{2\overset{\tiny \text{max}}{L}}\right\},\qquad 
0\ \le\ \tau\ <\ \tau^\star\ :=\ \min\left\{dt,\ \frac{\beta}{2\overset{\tiny \text{max}}{L}}\right\},
\label{eq:h_tau_bounds}
\end{equation}
where $dt^\star$ and $\tau^\star$ respectively denote the \emph{maximum allowable sampling period} and the \emph{maximum allowable delay} that preserve the discrete-time contraction property. With these choices, the contraction factor satisfies $1-\beta dt\in(0,1)$, and the perturbation term $\overset{\tiny \text{max}}{L}(dt^2+dt\tau)$ remains dominated by the exponential decay. This bound is conservative and is only used to guarantee preservation of the discrete-time contraction property.
Under the experimental setting ($dt=0.0025\,$s, $\tau<dt$), these bounds are readily satisfied.

\textbf{\textit{Remark \ref{supp:extended-proofs}.5:}} The sampling bound in \eqref{eq:h_tau_bounds} reflects a trade-off between the continuous-time convergence rate and the robustness of the implementation of the sampled-data. 
If the convergence rate $\beta$ is too small, the system has a limited ability to compensate for discretization errors. In this case, the resistance of the closed-loop system to sampling-induced perturbations becomes weaker, and a smaller sampling period is required to maintain stability.
On the other hand, if $\beta$ is excessively large, although continuous-time convergence becomes faster, the allowable sampling period may also decrease since the discrete implementation must capture the faster system dynamics. This reflects the classical trade-off between control bandwidth and sampling frequency in digital control systems.
Therefore, the admissible sampling period depends on a balance between the nominal exponential decay and the perturbation growth induced by the implementation of the sampled-data.

\subsubsection{Input-to-State Practical Stability (ISpS)}
Inequality \eqref{eq:ISpS_exp_estimate} is precisely an ISpS estimate for the sampled-data closed loop with respect to the bounded residuals that define $\mathbf{C}_{\bm{\textit R}}$ (see~\eqref{C_R}). Therefore, the sampled implementation preserves the exponential convergence up to a small ball whose radius scales with $\varepsilon_{\bm{\textit R}}$, $\varepsilon_{\mathbf M}$, and $(dt,\tau)$, thereby establishing the discrete-time ISpS property of the sampled-data closed loop. Thus, \textbf{\textit{Proposition 4}} is established.

\section{Extended Information for Experiment 1}
\label{supp:Experiment-1}
The numerical simulations were conducted in \textit{MATLAB R2023b} on a personal computer equipped with a \textit{12th Gen Intel Core i7-12700KF} CPU and 32 GB RAM, running \textit{Windows 10 (64-bit)}. The simulation used a fixed-step ODE3 (Bogacki–Shampine) solver with a step size of $dt = 0.0025$ s. All controller modules were implemented in Simulink. The adaptive law \textit{``slices"} and neural network \textit{``slices"} were realized via \textit{S-Function} blocks, while the remaining control logic was implemented using \textit{MATLAB Function} blocks.

When no model uncertainties introduced (Fig.~\ref{fig:exp1_no_disturb}), the parameters of the reference inertia tensor and the real inertia tensor were as follows: 
\begin{equation}
{\small
   \begin{array}{cc}
   &\bm{J}_{\textrm{ref}}=10^{-2}\mathrm{diag}[1.1,2.0,2.3]~\mathrm{kg}\mathrm{m}^{2},\\
   &\bm{J}_{\textrm{real}}=10^{-2}\mathrm{diag}[1.1,2.0,2.3]~\mathrm{kg}\mathrm{m}^{2},
   \notag
\end{array} 
}
\label{S:Parameters of inertia tensor}
\end{equation}
where the reference value was used to construct the baseline controller (SANM off).

When model uncertainties were introduced (Fig.~\ref{fig:exp1_with_disturb}), the parameters of the reference inertia tensor remained unchanged, whereas those of the real inertia tensor were increased to:
\begin{equation}
{\small
   \begin{array}{cc}
   &\bm{J}_{\textrm{ref}}=10^{-2}\mathrm{diag}[1.1,2.0,2.3]~\mathrm{kg}\mathrm{m}^{2},\\
   &\bm{J}_{\textrm{real}}=10^{-2}\mathrm{diag}[2.1,3.1,4.2]~\mathrm{kg}\mathrm{m}^{2}.
   \notag
\end{array} 
}
\label{S:Parameters of inertia tensor2}
\end{equation}
\paragraph{Control Parameters}\label{S:sec:control_parameters}
The attitude controller was tested under \textbf{\textit{Scenario 2 ($\bm{J}$ is unknown)}} and the PD gains were chosen as $k_{R}=100$, $ k_{\Omega}=80$. The parameters of adaptive law \textit{``slices"} in SANM were selected as
\begin{equation}
{\small
   \begin{array}{cc}
   &\eta_{1}=0.01,\, \eta_{2}=0.01,\, \eta_{3}=0.05, 
\, c_R=0.6,\\
&\mathfrak{s}_{1}=0.02, \,\mathfrak{s}_{2}=0.02, \,\mathfrak{s}_{3}=0.02, \,\\
   &\bm{\bar{J}}^{\text{vec}}(0)=10^{-2}(1,2,2)^{\top}~\mathrm{kg}\mathrm{m}^{2},\overset{\tiny \text{max}}{J}_{1}=0.03, \overset{\tiny \text{max}}{J}_{2}=0.03, \overset{\tiny \text{max}}{J}_{3}=0.04. 
   \notag
\end{array} 
}
\label{s:Parameters of AL slices}
\end{equation}

In the test of Fig.~\ref{fig:exp1_no_disturb}, each neural network \textit{``slice"} employed a hidden layer with $l=5$ neurons. The bound of the network weights ($\overset{\tiny \text{max}}{W}_{j}$) was treated as a sensitivity variable, and its value was selected from ${1,10,100}$ to examine the effect of weight saturation on the convergence speed. Other parameters were selected as follows.
\begin{equation}
{\small
   \begin{array}{cc}
\textbf{centers:}\,\,&\begin{bmatrix}\textbf{c}_{11},\textbf{c}_{21},\textbf{c}_{31},\textbf{c}_{41},\textbf{c}_{51}\end{bmatrix}=\begin{bmatrix}-1&-0.5&0&0.5&1\\ -10&-5&0&5&10\\\end{bmatrix},\\
&\begin{bmatrix}\textbf{c}_{12},\textbf{c}_{22},\textbf{c}_{32},\textbf{c}_{42},\textbf{c}_{52}\end{bmatrix}=\begin{bmatrix}-1&-0.5&0&0.5&1\\ -10&-5&0&5&10\\\end{bmatrix},\\
&\begin{bmatrix}\textbf{c}_{13},\textbf{c}_{23},\textbf{c}_{33},\textbf{c}_{43},\textbf{c}_{53}\end{bmatrix}=\begin{bmatrix}-1&-0.5&0&0.5&1\\ -6&-3&0&3&6\\\end{bmatrix},\\
    \textbf{widths:}\,\,&b_{11}=b_{21}=b_{31}=b_{41}=b_{51}=2,\\
   &b_{12}=b_{22}=b_{32}=b_{42}=b_{52}=2,\\
   &b_{13}=b_{23}=b_{33}=b_{43}=b_{53}=3,\\
 \textbf{dead-zone thresholds:}\,\,&\zeta_1=\zeta_2=\zeta_3=0.0005,\\
\textbf{learning rates:}\,\,&\gamma_{\bm{\textit{R}}1}=120,\,\gamma_{\bm{\textit{R}}2}=120,\,\gamma_{\bm{\textit{R}}3}=50.
   \notag
   \end{array}
}
\label{S:Parameters of NN slices}
\end{equation}

In the test of Fig.~\ref{fig:exp1_with_disturb}, each neural network \textit{``slice''} also employed a hidden layer with $l=5$ neurons. The learning rate of the \textit{``slice"} along the $\bm{\vec{b}}_{1}$-axis ($\gamma_{\bm{\textit{R}}1}$) was treated as a sensitivity variable, and its value was selected from $\{10,100,500\}$ to examine the effect of the learning rate on the performance against time-varying disturbances. Other parameters were selected as follows.
\begin{equation}
{\small
   \begin{array}{cc}
\textbf{centers:}\,\,&\begin{bmatrix}\textbf{c}_{11},\textbf{c}_{21},\textbf{c}_{31},\textbf{c}_{41},\textbf{c}_{51}\end{bmatrix}=\begin{bmatrix}-1&-0.5&0&0.5&1\\ -10&-5&0&5&10\\\end{bmatrix},\\
&\begin{bmatrix}\textbf{c}_{12},\textbf{c}_{22},\textbf{c}_{32},\textbf{c}_{42},\textbf{c}_{52}\end{bmatrix}=\begin{bmatrix}-1&-0.5&0&0.5&1\\ -10&-5&0&5&10\\\end{bmatrix},\\
&\begin{bmatrix}\textbf{c}_{13},\textbf{c}_{23},\textbf{c}_{33},\textbf{c}_{43},\textbf{c}_{53}\end{bmatrix}=\begin{bmatrix}-1&-0.5&0&0.5&1\\ -6&-3&0&3&6\\\end{bmatrix},\\
    \textbf{widths:}\,\,&b_{11}=b_{21}=b_{31}=b_{41}=b_{51}=2,\\
   &b_{12}=b_{22}=b_{32}=b_{42}=b_{52}=2,\\
   &b_{13}=b_{23}=b_{33}=b_{43}=b_{53}=3,\\
   \textbf{dead-zone thresholds:}\,\,&\zeta_1=\zeta_2=\zeta_3=0.0005,\\
\textbf{weight bound:}\,\,&\overset{\tiny \text{max}}{W}_{j}=200.
   \notag
   \end{array}
}
\label{S:Parameters of NN slices}
\end{equation}

\paragraph{Phenomenon: (Larger Weight Bound Slows Convergence)}\label{S:sec:Phenomenon_1}

To investigate the influence of the network-weight bound on the transient behavior of the SANM-augmented controller, a variant experiment was conducted in which the upper bound of each neural network \textit{``slice"}, $\overset{\tiny \text{max}}{W}_{j}$, was varied among $\{1,10,100\}$ while keeping other parameters fixed. Each slice employed a hidden layer with $l=5$ neurons, and the baseline geometric controller on $\mathbf{SO}(3)$ was executed under the same initial condition without any external disturbance, ensuring that the observed behavior originates purely from the adaptive-neural component.

Fig.~\ref{fig:exp1_no_disturb} demonstrates a clear monotonic tendency: \textbf{larger weight bounds may lead to slower convergence of the attitude-configuration error} $\Psi_{\textit R}$. When $\overset{\tiny \text{max}}{W}_{j}$ is small, the projection operator confines the weight update laws within a tighter region, producing stronger regularization on the neural output and a faster transient decay. Conversely, when the admissible weight space is enlarged, the adaptive parameter trajectories explore a broader region before settling, resulting in slower error reduction even though the same asymptotic equilibrium is reached. 

This phenomenon aligns with the Lyapunov analysis presented in Supp.~\ref{supp:extended-proofs}, where the derivative of the composite function satisfies
\begin{equation}
\dot{\bm{\mathcal V}}_{\bm{\textit R}} \le -2\beta\,\bm{\mathcal V}_{\bm{\textit R}}+\mathbf{C}_{\bm{\textit R}},
\end{equation}
and the residual term $\mathbf{C}_{\bm{\textit R}}$ increases with the bound $\overset{\tiny \text{max}}{W}_{j}$ due to the enlarged estimation region. Hence, a larger weight bound weakens the effective negative definiteness of $\dot{\bm{\mathcal V}}_{\bm{\textit R}}$, thereby \textbf{reducing the exponential convergence rate} while preserving stability.

In practice, this trade-off suggests that the weight bound should be tuned moderately: overly small bounds restrict the network’s expressive ability, while overly large ones can slow the convergence under large commanded rotations (even though the convergence rate under disturbances will increase, the nominal no-disturbance convergence may become slower).

\section{Extended Information for Experiment 2}
\label{supp:sec:Experiment-2}

\begin{figure}[ht]
    \centering
    \begin{subfigure}[b]{0.32\linewidth}
        \centering
        \includegraphics[width=\linewidth]{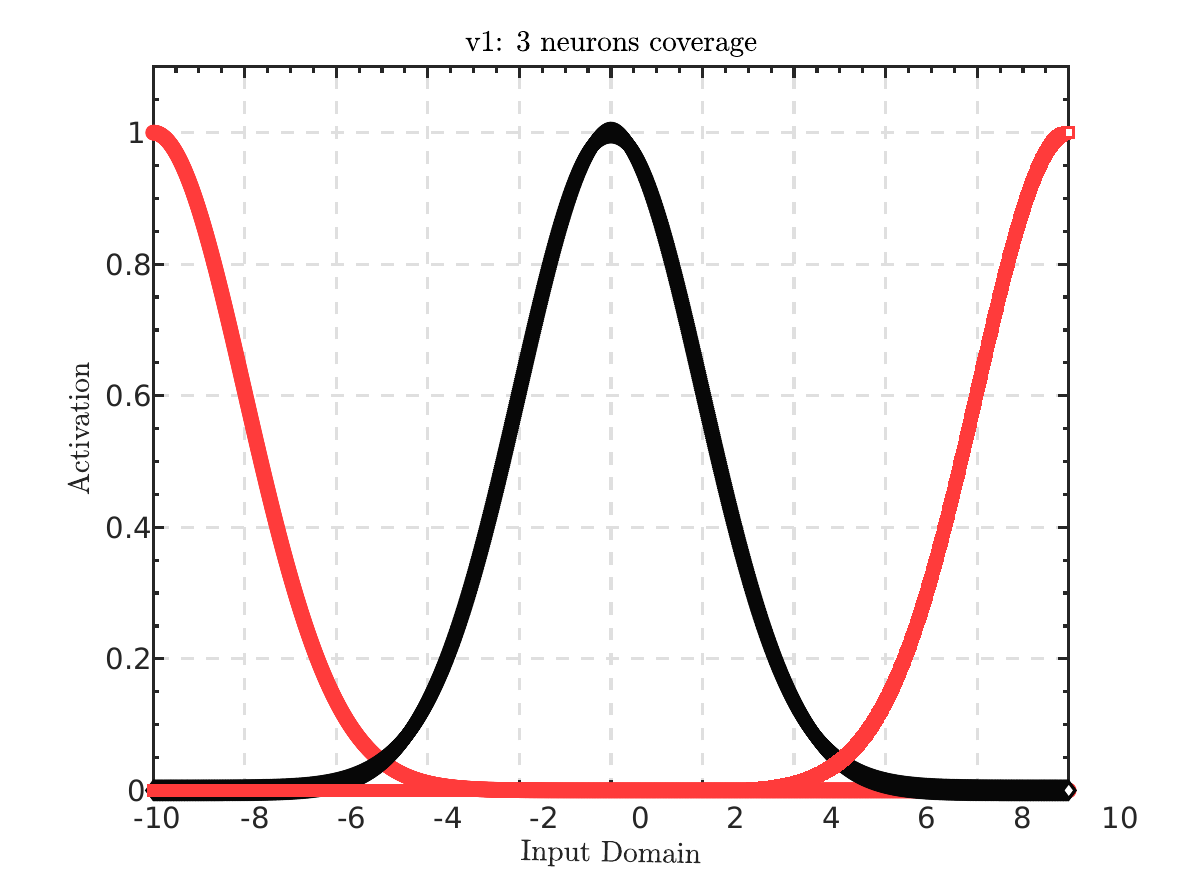}
        \caption{\footnotesize SANM.v1 (3 neurons)}
        \label{fig:RBF_v1}
    \end{subfigure}
    \hfill
    \begin{subfigure}[b]{0.32\linewidth}
        \centering
        \includegraphics[width=\linewidth]{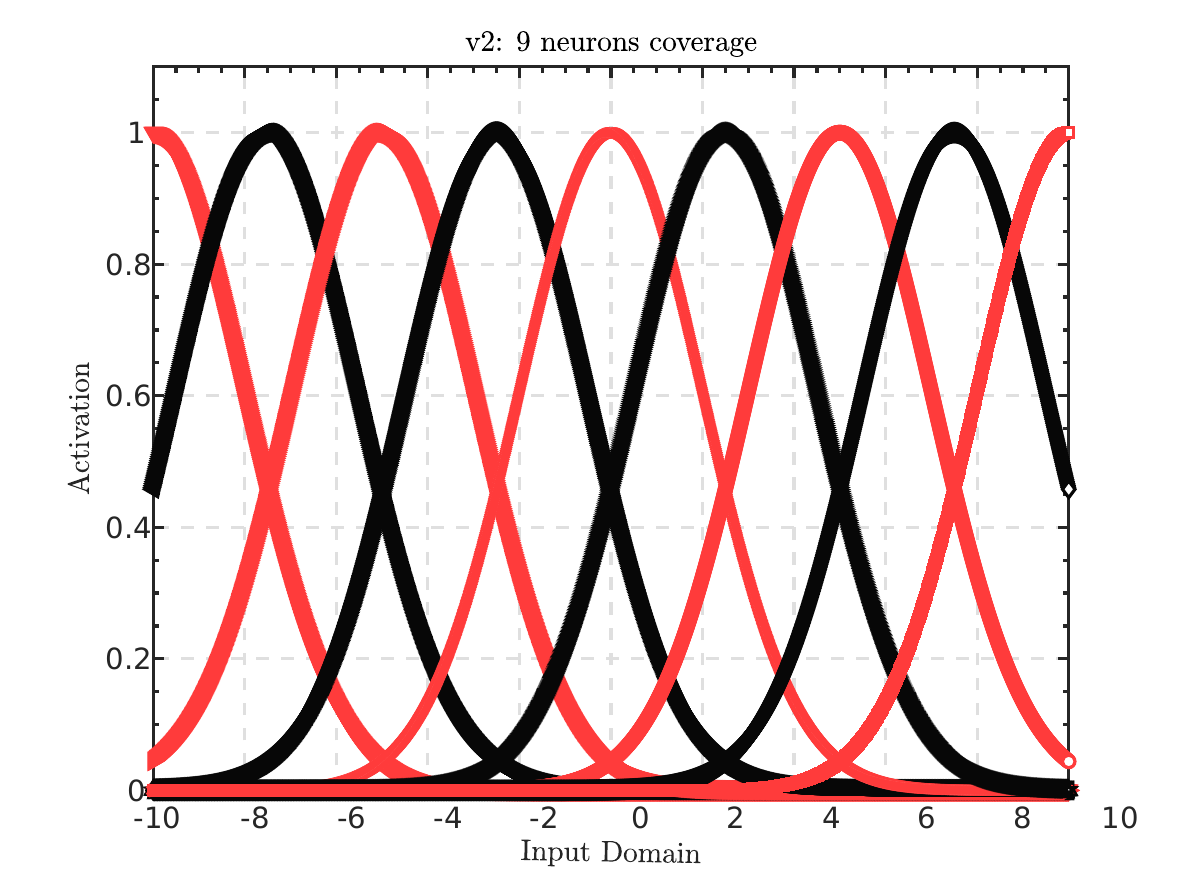}
        \caption{\footnotesize SANM.v2 (9 neurons)}
        \label{fig:RBF_v2}
    \end{subfigure}
    \hfill
    \begin{subfigure}[b]{0.32\linewidth}
        \centering
        \includegraphics[width=\linewidth]{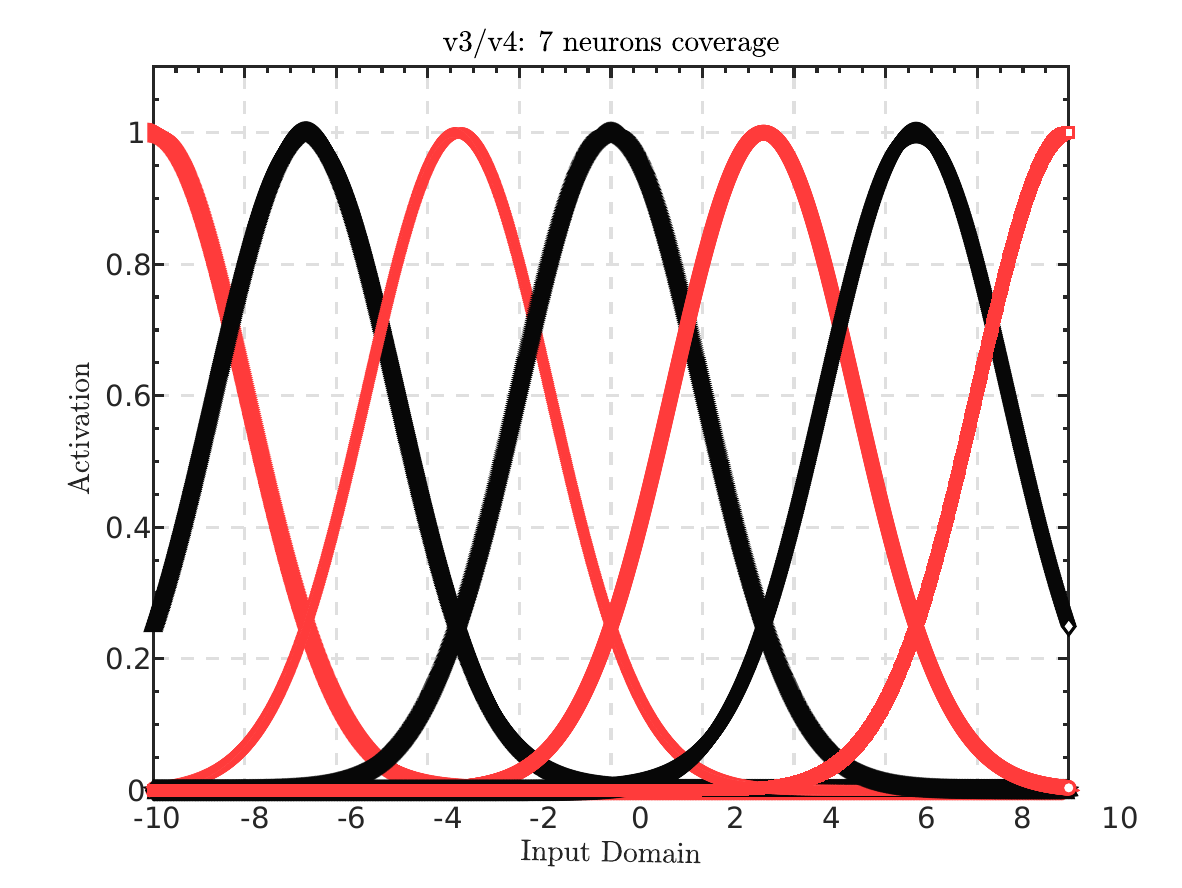}
        \caption{\footnotesize SANM.v3/v4 (7 neurons)}
        \label{fig:RBF_v3v4}
    \end{subfigure}

    \caption{\footnotesize Visualization of the RBF coverage densities for the SANM variants used in \textit{Experiment 2}. Each subfigure shows the Gaussian activation functions along the angular-velocity–error input domain $[-10,10]$, and illustrates how, within the same input domain and under identical kernel widths, the number of neurons determines the coverage density of the RBF kernels used in each neural network \textit{``slice"}.}
    \label{fig:RBF-coverage}
\end{figure}

\begin{table}[ht]
\centering
\caption{Summary of Variables for SANM Variants (v1--v4) in \textit{Experiment 2}.}
\label{tab:SANM_variants}
\renewcommand{\arraystretch}{1.2}
\begin{tabular}{c|c|c|c|c}
\hline
\textbf{Variant} & \textbf{Neurons} & \textbf{Coverage Density} &
\textbf{Learning Rate $\gamma_{\bm{\textit{R}}j}$} & \textbf{Notes} \\ 
\hline
\textbf{v1} & 3  & Sparse (widely spaced RBF centers) & Low $\{35,35,10\}$ & Baseline sparse configuration \\
\hline
\textbf{v2} & 9  & Dense (high kernel overlap) & Low $\{35,35,10\}$ & Tests effect of excessive overlap \\
\hline
\textbf{v3} & 7  & Moderate (balanced spacing) & Low $\{35,35,10\}$ & Sweet-spot density candidate \\
\hline
\textbf{v4} & 7  & Moderate (balanced spacing) & High $\{120,120,50\}$ & Effect of learning rate on error ball \\
\hline
\end{tabular}
\end{table}

This subsection provides a detailed summary of the parameter configurations used for all SANM variants (v1–v4). 
The Table~\ref{tab:SANM_variants} summarizes the sensitivity variables for each variant.
As described in the main text, the adaptive-law slices were intentionally disabled in these variants so that disturbance identification relied solely on the neural-network slices.
The variants differ only in their neural-network structures and adaptation-related parameters. Specifically, the number of neurons (coverage densities) and the learning rates. Other parameters unrelated to the variant factors, such as the widths, weight bound and dead-zone settings, are summarized as follows.
\begin{equation}
{
   \begin{array}{cc}
    \textbf{widths:}\,\,&b_{11}=b_{21}=b_{31}=b_{41}=b_{51}=2,\\
   &b_{12}=b_{22}=b_{32}=b_{42}=b_{52}=2,\\
   &b_{13}=b_{23}=b_{33}=b_{43}=b_{53}=3,\\
 \textbf{dead-zone thresholds:}\,\,&\zeta_1=\zeta_2=\zeta_3=0.0005,\\
 \textbf{weight bound:}\,\,&\overset{\tiny \text{max}}{W}_{j}=500.\\
   \end{array}
}
\label{Parameters-in-Wind-disturbance-Experiment}
\end{equation}

In \textit{Experiment 2}, the variants v1–v3 were constructed as post-ablation sensitivity variants by modifying the RBF coverage density. This was achieved by assigning different numbers of neurons within the same input domain while keeping all RBF centers uniformly distributed and all kernel widths identical.
This design ensures that changes in locality arise solely from the neuron count: denser configurations (more neurons) yield stronger overlap among RBF kernels, whereas sparser configurations (fewer neurons) produce more localized responses.
These three variants therefore enable a controlled investigation of how coverage density alone influences the behavior of neural network \textit{``slices"} and the closed-loop stability under wind disturbances.
Variant v4, in contrast, adopts the same RBF density as v3 but employs a higher learning rate to examine the effect of the learning rate on convergence behavior.
The detailed center settings for each variant are listed as follows.
\begin{equation}
{
   \begin{array}{cc}
\,\,&\begin{bmatrix}\textbf{c}_{11},\textbf{c}_{21},\textbf{c}_{31}\end{bmatrix}
   =\begin{bmatrix}-1 & 0 & 1\\ -10 & 0 & 10\\\end{bmatrix},\\[0.6em]

\textbf{v1 (3 neurons):}&\begin{bmatrix}\textbf{c}_{12},\textbf{c}_{22},\textbf{c}_{32}\end{bmatrix}
   =\begin{bmatrix}-1 & 0 & 1\\ -10 & 0 & 10\\\end{bmatrix},\\[0.6em]

&\begin{bmatrix}\textbf{c}_{13},\textbf{c}_{23},\textbf{c}_{33}\end{bmatrix}
   =\begin{bmatrix}-1 & 0 & 1\\ -6 & 0 & 6\\\end{bmatrix},
   \notag
   \end{array}
}
\label{Parameters-in-Wind-disturbance-Experiment-v1}
\end{equation}
\begin{equation}
{
   \begin{array}{cc}
\,\,&\begin{bmatrix}
\textbf{c}_{11},\textbf{c}_{21},\textbf{c}_{31},\textbf{c}_{41},\textbf{c}_{51},
\textbf{c}_{61},\textbf{c}_{71},\textbf{c}_{81},\textbf{c}_{91}
\end{bmatrix}
=
\begin{bmatrix}
-1 & -0.75 & -0.5 & -0.25 & 0 & 0.25 & 0.5 & 0.75 & 1 \\
-10 & -7.5 & -5 & -2.5 & 0 & 2.5 & 5 & 7.5 & 10
\end{bmatrix},
\\[0.9em]

\textbf{v2 (9 neurons):}&
\begin{bmatrix}
\textbf{c}_{12},\textbf{c}_{22},\textbf{c}_{32},\textbf{c}_{42},\textbf{c}_{52},
\textbf{c}_{62},\textbf{c}_{72},\textbf{c}_{82},\textbf{c}_{92}
\end{bmatrix}
=
\begin{bmatrix}
-1 & -0.75 & -0.5 & -0.25 & 0 & 0.25 & 0.5 & 0.75 & 1 \\
-10 & -7.5 & -5 & -2.5 & 0 & 2.5 & 5 & 7.5 & 10
\end{bmatrix},
\\[0.9em]

&
\begin{bmatrix}
\textbf{c}_{13},\textbf{c}_{23},\textbf{c}_{33},\textbf{c}_{43},\textbf{c}_{53},
\textbf{c}_{63},\textbf{c}_{73},\textbf{c}_{83},\textbf{c}_{93}
\end{bmatrix}
=
\begin{bmatrix}
-1 & -0.75 & -0.5 & -0.25 & 0 & 0.25 & 0.5 & 0.75 & 1 \\
-6 & -4.5 & -3 & -1.5 & 0 & 1.5 & 3 & 4.5 & 6
\end{bmatrix},
\notag
   \end{array}
}
\label{Parameters-in-Wind-disturbance-Experiment-v2}
\end{equation}

\begin{equation}
{
   \begin{array}{cc}
\,\,&
\begin{bmatrix}
\textbf{c}_{11},\textbf{c}_{21},\textbf{c}_{31},\textbf{c}_{41},\textbf{c}_{51},\textbf{c}_{61},\textbf{c}_{71}
\end{bmatrix}
=
\begin{bmatrix}
-1 & -0.6667 & -0.3333 & 0 & 0.3333 & 0.6667 & 1 \\
-10 & -6.6667 & -3.3333 & 0 & 3.3333 & 6.6667 & 10
\end{bmatrix},
\\[1.0em]

\textbf{v3,v4 (7 neurons):}&
\begin{bmatrix}
\textbf{c}_{12},\textbf{c}_{22},\textbf{c}_{32},\textbf{c}_{42},\textbf{c}_{52},\textbf{c}_{62},\textbf{c}_{72}
\end{bmatrix}
=
\begin{bmatrix}
-1 & -0.6667 & -0.3333 & 0 & 0.3333 & 0.6667 & 1 \\
-10 & -6.6667 & -3.3333 & 0 & 3.3333 & 6.6667 & 10
\end{bmatrix},
\\[1.0em]

&
\begin{bmatrix}
\textbf{c}_{13},\textbf{c}_{23},\textbf{c}_{33},\textbf{c}_{43},\textbf{c}_{53},\textbf{c}_{63},\textbf{c}_{73}
\end{bmatrix}
=
\begin{bmatrix}
-1 & -0.6667 & -0.3333 & 0 & 0.3333 & 0.6667 & 1 \\
-6 & -4 & -2 & 0 & 2 & 4 & 6
\end{bmatrix}.
\notag
   \end{array}
}
\label{Parameters-in-Wind-disturbance-Experiment-v3v4}
\end{equation}

From these settings, it can be observed that when the RBF centers are uniformly distributed, keeping the kernel widths identical across neurons implies that, for the same input domain, different numbers of neurons naturally lead to different coverage densities, as shown in Fig.~\ref{fig:RBF-coverage}.
In other words, increasing the number of neurons results in a denser overlap of the RBF kernels, whereas fewer neurons produce a sparser coverage pattern.
This structural difference directly affects the locality of the neural-network \textit{``slices"}, which is a key factor examined in the comparison studies in \textit{Experiment 2}.

\section{Extended Information for Experiment 3}
\label{supp:sec:Experiment-3}

\begin{figure}[ht]
      \centering
      \includegraphics[scale=0.32]{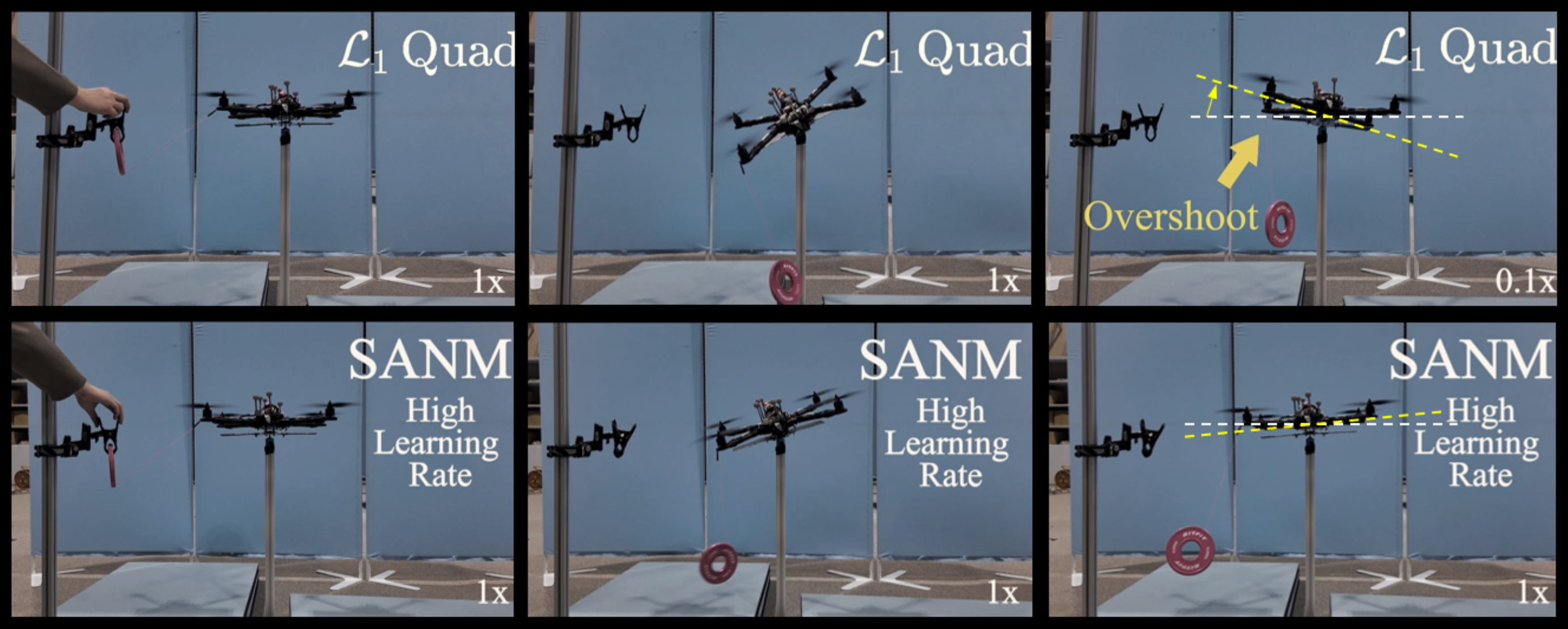}
      \caption{\footnotesize Time-sequenced snapshots from the real-world impact experiment, illustrating the non-overshoot exponential convergence achieved by the SANM-augmented geometric attitude controller.  In contrast, the $\mathcal{L}_1$-adaptive geometric controller exhibits overshoot, as $\mathcal{L}_1$-adaptive geometric control does not provide exponential convergence guarantees under time-varying disturbances. For the full video demonstration, refer to the supplementary video: \url{https://youtu.be/kDE5079TgCI}. }
      \label{fig:OVERSHOOT}
\end{figure}

\begin{table}[ht]
\centering
\caption{Summary of Variables for SANM Variants (v5--v7) in \textit{Experiment 3}.}
\label{tab:SANM_variants_exp3}
\renewcommand{\arraystretch}{1.2}
\begin{tabular}{c|c|c|c|c}
\hline
\textbf{Variant} & \textbf{Neurons} & \textbf{Coverage Density} &
\textbf{Learning Rate $\gamma_{\bm{\textit{R}}j}$} & \textbf{Notes} \\ 
\hline
\textbf{v5} & 7 & Moderate (same as v4) & $\{35,35,10\}$ (Low) 
& Baseline learning-rate setting \\ 
\hline
\textbf{v6} & 7 & Moderate  & $\{80,80,30\}$ (Medium) 
& Tests effect of increased learning rate \\ 
\hline
\textbf{v7} & 7 & Moderate & $\{120,120,50\}$ (High) 
& Effect of High learning rate on convergence rate \\ 
\hline
\end{tabular}
\end{table}

This section provides additional details for \textit{Experiment 3}, which evaluates the transient-response performance of the proposed SANM-augmented geometric attitude controller under real-world impact disturbances. As described in the main text, the experiment was conducted on a testbed where the position dynamics were isolated so that only the rotational subsystem was subject to the disturbance induced by a $0.25\,\mathrm{kg}$ payload drop attached to one arm of the quadrotor. This setup produced a sharp, time-varying nonlinear disturbance moment, allowing a controlled assessment of the anti-impact robustness of SANM.

Three variants of SANM (v5–v7) were designed as variant groups for this experiment. Unlike \textit{Experiment 2}, where the sensitivity variable was the RBF coverage density, all variants in \textit{Experiment 3} used the same neural network configuration: each slice
contained 7 RBF neurons with identical centers, widths, and bounds as given in \eqref{Parameters-in-Wind-disturbance-Experiment}, resulting in the same coverage density across variants. The adaptive-law \textit{``slices"}
were fully enabled and adaptive rates were kept identical $\{1/\eta_{j}\}_{j=1,2,3}:=\{10,10,10\}$ in v5–v7. Therefore, the only variant factor
was the learning-rate vector $\{\gamma_{\bm{\textit{R}} j}\}_{j=1,2,3}$, which was
varied across the three variants to examine its influence on the convergence rate under impact disturbances, which act as rapidly time-varying nonlinear disturbance inputs.

Table~\ref{tab:SANM_variants_exp3} summarizes the controlled differences among the three variants. The variant v5 uses the baseline learning rate configuration, whereas v6 and v7 progressively increase the learning rates to accelerate the response. This design allows a clean comparison of how the learning rate influences the disturbance-rejection performance without modifying any other structural component of SANM.

As shown in Fig.~\ref{fig:OVERSHOOT}, the SANM-augmented controller maintained a high-damping, non-overshoot transient response, consistent with the theoretically established exponential convergence property under time-varying disturbances (Supp.~\ref{supp:extended-proofs}). By contrast, the $\mathcal{L}_1$-adaptive geometric controller exhibited clear overshoot during recovery from the impact disturbance, which is consistent with the fact that the $\mathcal{L}_1$-adaptive geometric control does not provide guaranties of exponential convergence in the presence of time-varying disturbances.

The variant comparison among v5–v7 further shows that increasing the learning rate monotonically accelerates the convergence of the attitude error after impact. This observation is consistent with the theoretical result that the exponential decay rate of the error dynamics scales with the learning rates, provided that the estimation errors remain uniformly bounded. Together, these data confirm that the SANM plays a critical role in shaping the transient recovery behavior under impact disturbances, and that SANM exhibits superior robustness and non-overshoot convergence compared with benchmark geometric controllers.

\section{Extended Information for Experiment 4}
\label{supp:sec:Experiment-4}
In this experiment, the parameters of the adaptive law \textit{``slices"} in SANM were selected as
\begin{equation}
{
   \begin{array}{cc}
   &\eta_{1}=0.01,\, \eta_{2}=0.01,\, \eta_{3}=0.05, 
\, c_R=0.6,\\
&\mathfrak{s}_{1}=0.02, \,\mathfrak{s}_{2}=0.02, \,\mathfrak{s}_{3}=0.02, \,\\
   &\overset{\tiny \text{max}}{J}_{1}=0.03, \overset{\tiny \text{max}}{J}_{2}=0.03, \overset{\tiny \text{max}}{J}_{3}=0.04. 
   \notag
\end{array} 
}
\label{Parameters of slices}
\end{equation}
Each  neural network \textit{``slice"} employed a hidden layer with $l=5$ neurons and their parameters were selected as
\begin{equation}
{
   \begin{array}{cc}
\textbf{centers:}\,\,&\begin{bmatrix}\textbf{c}_{11},\textbf{c}_{21},\textbf{c}_{31},\textbf{c}_{41},\textbf{c}_{51}\end{bmatrix}=\begin{bmatrix}-1&-0.5&0&0.5&1\\ -10&-5&0&5&10\\\end{bmatrix},\\
&\begin{bmatrix}\textbf{c}_{12},\textbf{c}_{22},\textbf{c}_{32},\textbf{c}_{42},\textbf{c}_{52}\end{bmatrix}=\begin{bmatrix}-1&-0.5&0&0.5&1\\ -10&-5&0&5&10\\\end{bmatrix},\\
&\begin{bmatrix}\textbf{c}_{13},\textbf{c}_{23},\textbf{c}_{33},\textbf{c}_{43},\textbf{c}_{53}\end{bmatrix}=\begin{bmatrix}-1&-0.5&0&0.5&1\\ -6&-3&0&3&6\\\end{bmatrix},\\
    \textbf{widths:}\,\,&b_{11}=b_{21}=b_{31}=b_{41}=b_{51}=2,\\
   &b_{12}=b_{22}=b_{32}=b_{42}=b_{52}=2,\\
   &b_{13}=b_{23}=b_{33}=b_{43}=b_{53}=3,\\
 \textbf{dead-zone thresholds:}\,\,&\zeta_1=\zeta_2=\zeta_3=0.0005,\\
 \textbf{weight bound:}\,\,&\overset{\tiny \text{max}}{W}_{j}=500,\\
\textbf{learning rates:}\,\,&\gamma_{\bm{\textit{R}}1}=120,\,\gamma_{\bm{\textit{R}}2}=120,\,\gamma_{\bm{\textit{R}}3}=50.
   \notag
   \end{array}
}
\label{Parameters of slices2}
\end{equation}

The quadrotor started from rest on the ground with zero initial attitude and angular velocity errors. The weights of the neural network were initialized to zero and the estimated inertia feature vector was initialized to $\bm{\bar{J}}^{\text{vec}}(0)=10^{-2}(1,2,2)^{\top}~\mathrm{kg}\mathrm{m}^{2}$. 

\section{Extended Information for Experiment 5}
\label{supp:sec:Experiment-5}
In this experiment, most of the parameters were inherited from \textit{Experiment 4}. However, due to the sim-to-real gap, such as differences in motor thrust and response, the PD gains were reduced to  $k_{R}=40$, $ k_{\Omega}=80$ and some parameters of \textit{``slices"} were adjusted as follows:
\begin{equation}
{\small
   \begin{array}{cc}
   &\eta_{1}=0.05,\, \eta_{2}=0.05,\, \eta_{3}=0.05, 
\,\\
&\gamma_{\bm{\textit{R}}1}=80,\,\gamma_{\bm{\textit{R}}2}=80,\,\gamma_{\bm{\textit{R}}3}=50.
   \notag
\end{array} 
}
\label{Parameters changed}
\end{equation}
In addition, the estimated inertia feature vector was initialized to $\bm{\bar{J}}^{\text{vec}}(0)=10^{-2}(1,1,2)^{\top}~\mathrm{kg}\mathrm{m}^{2}$ to reflect the centrosymmetric X-configuration of the real quadrotor.

\section{Guidance on Parameter Tuning}
\label{supp:Guidance on Parameter Tuning}
\subsection{Parameter Tuning Guidelines}
\subsubsection{Tuning the baseline geometric PD controller}
Begin by tuning the \textbf{proportional–derivative (PD) gains} ${k_{R}, k_{\Omega}}$ with SANM disabled.
The objective is to obtain a stable and non-oscillatory attitude response under nominal conditions.
A well-tuned PD controller ensures that the closed-loop dynamics provide a reliable foundation for the SANM adaptation that will be enabled later.

\subsubsection{Tuning the neural network \textit{``slices"}}

After the PD gains are fixed, enable SANM and start by activating only the neural network \textit{``slices"}.
Next, following the constraint derived in the stability proof (see Eq.~\eqref{cR bound}), select a sufficiently small $c_R$ that satisfies
\begin{equation}
{
    \begin{aligned}
    c_R\!< \min\left\{\frac{k_Rk_{\Omega}}{k_{\Omega}^2+k_R},\sqrt{k_R},\sqrt{\frac{2k_{R}}{2-\psi_{\textit{R}}}},k_\Omega\right\},
\end{aligned}
}
\notag
\end{equation}
where $c_R$ determines the relative weighting of the attitude error within the neural-network inputs.
Then, use a small initial learning rate, such as $\{10,10,5\}$, to avoid overly aggressive adaptation.
With this learning rate fixed, select an appropriate \textbf{RBF coverage density}. A practical approach is to choose 5 or 7 neurons and distribute their centers uniformly over the input domain. Then adjust the kernel width to identify a suitable coverage density (\textit{``sweet spot”}).
This step follows the same principles discussed in the extended information of \textit{Experiment 2} (Supp.~\ref{supp:sec:Experiment-2}).
To assist this process, it is useful to monitor the neural network output $\bm{\bar{\phi}}_{\bm{\textit{R}}}^{[j]}$: pronounced oscillations typically indicate overly sparse coverage (excessive localization), whereas amplified or diverging behavior suggests excessively dense coverage that suppresses local feature extraction.

\textbf{\textit{Remark \ref{supp:Guidance on Parameter Tuning}.1:}}
Since \textit{Sliced Learning} adopts a \textit{learning-from-error} strategy, it benefits from \textbf{\textit{$\mathbf{SO}(3)$-Preserving}}. The suitable coverage density is easy to identify in practice. Even when the RBFs are too sparse, the learning process does not diverge. Instead, it may exhibit localized oscillation due to insufficient overlap, which provides a clear diagnostic signal for adjusting the width or number of neurons.

After identifying a suitable coverage density, observe whether the network weights exhibit drift.
Such drift may arise from sensor biases or small model mismatches and can accumulate during long-term operation.
If persistent weight drift is detected, consider increasing the \textbf{dead-zone thresholds} $\zeta_j$ to suppress unnecessary updates driven by very small errors.
Additionally, if it is desirable to limit the maximum neural network output, the \textbf{weight bound} $\overset{\tiny\text{max}}{W}_{j}$ can be reduced to constrain the overall output magnitude of the neural network.

\textbf{\textit{Remark \ref{supp:Guidance on Parameter Tuning}.2 (Selection of the Dead-Zone Threshold \(\zeta_j\)):}}
The dead-zone threshold \(\zeta_j\) is mainly introduced to suppress unnecessary neural-network updates caused by small bias-induced error offsets near the equilibrium, such as those originating from gyroscope zero bias or minor sensor mismatch. Since the proposed SANM adopts a \textit{learning-from-error} strategy, \(\zeta_j\) is not intended to play a dominant role in shaping the essential closed-loop stability, but rather to prevent slow weight drift in the near-equilibrium region.
In practice, \(\zeta_j\) should be selected slightly above the measured or observed sensor-bias-induced error level. This ensures that the dead-zone suppresses bias-driven drift while still allowing normal disturbance-learning behavior outside the near-equilibrium region. At the same time, \(\zeta_j\) should remain sufficiently small relative to the practical residual error ball around the attitude equilibrium, so that it does not interfere with the stability behavior established in the theoretical analysis.
For example, in our experiments, the sensor-bias-induced attitude-error offset was observed to be approximately \(0.0003\), and we selected \(\zeta_j=0.0005\), which is slightly above this bias level while remaining significantly smaller than the practical residual error ball around the equilibrium. Only when the UAV hardware is not properly calibrated, leading to a relatively large IMU zero bias, may a significantly larger dead-zone be required to suppress drift. In such cases, its influence on the local tracking accuracy should be considered more carefully.

Finally, following the procedure used for variants v5–v7 in \textit{Experiment 3}, gradually increase the \textbf{learning rates} until a suitable magnitude is reached.
It is important to note that faster convergence is not always better.
The rotational dynamics of the three axes $\{\bm{\vec{b}}_{j}\}_{j=1,2,3}$ (corresponding to roll, pitch, and yaw) often have different practical priorities because a quadrotor allocates control moments through motor-speed allocation rather than independent actuators.
For example, under the impact disturbance in \textit{Experiment 3}, the quadrotor should preferentially allocate motor authority to the $\bm{\vec{b}}_{1}$ (roll) and $\bm{\vec{b}}_{2}$ (pitch) axes to rapidly restore a stable attitude.
In such cases, the $\bm{\vec{b}}_{3}$ (yaw)-axis learning rate should be set smaller, giving it lower priority in moment allocation.
Conversely, in applications such as consumer-grade flying cameras, yaw stability is more critical for maintaining consistent heading and smooth image capture. Therefore, a higher $\bm{\vec{b}}_{3}$ (yaw)-axis learning rate may be preferred to enhance disturbance rejection in that channel.

\textbf{\textit{Remark \ref{supp:Guidance on Parameter Tuning}.3:}}
It is worth noting that increasing the learning rate accelerates the convergence rate, but for a discrete-time system such as a quadrotor, the convergence rate cannot be arbitrarily high.
Therefore, the learning rates must be chosen carefully to ensure that the resulting closed-loop behavior remains within the allowable bounds established in the theoretical analysis (see \eqref{eq:h_tau_bounds}).

\subsubsection{Tuning the adaptive law \textit{``slices"}}
The adaptive-law \textit{``slices”} primarily act as a global gain amplifier for the overall controller response.
Under non-extreme disturbances, these terms can often remain disabled.
Unlike the neural network \textit{``slices"}, the adaptive laws do not possess universal approximation capability, and thus their updates should remain smooth and gradual.
To avoid overly aggressive adaptation, ensure that the \textbf{adaptive rates} $\{1/\eta_{j}\}_{j=1,2,3}$ are not set too high, which would otherwise cause abrupt variations in the estimated parameters.

After identifying a suitable adaptive rate, configure the \textbf{pull-back factors} $\{\mathfrak{s}_{j}\}_{j=1,2,3}$ to ensure that the adaptive estimates are always driven back once they reach their prescribed upper bounds. This prevents drift and maintains boundedness of the adaptive parameters throughout flight. In addition, the \textbf{adaptive upper bounds} $\{\overset{\tiny \text{max}}{J}_{j}\}_{j=1,2,3}$ should not be set excessively large. Otherwise, the adaptive law \textit{``slice"} may dominate the SANM module and overshadow the contribution of the neural network \textit{``slice"}. In practice, the adaptive law is intended only as an auxiliary component, primarily for compensating slowly varying or quasi-static inertia-related deviations. Whereas the neural network \textit{``slice"} identifies the dominant time-varying disturbance features at the acceleration level, it can in principle also compensate for inertia-related deviations.
However, the adaptive law serves only as a supplementary mechanism that offloads part of these quasi-static inertia effects from the neural network, preventing them from being absorbed into the general disturbance representation.

\end{supplementary}

\end{document}